%% file: EOT-TSP-2021-V19.tex
\documentclass[10pt, twoside, romanappendices]{IEEEtran}
\usepackage{amsmath}
\usepackage{amssymb}
\usepackage{graphicx}
\usepackage{color}
\usepackage{cite}
\usepackage{bm}
\usepackage{epsfig,psfrag}
\usepackage{tabularx}
\usepackage{acronym}
\usepackage[yyyymmdd,hhmmss]{datetime}
\usepackage{bbm}
\usepackage{notation}
\usepackage{mathrsfs} 
\usepackage{tabularx}
\usepackage{multirow}
\usepackage{colortbl,pgfplotstable}
\usepackage{subfig}
\usepackage{multirow}
\usepackage[linesnumbered,ruled,vlined]{algorithm2e}

\allowdisplaybreaks
\newcolumntype{L}[1]{>{\raggedright\arraybackslash}p{#1}}
\newcolumntype{C}[1]{>{\centering\arraybackslash}p{#1}}
\newcolumntype{R}[1]{>{\raggedleft\arraybackslash}p{#1}}
\definecolor{BLUE}{rgb}{0,0,1}

\definecolor{niceblue}{RGB}{0,153.7,255}

\providecommand{\rd}{\textcolor{black}}

\providecommand{\ist}{\hspace*{.3mm}}
\providecommand{\rmv}{\hspace*{-.3mm}}
\providecommand{\iist}{\hspace*{1mm}}
\providecommand{\rrmv}{\hspace*{-1mm}}
\providecommand{\nn}{\nonumber}
\providecommand{\sist}{\hspace*{.15mm}}
\newcommand{\T}{\text{T}}

\acrodef{pnt}[PNT]{positioning, navigation and timing}
\acrodef{d2d}[D2D]{device-to-device}
\acrodef{aoa}[AOA]{angle-of-arrival}
\acrodef{nls}[NLS]{network localization and synchronization}
\acrodef{bp}[BP]{belief propagation}
\acrodef{spbp}[SPBP]{sigma point belief propagation}
\acrodef{fg}[FG]{factor graph}
\acrodef{pdf}[PDF]{probability density function}
\acrodef{mmse}[MMSE]{minimum mean-square error}
\acrodef{sp}[SP]{sigma point}
\acrodef{trwspa}[TRW-SPA]{tree-reweighed sum-product algorithm}
\acrodef{armse}[ARMSE]{average root mean square error}
\acrodef{uwb}[UWB]{ultra-wide bandwidth}
\acrodef{gnss}[GNSS]{global navigation satellite system}
\acrodef{spa}[SPA]{sum-product algorithm}
\acrodef{trw}[TRW]{tree-reweighed}
\acrodef{eap}[EAP]{edge appearance probability}
\acrodef{pmf}[PMF]{probability mass function}
\acrodef{eop}[EOP]{error outage probability}
\acrodef{ospa}[OSPA]{optimal subpattern assignment}
\acrodefplural{eap}[EAPs]{edge appearance probabilities}
\acrodef{po}[PO]{potential object}
\acrodef{spada}[SPADA]{sum-product algorithm for data association}
\acrodef{pda}[PDA]{probabilistic data association}
\acrodef{jpda}[JPDA]{joint probabilistic data association}
\acrodef{phd}[PHD]{probability hypothesis density}
\acrodef{cphd}[CPHD]{cardinalized \ac{phd}}
\acrodef{mht}[MHT]{multi-hypothesis tracking}
\acrodef{rfs}[RFS]{random finite sets}
\acrodef{slam}[SLAM]{simultaneous localization and mapping}
\acrodef{iid}[iid]{independent and identically distributed}
\acrodef{radar}[RADAR]{radio detection and ranging}
\acrodef{mot}[MOT]{multiobject tracking}
\acrodef{eot}[EOT]{extended object tracking}

\DeclareMathAlphabet{\mathpzc}{OT1}{pzc}{m}{it}

\newdateformat{monthyeardate}{\monthname[\THEMONTH] \THEDAY, \THEYEAR} 

\newcommand{\paperTitle}{Scalable Detection and Tracking\\ of Geometric Extended Objects}

\pgfplotsset{compat=1.14}

\begin{document}
\title{\paperTitle \vspace*{0mm}}

\author{\normalsize Florian Meyer,~\IEEEmembership{Member,~IEEE} and Jason L. Williams,~\IEEEmembership{Senior Member,~IEEE}  \thanks{
%

This work was supported in part by the University of California San Diego and in part by the Office of Naval Research under Grant N00014-21-1-2267. This work was presented at the IEEE ICASSP-20, Barcelona, Spain, May 2020.}

\thanks{Florian~Meyer is with the Scripps Institution of Oceanography and the Department of Electrical and Computer Engineering, University of California San Diego, La Jolla, CA, USA (e-mail: \texttt{flmeyer@ucsd.edu}).}

\thanks{Jason~L.~Williams is with Data61, Commonwealth Scientific and Industrial Research Organisation, Brisbane, Australia (e-mail: \texttt{jason.} \texttt{williams@data61.csiro.au}).}




\vspace*{-1mm}
}

\maketitle

\begin{abstract}
\rd{Multiobject tracking provides situational awareness that enables new applications for modern convenience, public safety, and homeland security. This paper presents a factor graph formulation and a particle-based \ac{spa} for scalable detection and tracking of extended objects. The proposed method dynamically introduces states of newly detected objects, efficiently performs probabilistic multiple-measurement to object association, and jointly infers the geometric shapes of objects. Scalable \ac{eot} is enabled by modeling association uncertainty by measurement-oriented association variables and newly detected objects by a Poisson birth process. Contrary to conventional \ac{eot} methods, a fully particle-based approach makes it possible to describe different geometric object shapes. The proposed method can reliably detect, localize, and track a large number of closely-spaced extended objects without gating and clustering of measurements. We demonstrate significant performance advantages of our approach compared to the recently introduced Poisson multi-Bernoulli mixture filter. In particular, we consider a simulated scenarios with up to twenty closely-spaced objects and a real autonomous driving application where measurements are captured by a lidar sensor.}
\end{abstract}



\section{Introduction }
\label{sec:introduction}

\rd{Emerging sensing technologies and innovative signal processing methods will lead to new capabilities for autonomous operation in a wide range of problems such as applied ocean sciences and ground navigation. A technology that enables new capabilities in this context is \ac{eot}. \ac{eot} methods make it possible to detect and track an unknown number of objects with unknown shapes by using modern sensors such as high-resolution radar, sonar, and lidar \cite{GraBauReu:J17}.}

Due to the high resolution of these sensors, the point object assumption used in conventional multiobject tracking algorithms \cite{BarWilTia:B11,Mah:B07,MeyKroWilLauHlaBraWin:J18} is no longer valid. Objects have an unknown shape that must be inferred together with their kinematic state. In addition, data association is particularly challenging due to the overwhelmingly large number of possible association events \cite{GraFatSve:J19}. Even if only a single extended object is considered, the number of possible measurement-to-object association events scales combinatorially in the number of measurements. In the more realistic case of multiple extended objects, this ``combinatorial explosion'' of possible events is further exacerbated.

\subsection{State of the Art }
\label{sec:SoA}
 
An important aspect of \ac{eot} is the modeling and inference of object shapes. The shape of an object determines how the measurements that originate from the object are spatially distributed around its center. An important and widely used model \cite{SchReuWan:C15,GraLunOrg:J12,BeaReuGraVoVoSch:J16,GraFatSve:J19,XiaGraSve:C19} based on random matrix theory has been introduced in \cite{Koc:J08}. This model considers elliptically shaped objects. Here, unknown reflection points of measurements on extended objects are modeled by a random matrix that acts as the covariance matrix in the measurement model. This random extent state is estimated sequentially together with the kinematic state. \rd{The approach proposed in \cite{Koc:J08} introduces a closed-form update step for linear dynamics. It exploits that the Gaussian inverse Wishart distribution is a conjugate prior for linear Gaussian measurement models.}

 A major drawback of the original random matrix-based tracking method is that the driving noise variance in the state transition function of the kinematic state must be proportional to the extent of the object \cite{Koc:J08}. Furthermore, the orientation of the random matrix is constant. These limitations have been addressed by improved and refined random matrix-based extent models introduced in \cite{FelFraKoc:J11,GraOrg:J14a,YanBau:J19} and \rd{particle-based methods \cite{VivGraBraWil:J17}.} 
 A significant number of additional application-dependent object extent models have been proposed recently (see \cite{GraBauReu:J17} and references therein). The state-of-the-art algorithm for tracking an unknown number of objects with elliptical shape is the Poisson multi-Bernoulli mixture (PMBM) filter \cite{GraFatSve:J19}. The PMBM filter is based on a  model \cite{Wil:J15} where objects that have produced a measurement are described by a multi-Bernoulli probability distribution and objects that exist but have not yet produced a measurement yet are described by a Poisson distribution. This PMBM model is a conjugate prior with respect to the prediction and update steps of the extended object tracking problem \cite{GraFatSve:J19}. 

Important methods for the tracking of point objects include \ac{jpda} \cite{BarWilTia:B11}, \ac{mht} \cite{Rei:J79,Kur:B90,CorCar:J05}, and approaches based on \ac{rfs} \cite{Mah:B07,VoSinDou:J05,VoVoCan:J07,VoVoPhu:J14,Wil:J15}. Many of these point object tracking methods have been adapted for extended object tracking. In \cite{HabThaKirGriWak:C11}, a probabilistic data association algorithm for the tracking of a single object that can produce more than one measurement has been introduced. A multiple-measurement \ac{jpda} filter for the tracking of multiple objects has been independently developed in \cite{HamSveSanSor:J12} and \cite{HabThaThaMalKir:J13}. The method in \cite{SchReuWan:C15} combines a random matrix extent model with multisensor \ac{jpda} for the tracking of multiple extended objects. The computational complexity of these methods scales combinatorially in the number of objects and the number of measurements. They rely on gating, a suboptimal preprocessing technique that excludes unlikely data association events. Thus, these methods are only suitable in scenarios where objects produce few measurements and no more than four objects come into close proximity at any time \cite{MeyWin:J20}.

\rd{Another widely used approach to limit computational complexity for data association with extended objects is to perform clustering of spatially close measurements. Here, based on the assumption that all measurements in a cluster correspond to the same object, the majority of possible data association events is discarded in a suboptimal preprocessing step. Based on this approach, tracking methods} for objects that produce multiple measurements have been developed in the \ac{mht} \cite{CorCar:J18} and the \ac{jpda} \cite{GenVivBraSolAmi:J15,VivBra:J16} frameworks. Practical implementations of the update step of \ac{rfs}-based methods for \ac{eot} \cite{GraLunOrg:J12,BeaReuGraVoVoSch:J16,GraFatSve:J19,XiaGraSve:C19} including the one of the PMBM filter \cite{GraFatSve:J19} also rely on gating and clustering and suffer from a reduced performance if objects are in close proximity. \rd{An alternative approach is to perform data association with extended objects using sampling methods \cite{GraSveReuXiaFat:J18,BoeBauWirReu:C19}. Here, in scenarios with objects that are in proximity, a better estimation performance can be achieved compared to methods based on clustering. These methods based on sampling, however, also discard the majority of possible data association events, and thus eventually fail if the number of relevant events becomes too large \cite{KroMeyHla:J20}.}

An innovative approach to high-dimensional estimation problems is the framework probabilistic graphical models \cite{KolFri:B09}. \rd{In particular, the \ac{spa} \cite{KscFreLoe:01,YedFreWei:05,KolFri:B09} can provide scalable solutions to high-dimensional estimation problems. The \ac{spa} performs local operations through ``messages'' passed along the edges of a factor graph.} Many traditional sequential estimation methods such as the Kalman filter, the particle filter, and \ac{jpda} can be interpreted as instances of  \ac{spa}. In addition, \ac{spa} has led to a variety of new estimation methods in a wide range of applications \cite{FosMihIma:J99,RicUrb:J01,MeyEtzLiuHlaWin:J18}. For \ac{pda} with point objects, an \ac{spa}-based method referred to as the \ac{spada} is obtained by executing the \ac{spa} on a bipartite factor graph and simplifying the resulting \ac{spa} messages \cite{WilLau:J14,MeyKroWilLauHlaBraWin:J18,KroMeyHla:J20}. The computational complexity of this algorithm scales as the product of the number of measurements and the number of objects. Sequential estimation methods that embed the \ac{spada} to reduce computational complexity have been introduced for \ac{mot} \cite{Wil:J15,KroMeyHla:C16,MeyBraWilHla:J17,SolMeyBraHla:J19,LauWil:C16,ZhaMey:C21}, indoor localization \cite{LeiMeyMeiWitHla:C16,LeiMeyHlaWitTufWin:J19,MenMeyBauWin:JS19,LiLeiVenTuf:J21}, as well as simultaneous localization and tracking \cite{MeyWin:C18,MeyHliHla:J16,MeyLiuWin:C18}. 

\ac{spa} methods for probabilistic data association based on a bipartite factor graph have recently been investigated for the tracking of extended objects \cite{MeyZheWin:C19,MeyWin:J20,LiWeiCheWeiZha:C21}. Here, the number of measurements produced by each object is modeled by an arbitrary truncated \ac{pmf}. An \ac{spa} for data association with extended objects with a computational complexity that scales as the product of the number of measurements and the number of objects has been introduced \cite{MeyZheWin:C19,MeyWin:J20}. This method can track multiple objects that potentially generate a large number of measurements but was found unable to reliably determine the probability of existence of objects detected for the first time. An alternative method for scalable data association with extended objects has been developed independently and has been presented \vspace{-.5mm} in \cite{YanThoBau:C18,YanWolBau:C20}. \rd{These methods however either assume that the number of objects are known, or that a certain prior information of new object locations is available. They are thus unsuitable for scenarios with spontaneous object birth.}

\subsection{Contributions and Notations}

In this paper, we introduce a Bayesian particle-based \ac{spa} for scalable detection and tracking of an unknown number of extended objects. Object birth and the number of measurements generated by an object are described by a Poisson \ac{pmf}. The proposed method is derived on a factor graph that depicts the statistical model of the extended object tracking problem and  is based on a representation of data association uncertainty by means of measurement-oriented association variables. Contrary to the factor graph used in \cite{MeyZheWin:C19}, it makes it possible to reliably determine the existence of objects by means of the \ac{spa}. In our statistical model, the object extent is modeled as a  basic geometric shape that is represented by a positive semidefinite extent state. \rd{Contrary to conventional \ac{eot} algorithms with a random-matrix model, the proposed fully particle-based method is not limited to elliptical object shapes. A fully particle-based approach also makes it possible to consider a uniform distribution of measurements on the object extent. Furthermore, our method has a computational complexity that scales only quadratically in the number of objects and the number of measurements. It can thus accurately detect, localize, and track multiple closely-spaced extended objects that generate a large number of measurements.} The contributions of this paper are as \vspace{.1mm} follows.
\begin{itemize}
\item We derive a new factor graph for the problem of detection and tracking of multiple extended objects and introduce the corresponding \vspace*{.5mm}message passing equations.
\item We establish a particle-based scalable message passing algorithm for the detection and tracking of an unknown number of extended objects.
\item We demonstrate performance advantages in a challenging simulated scenario and by using real lidar measurements acquired by an autonomous vehicle.
\end{itemize}

This paper advances over the preliminary account of our method provided in the conference publication \cite{MeyWil:C20} by (i) simplifying the representation of data association uncertainty, (ii) including object extents in the statistical model, (iii) presenting a detailed derivation of the factor graph, (iv) establishing a particle-based implementation of the proposed method, (v) discussing scaling properties, (vi) demonstrating performance advantages compared to the recently proposed Poisson multi-Bernoulli mixture filter \cite{GraFatSve:J19} using synthetic and real measurements.

The probabilistic model established in this paper is closely related to the probabilistic model employed by the PMBM filter (cf.~\cite{MeyKroWilLauHlaBraWin:J18}). Our focus is on obtaining a factor graph formulation for which the SPA scales well, and on exploiting the accuracy and flexibility of a particle-based implementation.

\emph{Notation:} Random variables are displayed in sans serif, upright fonts, while their realizations are in serif, italic fonts. 
Vectors and matrices are denoted by bold lowercase and uppercase letters, respectively. For example, a random variable and its realization are denoted by $\rv x$ and $x$; a random vector and its realization 
by $\RV x$ and $\V x$; and a random matrix and its realization are denoted by $\RM{X}$ and $\M{X}$. 
Furthermore, ${\V{x}}^{\text T}$ denotes the transpose of vector $\V x$; 
$\propto$ indicates equality up to a normalization factor;
$f(\V x)$ denotes the \ac{pdf} of 
random  vector $\RV{x}$. $\Set{N}(\V{x}; \V{\mu},\M{\Sigma})$ denotes the Gaussian \ac{pdf}  (of random vector $\RV{x}$) with mean $\V{\mu}$ and covariance \vspace{0mm} matrix $\M{\Sigma}$, $\Set{U}(\V{y}; \Set{S})$ denotes the uniform \ac{pdf}   (of random vector $\RV{y}$) with support $\Set{S}$, and $\Set{W}\big(\M{X}; q, \M{Q} \big)$  denotes the Wishart distribution (of random matrix $\RM{X}$) with degrees of freedom $q$ and mean $q \ist \M{Q}$. The determinant of matrix $\M{Q}$ is denoted $|\M{Q}|$. Finally, $\mathrm{bdiag}(\M{M}_1, \dots, \M{M}_I)$ denotes the block diagonal matrix that consists of submatrices\vspace{0mm} $\M{M}_1, \dots, \M{M}_I$.

\section{System Model \vspace{-0mm} }
\label{sec:systemModel}

\begin{figure}
\centering
\vspace*{-5mm}
\subfloat[\label{fig:figure1a}]{\hspace{2mm}\scalebox{0.32}{\includegraphics[scale=2.2]{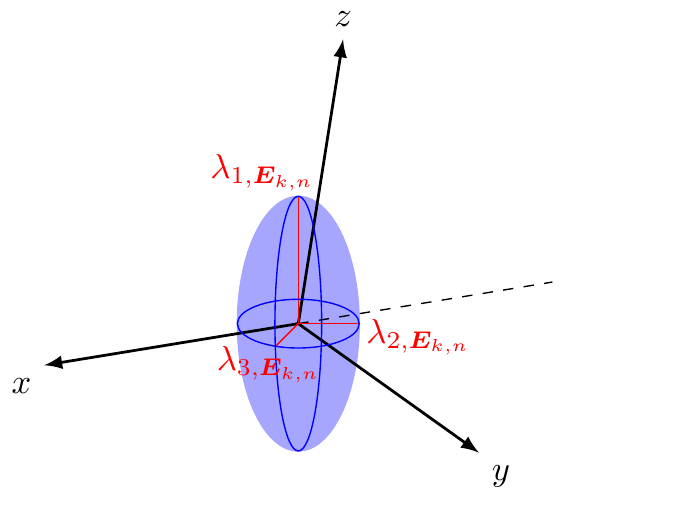}}}
\subfloat[\label{fig:figure1b}]{\hspace{-2mm}\scalebox{0.32}{\includegraphics[scale=2.2]{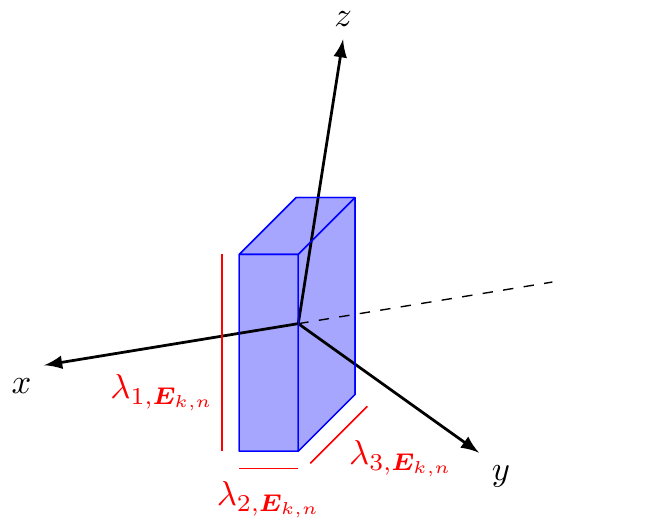}}}
\captionsetup{singlelinecheck = false, justification=justified}
\vspace{.5mm}
\caption{Object shapes represented by extent state $\M{E}_{k,n}$ in a 3-D scenario. An ellipsoid (a) or cube (b) is defined by the eigenvalues $\lambda_{1,\M{E}_{k,n}} > \lambda_{2,\M{E}_{k,n}} > \lambda_{3,\M{E}_{k,n}} \geq 0 $ of the $3 \times 3$ symmetric, positive-semidefinite matrix $\M{E}_{k,n}$. The local reference frame and thus the orientation of the object is defined by the eigenvectors $\protect\V{\lambda}_{1,\M{E}_{k,n}}$, $\protect\V{\lambda}_{2,\M{E}_{k,n}}$, and $\protect\V{\lambda}_{3,\M{E}_{k,n}}$ of $\M{E}_{k,n}$.}
\label{fig:figure1}
\end{figure}

At time $n$, object $k$ is described by a kinematic state and an extent state. The kinematic state $\RV{x}_{k,n} \triangleq \big[\RV{p}^{\T}_{k,n} \ist \RV{m}^{\T}_{k,n} \big]^{\T}$ consists of the object's position $\RV{p}_{k,n}  \rmv\in\rmv \mathbb{R}^{d_{\V{p}}}$ and possibly further parameters $\RV{m}_{k,n}  \in \mathbb{R}^{d_{\V{m}}}$ such as velocity or turn rate.  The extent state $\RM{E}_{k,n} \rmv\in\rmv \mathbb{R}^{d_{\V{p}} \rmv\times\rmv d_{\V{p}}}$ is a symmetric, positive semidefinite random matrix that can either model an ellipsoid or a cube. Example realizations of $\RM{E}_{k,n}$ and corresponding object shapes are shown in Fig.~\ref{fig:figure1}. Formally, we also introduce the vector notation $\RV{e}_{k,n}\rmv$ of extent state $\RM{E}_{k,n}\rmv$, which is the concatenation of diagonal and unique off-diagonal elements of $\RM{E}_{k,n}$, e.g., in a 2-D tracking scenario the vector notation of extent matrix $\RM{E}_{k,n} = \big[\ist [\rv{e}^{(11)}_{k,n}  \ist\ist\ist \rv{e}^{(21)}_{k,n}]^{\T} \ist\ist [\rv{e}^{(12)}_{k,n}  \ist\ist\ist \rv{e}^{(22)}_{k,n} ]^{\T} \big]$ is given by $\RV{e}_{k,n} = \big[ \rv{e}^{(11)}_{k,n}  \ist\ist\ist \rv{e}^{(21)}_{k,n}  \ist\ist\ist \rv{e}^{(22)}_{k,n}  \big]^{\T}$\rmv\rmv\rmv. \rd{Note that the support of $\RV{e}_{k,n}$ corresponds to all positive-semidefinite matrices in $\mathbb{R}^{d_{\mathbf{p}} \rmv\times\rmv d_{\mathbf{p}}}$.} In what follows, we will use extent matrix $\RM{E}_{k,n}$ and its vector notation $\RV{e}_{k,n}$ interchangeably. 

As in \cite{MeyBraWilHla:J17,MeyLiuWin:C18,MeyKroWilLauHlaBraWin:J18}, we account for the time-varying and unknown number of extended objects by introducing \acp{po} $k \in \{ 1,\dots, K_n \}$. \rd{The number of \acp{po} $K_n$ is the maximum possible number of actual objects that have produced a measurement so far \cite{MeyKroWilLauHlaBraWin:J18}. (Note that $K_n$ increases with time $n$.)} The existence/non-existence of \ac{po} $k$ is modeled by the binary existence variable $\rv{r}_{k,n} \rmv\in \{0,1\}$ in the sense that \ac{po} $k$ represents an actual object if and only if $r_{k,n} \!=\! 1$. Augmented \ac{po} states are denoted as $\RV{y}_{k,n} \rmv=\rmv [\RV{x}^{\T}_{k,n} \; \RV{e}^{\T}_{k,n} \;  \rv{r}_{k,n} ]^\T\rmv\rmv$.

Formally, \ac{po} $k$ is also considered if it is non-existent, i.e., if $r_{k,n} \!=\rmv 0$.  The states $\RV{x}_{k,n}$ of non-existent \acp{po} are obviously irrelevant and have no impact on the estimation solution. Therefore, all hybrid continuous/discrete \acp{pdf} defined on augmented \ac{po} states, $f(\V{y}_{k,n}) =\rmv f(\V{x}_{k,n}\rmv, \V{e}_{k,n}\rmv, r_{k,n})$, are of the form $f(\V{x}_{k,n}\rmv, \V{e}_{k,n}\rmv, 0 )$ $=\rmv f_{k,n} f_{\mathrm{d}}(\V{x}_{k,n},\V{e}_{k,n})$, where $f_{\mathrm{d}}(\V{x}_{k,n},\V{e}_{k,n})$ is an arbitrary ``dummy \ac{pdf}'' and $f_{k,n} \!\rmv\in [0,1]$ is the probability that the \ac{po} does not exist.\footnote{\rd{Note that representing objects states and existence variables by this type of \acp{pdf} is analogous to a multi-Bernoulli formulation in the \ac{rfs} framework \cite{Wil:J15,MeyKroWilLauHlaBraWin:J18,GraFatSve:J19}.}} For every \ac{po} state at previous times, there is one ``legacy'' \ac{po} state $\underline{\RV{y}}_{k,n}$ at current time $n$. The \vspace{0mm} number of legacy objects and the joint legacy \ac{po} state at time $n$ are \vspace{-.5mm} introduced as $\underline{K}_n \rmv=\rmv K_{n-1}$ and $\underline{\RV{y}}_n \rmv\triangleq\rmv \big[ \underline{\RV{y}}^{\T}_{1,n} \rmv\cdots\ist \underline{\RV{y}}^{T}_{\underline{K}_{n},n} \big]^{\T}\rmv\rmv$, respectively.


\subsection{State-Transition Model}
\label{sec:StateTransitionModel}

The \vspace{-.5mm} state-transition pdf for legacy \ac{po} state  $\underline{\RV{y}}_{n}$ factorizes as
\vspace{-.5mm}
\begin{equation}
f(\underline{\V{y}}_{n}|\V{y}_{n-1}) = \prod^{\underline{K}_{n}}_{k=1} f\big( \underline{\V{y}}_{k,n} \big| \V{y}_{k,n-1} \big)
\label{eq:stateTransitionU}
\end{equation}
where the single-object augmented state-transition pdf 
$f\big( \underline{\V{y}}_{k,n} \big| \V{y}_{k,n-1} \big) = f\big( \underline{\V{x}}_{k,n}\rmv, \underline{\V{e}}_{k,n}\rmv,  \underline{r}_{k,n} \big| \V{x}_{k,n-1}, \V{e}_{k,n-1}\rmv, r_{k,n-1}\big)$ is given as \vspace{.3mm} follows. 
If \ac{po} $k$ does not exist at time $n-1$, i.e., $\rv{r}_{k,n-1} \!=\rmv 0$, then it does not exist at time $n$ either, i.e., $\underline{\rv{r}}_{k,n} \!=\! 0$, and thus its state pdf is 
$f_{\mathrm{d}}\big( \underline{\V{x}}_{k,n},  \underline{\V{e}}_{k,n}\big)$.
This means \vspace{1.2mm} that 
\begin{align}
 &f\big( \underline{\V{x}}_{k,n},  \underline{\V{e}}_{k,n}, \underline{r}_{k,n} \big| \V{x}_{k,n-1}, \V{e}_{k,n-1}, r_{k,n-1} \!\rmv=\! 0 \big) \nn\\[2mm]
 &\hspace{4mm}= \begin{cases}
   f_{\mathrm{d}}\big( \underline{\V{x}}_{k,n},  \underline{\V{e}}_{k,n}\big) \ist, & \!\! \underline{r}_{k,n} \!\rmv=\! 0 \\[1mm]
   0 \ist, & \!\! \underline{r}_{k,n} \!\rmv=\! 1 . 
   \end{cases} \label{eq:singleTargetStateTrans_0} \\[-8.5mm]
   \nn\\
   \nn
\end{align}
On the other hand, if \ac{po} $k$ exists at time $n-1$, i.e., $\rv{r}_{k,n-1} \!=\! 1$, then the probability that it still exists at time $n$, i.e., $\underline{\rv{r}}_{k,n} \!=\! 1$, 
is given by the survival probability $p_{\mathrm{s}}$, and if it still exists at time $n$, its state $\underline{\RV{x}}_{k,n}$ is distributed according to a single-object
state-transition pdf $f\big( \underline{\V{x}}_{k,n}, \underline{\V{e}}_{k,n} \big| \V{x}_{k,n-1}, \V{e}_{k,n-1}\big)$. \vspace{1mm} Thus,
\begin{align}
&f\big( \underline{\V{x}}_{k,n},  \underline{\V{e}}_{k,n}, \underline{r}_{k,n} \big| \V{x}_{k,n-1}, \V{e}_{k,n-1}, r_{k,n-1} \!\rmv=\! 1 \big) \nn\\[1.5mm]
&\hspace{0mm}=  
   \begin{cases}
		\big( 1 \!\rmv-\rmv p_{\mathrm{s}} \big) \ist f_{\mathrm{d}}\big( \underline{\V{x}}_{k,n}, \underline{\V{e}}_{k,n} \big) \ist, & \!\!\underline{r}_{k,n} \!\rmv=\! 0 \\[1.5mm] 
		p_{\mathrm{s}} \ist f\big( \underline{\V{x}}_{k,n}, \underline{\V{e}}_{k,n} \big| \V{x}_{k,n-1}, \V{e}_{k,n-1}\big) \ist, & \!\!\underline{r}_{k,n} \!\rmv=\! 1.
   \end{cases} \label{eq:singleTargetStateTrans_1}\\[-.5mm]
    \nn\\[-9.5mm]
\nn
\end{align}
It is assumed that at time $n \rmv=\rmv 0$, the prior distributions $f(\V{y}_{k,0})$ are statistically independent across \acp{po} $k$. If no prior information is available, we have $K_0 \rmv=\rmv 0$.

A single-object state-transition pdf $f\big( \underline{\V{x}}_{k,n}, \underline{\V{e}}_{k,n} \big| \V{x}_{k,n-1},$ $\V{e}_{k,n-1}\big)$ that was found useful in many extended object tracking scenarios is given by \cite{GraOrg:J14a} 
\begin{align}
& f\big( \underline{\V{x}}_{k,n},  \underline{\V{e}}_{k,n} \big|\V{x}_{k,n-1}, \V{e}_{k,n-1} \big) \nn\\[2.5mm]
&= \Set{N}\big(\underline{\V{x}}_{k,n}; \mathpzc{f}(\V{x}_{k,n-1}), \M{\Sigma}_{k,n} \big) \ist \nn\\[1.5mm]
&\hspace{0mm}\times \Set{W}\Big(\underline{\M{E}}_{k,n}; q_{k,n},\frac{ \M{V}(\V{m}_{k,n-1}) \ist \M{E}_{k,n-1} \ist \M{V}(\V{m}_{k,n-1})^{\T}}{q_{k,n}}  \Big) \label{eq:stateTransition}
\end{align}
where $\mathpzc{f}(\V{x}_{k-1,n})$ is the state transition function of the kinematic state $\RV{x}_{k-1,n}$, $\M{\Sigma}_{k,n}$ is the kinematic driving noise covariance matrix, and $\M{V}(\V{m}_{k,n})$ is a rotation matrix. The degrees of freedom of the Wishart distribution determine the uncertainty of object extent prediction. \rd{A small $q_{k,n}$ implies a large state-transition \vspace{-3mm} uncertainty.}

\subsection{Measurement Model}
\label{sec:MeasurementModel}

\rd{At time $n$, a sensor produces measurements $\RV{z}_{l,n} \in \mathbb{R}^{d_{\V{z}}}$, $l \rmv\in \{1,$ $\dots,\mathsf{M}_n\}$. The joint measurement vector is denoted as $\RV{z}_n \triangleq [\RV{z}^{\T}_{1,n}$ $\cdots \ist \RV{z}^{\T}_{{\mathsf{M}_n,n}} ]^{\T}\rmv\rmv$.  (Note that the total number of measurements $\mathsf{M}_n$ is random.) Each measurement is either generated by a single object or is a false alarm.}

\rd{Let $\RV{z}_{l,n}$ be a measurement generated by the $k$th \ac{po}.\footnote{The fact that the \ac{po} with index $k$ generates a measurement implies $r_{k,n} \!=\! 1$.} In particular, let $\RV{v}^{(l)}_{k,n}$ be the relative position (with respect to $\RV{p}_{k,n}$) of the reflection point that generated $\RV{z}_{l,n}$. The considered general nonlinear measurement model is now given by
\begin{equation}
\RV{z}_{l,n} = \V{d}\big( \RV{p}_{k,n} \rmv+\rmv \RV{v}^{(l)}_{k,n}  \ist \big) + \RV{u}_{l,n} \label{eq:measModel}
\vspace{.4mm}
\end{equation}
where $\V{d}(\cdot)$ is an arbitrary nonlinear function and $\RV{u}_{l,n} \sim \mathcal{N}(\V{u}_{l,n} ;\V{0},\M{\Sigma}_{\V{u}_{l,n}})$ is the measurement noise. (Note that the subspace $\RV{m}_{k,n}$ remains unobserved.) This measurement model directly determines the conditional \ac{pdf} $f\big(\V{z}_{l,n} |\V{x}_{k,n}, \V{v}^{(l)}_{k,n}\big)$.}

We consider three geometric object shape models. The extent state $\RM{E}_{k,n} \rmv=\rmv \M{E}_{k,n}$ \vspace{-.2mm} either defines (i) the covariance matrix of a Gaussian \vspace{-.05mm} \ac{pdf}, i.e., $\RV{v}_{k,n}^{(l)} \sim \mathcal{N}\big(\V{v}^{(l)}_{k,n};\V{0},\M{E}^2_{k,n}\big) $; (ii) the elliptical support $\Set{S}_{\mathrm{e}}(\M{E}_{k,n})$ of a uniform \ac{pdf}, i.e., $\RV{v}_{k,n}^{(l)} \sim \mathcal{U}\big(\V{v}_{k,n}^{(l)};\Set{S}_{\mathrm{e}}(\M{E}_{k,n})\big)$; or (iii) the cubical support $\Set{S}_{\mathrm{c}}(\M{E}_{k,n})$ of a uniform \ac{pdf}, i.e., $\RV{v}_{k,n}^{(l)} \sim \mathcal{U}\big(\V{v}_{k,n}^{(l)};\Set{S}_{\mathrm{c}}(\M{E}_{k,n})\big)$\textcolor{red}{.}\footnote{The cubical support $\Set{S}_{\mathrm{c}}(\M{E}_{k,n})$ and the elliptical support $\Set{S}_{\mathrm{e}}(\M{E}_{k,n})$ are defined by the eigenvalues and eigenvectors of $\M{E}_{k,n}$ as shown in Fig~\ref{fig:figure1}. For example, let us \vspace{.2mm} consider a 3-D scenario and let $\lambda_{1,\M{E}_{k,n}} > \lambda_{2,\M{E}_{k,n}} > \lambda_{3,\M{E}_{k,n}} \geq 0$ be \vspace{0mm} the eigenvalues of the $3 \times 3$ symmetric, positive-semidefinite matrix $\M{E}_{k,n}$. Length, width, and height of the cube $\Set{S}_{\mathrm{c}}(\M{E}_{k,n})$ are given by $\lambda_{1,\M{E}_{k,n}}$, $\lambda_{2,\M{E}_{k,n}}$, and  $\lambda_{3,\M{E}_{k,n}}$, respectively. Similarly, the lengths of the three principal semi-axes of the ellipsoid $\Set{S}_{\mathrm{e}}(\M{E}_{k,n})$, are given by $\lambda_{1,\M{E}_{k,n}}$, $\lambda_{2,\M{E}_{k,n}}$, and $\lambda_{3,\M{E}_{k,n}}.$ The orientation of  $\Set{S}_{\mathrm{c}}(\M{E}_{k,n})$ and $\Set{S}_{\mathrm{e}}(\M{E}_{k,n})$ is determined by the eigenvectors $\V{\lambda}_{1,\M{E}_{k,n}}$,$\V{\lambda}_{2,\M{E}_{k,n}}$, and $\V{\lambda}_{3,\M{E}_{k,n}}$. The eigenvalues $\lambda_{1,\M{E}_{k,n}}\rmv$, $\lambda_{2,\M{E}_{k,n}}\rmv$, and $\lambda_{3,\M{E}_{k,n}}$ are assumed distinct. This assumption is always valid for real sensor data and objects of practical interest.} The considered object shape model directly determines the conditional \ac{pdf} $f\big(\V{v}^{(l)}_{k,n} | \V{e}_{k,n}\big)$. The \ac{pdf} of measurement vector $\RV{z}_{l,n}$ conditioned on $\RV{x}_{k,n}$ and $\RV{e}_{k,n}$ can now\vspace{0mm}  be obtained as
\begin{align}
&f(\V{z}_{l,n} | \V{x}_{k,n}, \V{e}_{k,n}) \nn \\[1mm]
&\hspace{8mm}= \rmv \int \rmv f\big(\V{z}_{l,n} |\V{x}_{k,n}, \V{v}^{(l)}_{k,n}\big) \ist f\big(\V{v}^{(l)}_{k,n} | \V{e}_{k,n}\big) \ist \mathrm{d} \V{v}^{(l)}_{k,n}. \label{eq:Likelihood1}
\end{align}

An important special case of this model is the additive-Gaussian \vspace{.8mm} case, i.e.,
\begin{equation}
\RV{z}_{l,n} = \RV{p}_{k,n} \rmv+\rmv \RV{v}^{(l)}_{k,n} + \RV{u}_{l,n}. \label{eq:measModelLin}
\vspace{.8mm}
\end{equation}
For extent model (i), in this case, there exist closed-form expressions for  \eqref{eq:Likelihood1}. In particular, the likelihood function $f(\V{z}_{l,n} | \V{x}_{k,n}, \V{e}_{k,n})$ can be expressed as $\mathcal{N}\big(\V{z}_{l,n};\V{p}_{k,n}, \M{E}^2_{k,n} + \M{\Sigma}_{\V{u}_{l,n}}\big)$. Similarly, for models (ii) and (iii), there exists a simple approximation discussed in \cite[Section~2]{MeyWil:SM21}, that also results in a closed-form of \eqref{eq:Likelihood1}. 

\rd{If \ac{po} $k$ exists ($\rv{r}_{k,n} \rmv=\rmv 1$) it generates a random number of object-originated measurements $\RV{z}_{l,n}$, which are distributed according to the conditional \ac{pdf} $f(\V{z}_{l,n}|\V{x}_{k,n}, \V{e}_{k,n})$ discussed above. The number of measurements it generates is Poisson distributed with mean $\mu_{\mathrm{m}}(\RV{x}_{k,n},\RV{e}_{k,n})$. For example, this Poisson distribution can be determined by a spatial measurement density $\rho$ related to the sensor resolution and by the volume or surface area of the object extent, i.e., $\mu_{\mathrm{m}}(\RV{x}_{k,n},\RV{e}_{k,n}) = \rho \ist | \RM{E}_{k,n} |$. False alarm measurements  are modeled by a Poisson point process with mean $\mu_{\mathrm{fa}}$ and strictly positive \ac{pdf} $f_{\mathrm{fa}}(\V{z}_{l,n})$.}

\subsection{New \acp{po}  \vspace{-.6mm}}
\label{sec:MeasNewPOs}

Newly detected objects, i.e., actual objects that generated a measurement for the first time, are modeled by a Poisson point process with mean $\mu_{\mathrm{n}}$ and \ac{pdf} $f_{\mathrm{n}}(\overline{\V{x}}_{k,n},\overline{\V{e}}_{k,n})$. Following \cite{Wil:J15,MeyKroWilLauHlaBraWin:J18,GraFatSve:J19}, newly detected objects are represented by new \ac{po} states $\overline{\RV{y}}_{k,n}$, $k \in \{1,\dots,\mathsf{M}_n \}$. Each new \ac{po} state corresponds to a measurement $\RV{z}_{k,n}$; $\overline{\rv{r}}_{k,n} \!=\! 1$ means that measurement $\RV{z}_{k,n}$  was generated by a newly detected object. Since newly detected objects can produce more than one measurement, we define a mapping from measurements to new \acp{po} by the following rule.\footnote{A detailed derivation and discussion is provided in the supplementary material \cite{MeyWil:SM21}.} At time $n$, if multiple measurements $l_1, \dots, l_L$ with $L \leq M_n$ are generated by the same newly detected object, we have $\overline{\rv{r}}_{k_{\mathrm{min}},n} \rmv=\rmv 1$ for $k_{\mathrm{min}} = \mathrm{min}(l_1, \dots, l_L)$ and $\overline{\rv{r}}_{k,n} \rmv=\rmv 0$ for all $k \in \big\{l_1,\dots,l_L\big\} \backslash \big\{k_{\mathrm{min}}\big\}$. As will be further discussed in  Section \ref{sec:DataAssocUncer}, with this mapping every association event related to newly detected objects can be represented by exactly one configuration of new existence variables $\overline{\rv{r}}_{k,n}$, $k \rmv\in\rmv\{1,\dots,\mathsf{M}_n\}$. We also introduce $\overline{\RV{y}}_n \triangleq [ \overline{\RV{y}}^{\T}_{1,n} \rmv\cdots\ist \overline{\RV{y}}^{\T}_{\mathsf{M}_n} ]^{\T}$ representing the joint vector of all new \ac{po} states.  Note that at time $n$, the total number of \acp{po} is given by \vspace{0mm} $\mathsf{K}_n\!=\underline{K}_n \!+\rmv \mathsf{M}_n$. 

Since new \acp{po} are introduced as new measurements are incorporated, the number of \ac{po} states would grow indefinitely. \rd{Thus, for the development of a feasible method, a suboptimal pruning step is employed. This pruning step removes unlikely \acp{po} and will be further discussed in Section \ref{sec:prob}.}

\subsection{Data Association Uncertainty \vspace{-.5mm}}
\label{sec:DataAssocUncer}

Tracking of multiple extended objects is complicated by the data association uncertainty: it is unknown which measurement $\RV{z}_{l,n}$ originated from which object $k$.  
To reduce computational complexity, following \cite{WilLau:J14,MeyBraWilHla:J17,MeyKroWilLauHlaBraWin:J18} we use \emph{measurement-oriented association variables}
\vspace{.5mm}
\[ 
\rv{b}_{l,n} \ist\triangleq \begin{cases} 
    k \rmv\in\rmv \{1,\dots,\underline{K}_n + l\} \ist , & \begin{minipage}[t]{40mm}if measurement $l$ is generated by \ac{po} $k$\end{minipage}\\[4mm]
   0 \ist, & \begin{minipage}[t]{40mm}if measurement $l$ is not generated by a \ac{po}\end{minipage}
  \end{cases}
\vspace{.5mm}
\]
and define the measurement-oriented association vector as $\RV{b}_n \rmv= [\rv{b}_{1,n} \ist \cdots \ist \rv{b}_{\mathsf{M}_n,n}]^{\mathrm{T}}\rmv\rmv$. This representation of data association makes it possible to develop scalable \acp{spa} for object detection and tracking. \rd{Note that the maximum value in the support of $\rv{b}_{l,n}$  is $\underline{K}_n + l$. (As discussed above, a measurement with index $l$ cannot be associated to a new PO with index $l' > l$, i.e., the event in which measurements $l'$ and $l$ are generated by the same newly detected object is represented through the new PO $l$.) This is a direct result of the mapping introduced in Section~\ref{sec:MeasNewPOs}.} In what follows, we write $\rv{b}_{l,n} \neq k$ short for $\rv{b}_{l,n} \rmv\in\rmv \{0,1,\dots,\underline{K}_n + l\} \rmv\backslash \{k\}$.

For a better understanding of new \acp{po} and measurement-oriented association vectors, we consider simple examples for fixed $M_n \rmv=\rmv 3$ and $\underline{K}_n \rmv=\rmv 0$. The event where all three measurements are generated by the same newly detected object, is represented by $\overline{r}_{1,n} \rmv=\rmv 1$, $\overline{r}_{2,n} \rmv=\rmv 0$, $\overline{r}_{3,n} \rmv=\rmv 0$, and $\V{b}_{n} \rmv=\rmv \big[\ist b_{1,n} \ist\ist b_{2,n}\ist\ist b_{3,n}\big]^{\mathrm{T}}  \rmv\rmv=\rmv [1\ist\ist1\ist\ist1]^{\mathrm{T}}\rmv\rmv$. Furthermore, the event where all three measurements are generated by three different newly detected objects, is represented by $\overline{r}_{1,n} \rmv=\rmv 1$,  $\overline{r}_{2,n} \rmv=\rmv 1$, $\overline{r}_{3,n} \rmv=\rmv 1$, and $\V{b}_{n} \rmv=\rmv [1\ist\ist2\ist\ist3]^{\mathrm{T}}\rmv\rmv$. Finally, the event where measurements $m \in \{2,3\}$ are generated by the same newly detected object and measurement $m = 1$ is a false alarm, is represented by $\overline{r}_{1,n} \rmv=\rmv 0$, $\overline{r}_{2,n} \rmv=\rmv 1$, $\overline{r}_{3,n} \rmv=\rmv 0$, and $\V{b}_{n} \rmv=\rmv [0\ist\ist2\ist\ist2]^{\mathrm{T}}\rmv\rmv$. Note how every event related to newly detected objects is represented by exactly one configuration of new existence variables $\overline{\rv{r}}_{n,k}$, $k \rmv\in\rmv \{1,2,3\}$ and association vector $\V{b}_{n}$.

\begin{figure}[t!]
\vspace{-2mm}
\centering
\hspace{-2.1mm}\includegraphics[scale=1.05]{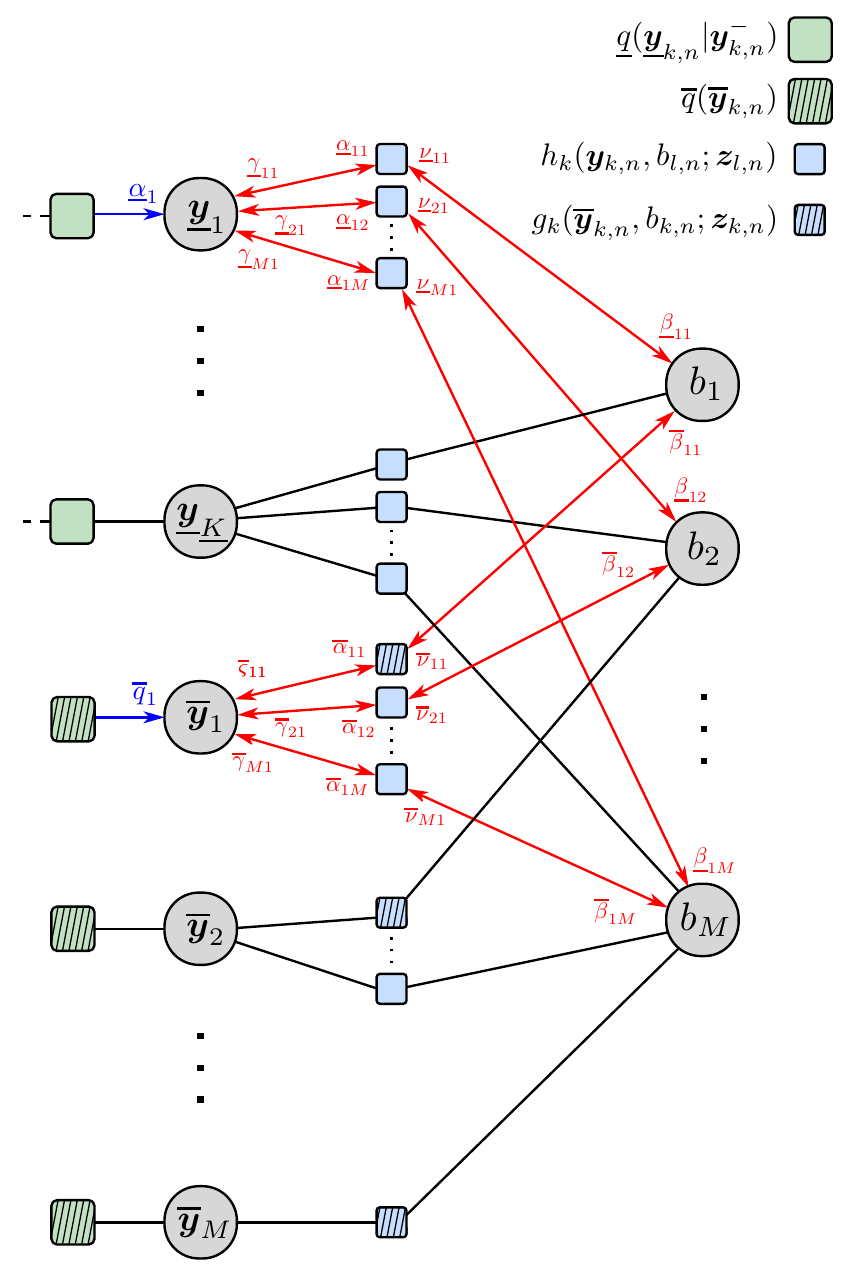}
\vspace{-1mm}
\caption{Factor graphs for EOT corresponding to the factorization \eqref{eq:factorGraph}. Factor nodes and variable nodes are depicted as boxes and
circles, respectively. Messages in blue are calculated only once and messages in red are calculated multiple times due to iterative message passing. The time index $n$ is omitted. The following short notations are used for the messages: $\underline{\alpha}_k \triangleq \underline{\alpha}(\underline{\V{y}}_{k})$, $\overline{q}_k \triangleq \overline{q}(\overline{\V{y}}_{k})$, $\underline{\gamma}_{lk} \triangleq \underline{\gamma}^{(p)}_{lk}(\underline{\V{y}}_{k})$, $\overline{\gamma}_{lk} \triangleq  \overline{\gamma}^{(p)}_{lk}(\overline{\V{y}}_{k})$, $\overline{\varsigma}_{kk} \triangleq  \overline{\varsigma}^{(p)}_{kk}(\overline{\V{y}}_{k})$, $\underline{\beta}_{kl}  \triangleq \underline{\beta}_{kl}^{(p)}(b_{l}) $,  $\overline{\beta}_{kl}  \triangleq \overline{\beta}_{kl}^{(p)}(b_{l}) $, $\underline{\nu}_{l k} \triangleq \underline{\nu}^{(p)}_{l k}(b_l)$, $\overline{\nu}_{l k} \triangleq \overline{\nu}^{(p)}_{l k}(b_l)$, $\overline{\alpha}_{kl} = \overline{\alpha}_{kl}^{(p)}\big(\overline{\V{y}}_{k}\big) $, and $\underline{\alpha}_{kl} = \underline{\alpha}_{kl}^{(p)}\big(\underline{\V{y}}_{k}\big) $.}
\label{fig:FactorGraph}
\end{figure}

\section{Problem Formulation and Factor Graph}
\label{sec:factorGraphProblem}

\vspace{.5mm}

In this section, we formulate the detection and estimation problem of interest and present the joint posterior pdf and factor graph underlying the proposed \ac{eot} method.

\vspace{-1mm}

 \subsection{Object Detection and State Estimation}
\label{sec:prob}

\vspace{.5mm}

We consider the problem of detecting legacy and new \acp{po} $k \!\in\! \{1,\dots,\underline{K}_n \rmv+\rmv M_n\}$ based on the binary existence variables $r_{k,n}$, as well as estimation of the object states $\RV{x}_{k,n}$ and extents $\RV{e}_{k,n}$ from the observed measurement history vector $\V{z}_{1:n} = \big[\V{z}_1^{\T} \cdots\ist \V{z}^{\T}_n \big]^{\T}\!$. Our Bayesian approach aims to calculate the marginal posterior existence probabilities $p(r_{k,n} \!=\! 1|\V{z}_{1:n})$ and the marginal posterior pdfs $f(\V{x}_{k,n}, \V{e}_{k,n}  | r_{k,n} \!=\! 1, \V{z}_{1:n} )$. We perform detection by comparing $p(r_{k,n} \!=\! 1|\V{z}_{1:n})$ with a threshold $P_{\text{th}}$, i.e., if $p(r_{k,n} \!=\! 1|\V{z}_{1:n}) > P_{\text{th}}$ \cite[Ch.~2]{Poo:B94}, \ac{po} $k$ is considered to exist. Estimates of $\RV{x}_{k,n}$ and $\RV{e}_{k,n}$ are  then obtained for the detected objects $k$ by the minimum mean-square error (MMSE) estimator \cite[Ch.~4]{Poo:B94}. In particular, an MMSE estimate of the kinematic state is obtained \vspace{1mm} as
\vspace{-1mm}
\begin{align}
\hat{\V{x}}^\text{MMSE}_{k,n} &\triangleq \int \rmv \V{x}_{k,n} \rmv  \int  f(\V{x}_{k,n}, \V{e}_{k,n}  | r_{k,n} \!=\! 1, \V{z}_{1:n} ) \mathrm{d}\V{e}_{k,n} \ist \mathrm{d}\V{x}_{k,n}. \nn\\[-1mm]
\label{eq:mmse}
\end{align}
Similarly, an MMSE estimate $\hat{\V{e}}^\text{MMSE}_{k,n}$ of the extent state is obtained by replacing $\V{x}_{k,n}$ with $\V{e}_{k,n}$ in \eqref{eq:mmse}.

We can compute the marginal posterior existence probability $p(r_{k,n} \!=\! 1 \ist | \ist \V{z}_{1:n})$ needed for object detection as discussed above,
from the marginal posterior pdf, $f(\V{y}_{k,n} | \V{z}_{1:n}) = f(\V{x}_{k,n}, \V{e}_{k,n},  r_{k,n}|\V{z}_{1:n})$, \vspace{0mm} as
\begin{align}
&p(r_{k,n} \!=\! 1|\V{z}_{1:n}) \nn\\
&\hspace{5mm} = \int \rmv\rmv\rmv \int f(\V{x}_{k,n}, \V{e}_{k,n},  r_{k,n} \!=\! 1 |\V{z}_{1:n}) \ist \mathrm{d}\V{x}_{k,n} \mathrm{d}\V{e}_{k,n}.
\label{eq:margpost_r} \\[-5.5mm]
\nn
\end{align}
Note that $p(r_{k,n} \!=\! 1|\V{z}_{1:n})$ is also needed for the suboptimal pruning step discussed in Section \ref{sec:resampling}.

Similarly, we can calculate the marginal posterior pdf  $f(\V{x}_{k,n}, \V{e}_{k,n} | r_{k,n} \!=\! 1, \V{z}_{1:n} )$ underlying MMSE state estimation (see \eqref{eq:mmse})
 from $f(\V{x}_{k,n}, \V{e}_{k,n}, r_{k,n}|\V{z}_{1:n})$ according to
\begin{align}
f(\V{x}_{k,n}, \V{e}_{k,n} \ist | \ist r_{k,n} \!=\! 1, \V{z}_{1:n} ) =  \frac{f(\V{x}_{k,n}, \V{e}_{k,n},  r_{k,n} \!=\! 1 |\V{z}_{1:n})}{p(r_{k,n} \!=\! 1|\V{z}_{1:n})} \,. \nn\\[-1.5mm]
\label{eq:margpost_x}\\[-6mm]
\nn
\end{align}
For this reason, the fundamental task is to compute the pdf $f(\V{y}_{k,n} | \V{z}_{1:n}) = f(\V{x}_{k,n}, \V{e}_{k,n},  r_{k,n}|\V{z}_{1:n})$. This pdf is a marginal density of $f(\V{y}_{0:n}, \V{b}_{1:n} | \V{z}_{1:n} )$, which involves all the augmented states and measurement-oriented association variables up to the current time $n$.

The main problem to be solved is to find a computationally feasible recursive (sequential) calculation of marginal posterior \acp{pdf} $f(\V{y}_{k,n} | \V{z}_{1:n})$. By performing message passing by means of the \ac{spa} rules \cite{KscFreLoe:01} on the factor graph that represents our statistical model discussed in Section \ref{sec:systemModel}, approximations \vspace{0mm} (``beliefs'') $\tilde{f}(\V{y}_{k,n})$ of this marginal posterior  pdfs  can be obtained in an efficient way for all legacy\vspace{-2mm} and new \acp{po}. 

 \subsection{The Factor Graph}
\label{sec:factorGraph}

\rd{We now assume that the measurements $\V{z}_{1:n}$ are observed and thus fixed.  By using common assumptions \cite{BarWilTia:B11,Mah:B07,MeyKroWilLauHlaBraWin:J18,GraFatSve:J19,GraBauReu:J17}, the joint posterior \ac{pdf} of $\RV{y}_{0:n}$ and $\RV{b}_{1:n}$, conditioned on $\V{z}_{1:n}$ is given \vspace{0mm} by\footnote{\rd{A detailed derivation of this joint posterior \ac{pdf} is provided in \cite[Section~1]{MeyWil:SM21}.}}}
\begin{align}
&f( \V{y}_{0:n}, \V{b}_{1:n}|\V{z}_{1:n})  \nn\\
&\hspace{1mm}\propto \bigg( \prod^{K_0}_{\ell = 1} f(\V{y}_{\ell,0})  \bigg) \rmv\rmv \prod^{n}_{n' = 1}   \bigg(  \prod^{M_{n'}}_{k = 1} \overline{q}(\overline{\V{y}}_{k,n'}) \ist\ist g_k\big( \overline{\V{y}}_{k,n'}, b_{k, n'}; \V{z}_{k,n'} \big) \nn\\
& \hspace{4.3mm} \times \prod^{M_{n'}}_{l = k \rmv+\rmv 1} \ist h_{\underline{K}_{n'} + k}  \big( \overline{\V{y}}_{k,n'}, b_{l,n'}; \V{z}_{l,n'} \big)  \rmv \bigg) \nn \\
&\hspace{4.3mm}\times \hspace{-.3mm} \prod^{\underline{K}_{n'}}_{k' = 1} \underline{q}\big(\underline{\V{y}}_{k'\rmv\rmv,n'} | \V{y}_{k'\rmv\rmv,n'-1}\big) \prod^{M_{n'}}_{l' = 1} h_{k'}\big( \underline{\V{y}}_{k'\rmv\rmv,n'}\rmv, b_{l'\rmv\rmv,n'}; \V{z}_{l'\rmv,n'} \big)  \label{eq:factorGraph}
\end{align}
where we introduced the functions $ h_k\big( \V{y}_{k,n}, b_{l,n}; \V{z}_{l,n} \big)$, $g_k\big( \overline{\V{y}}_{k,n}, b_{k,n}; \V{z}_{k,n} \big)$, $\underline{q}\big(\underline{\V{y}}_{k,n} | \V{y}_{k,n-1}\big)$, and $\overline{q}\big(\overline{\V{y}}_{k,n}\big)$ that will be discussed next. 

The \textit{pseudo likelihood functions} $h_k\big( \V{y}_{k,n}, b_{l,n}; \V{z}_{l,n} \big) = h_k\big( \V{x}_{k,n}, \V{e}_{k,n}, r_{k,n}, b_{l,n}; \V{z}_{l,n} \big)$ and $g_k\big( \overline{\V{y}}_{k,n}, b_{k,n}; \V{z}_{k,n} \big) = g_k\big( \overline{\V{x}}_{k,n}, \overline{\V{e}}_{k,n}, \overline{r}_{k,n}, b_{k,n}; \V{z}_{k,n} \big)$ are given by
\begin{align}
&h_k\big( \V{x}_{k,n},  \V{e}_{k,n}, 1, b_{l,n}; \V{z}_{l,n} \big)  \nn \\[.5mm]
&\hspace{13mm} \triangleq \begin{cases}
      \frac{\mu_{\mathrm{m}}(\V{x}_{k,n},\V{e}_{k,n}) f(\V{z}_{l,n}| \V{x}_{k,n}, \V{e}_{k,n})}{\mu_{\mathrm{fa}}  f_{\mathrm{fa}}(\V{z}_{l,n})}    \ist, 
       & \rmv\rmv b_{l,n} = k \\[1mm]
     1 \ist,  & \rmv\rmv b_{l,n} \neq k 
  \end{cases} \label{eq:pseudoLikelihood1}
\end{align}
and $h_k\big( \V{x}_{k,n}, \V{e}_{k,n}, 0,  b_{l,n}; \V{z}_{l,n} \big)  \rmv\triangleq\rmv 1 \rmv - \rmv \delta \big( b_{l,n} \rmv - \rmv k \big)$ as well as
\begin{align}
&\hspace{-2.5mm}g_k\big( \overline{\V{x}}_{k,n}, \overline{\V{e}}_{k,n}, 1, b_{k,n}; \V{z}_{k,n} \big) \nn \\[.5mm]
&\hspace{0mm} \triangleq \begin{cases}
      \frac{ \mu_{\mathrm{m}}(\overline{\V{x}}_{k,n},\overline{\V{e}}_{k,n}) \ist f(\V{z}_{k,n}| \overline{\V{x}}_{k,n},\overline{\V{e}}_{k,n})}{\mu_{\mathrm{fa}}  f_{\mathrm{fa}}(\V{z}_{k,n})}    \ist, 
       & \rmv\rmv b_{k,n} = \underline{K}_n + k \\[1mm]
     0 \ist,  & \rmv\rmv b_{k,n} \neq \underline{K}_n + k \
 \end{cases} \label{eq:pseudoLikelihood}
\end{align}
and $g_k\big( \overline{\V{x}}_{k,n}, \overline{\V{e}}_{k,n}, 0, b_{k,n}; \V{z}_{k,n} \big) \rmv\rmv\triangleq\rmv\rmv 1 \rmv - \rmv \delta \big( b_{k,n} \rmv - \rmv ( \underline{K}_n + k ) \big)$. Note that the second line in \eqref{eq:pseudoLikelihood} is zero because, as discussed in Section \ref{sec:DataAssocUncer}, the new \ac{po} with index $k$ exists ($\overline{r}_{k,n} \rmv=\rmv 1$) if and only if it produced measurement $k$.  

Furthermore, the \textit{factors containing prior distributions} for new \acp{po} $\overline{q}(\overline{\V{y}}_{k\rmv,n}) \rmv=\rmv \overline{q}(\overline{\V{x}}_{k,n}, \overline{\V{e}}_{k,n},  \overline{r}_{k,n})$, $k \rmv\in\rmv \{1,\dots,M_n\}$ are given by
\begin{align}
&\overline{q}(\overline{\V{x}}_{k,n},\overline{\V{e}}_{k,n},\overline{r}_{k,n})  \nn\\[1mm]
&\hspace{6mm}\,\triangleq \begin{cases}
      \mu_{\mathrm{n}} f_{\mathrm{n}}(\overline{\V{x}}_{k,n}, \overline{\V{e}}_{k,n}) \frac{\mathpzc{e}^{-\mu_{\mathrm{m}} (\overline{\V{x}}_{k,n},\overline{\V{e}}_{k,n})}}{1 - \mathpzc{e}^{-\mu_{\mathrm{m}} (\overline{\V{x}}_{k,n},\overline{\V{e}}_{k,n})}} \ist, 
       & \rmv\rmv \overline{r}_{k,n} = 1 \\[1mm]
     f_{\mathrm{d}}\big(\overline{\V{x}}_{k,n}, \overline{\V{e}}_{k,n}\big) \ist,  & \rmv\rmv \overline{r}_{k,n} = 0
  \end{cases} \nn\\[-9.5mm]
    \nn\\
  \nn
\end{align}
and the \textit{pseudo transition functions} (cf. \eqref{eq:singleTargetStateTrans_0} and \eqref{eq:singleTargetStateTrans_1}) for legacy \acp{po} $\underline{q}\big(\underline{\V{y}}_{k,n} | \V{y}_{k,n-1} \big) \triangleq \underline{q}\big(\underline{\V{x}}_{k,n}, \underline{\V{e}}_{k,n}, \underline{r}_{k,n} | \V{x}_{k,n-1}, \V{e}_{k,n-1},$ $ r_{k,n-1} \big)$ are given \vspace{0mm}  by $\underline{q}\big(\underline{\V{x}}_{k,n}, \underline{\V{e}}_{k,n}, \underline{r}_{k,n} \rmv\rmv= \rmv\rmv1 | \V{x}_{k,n-1}, \V{e}_{k,n-1},$ $r_{k,n-1} \big) \rmv\rmv=\rmv\rmv \mathpzc{e}^{-\mu_{\mathrm{m}}(\underline{\V{x}}_{k,n},\underline{\V{e}}_{k,n})} f\big(\underline{\V{x}}_{k,n}, \underline{\V{e}}_{k,n},\underline{r}_{k,n} \rmv\rmv\rmv=\rmv\rmv\rmv  1 | \V{x}_{k,n-1},$ $\V{e}_{k,n-1}, r_{k,n-1} \big) $ and $\underline{q}\big(\underline{\V{x}}_{k,n}, \underline{\V{e}}_{k,n}, \underline{r}_{k,n} \rmv\rmv\rmv =\rmv\rmv\rmv   0 | \V{x}_{k,n-1},\V{e}_{k,n-1},$ $r_{k,n-1} \big) \rmv\rmv= f\big(\underline{\V{x}}_{k,n},$ $\underline{\V{e}}_{k,n},$ $\underline{r}_{k,n} \rmv=\rmv  0 \ist | \ist \V{x}_{k,n-1}, \V{e}_{k,n-1}, r_{k,n-1} \big)$.

A detailed derivation of this factor graph is provided in the supplementary material \cite{MeyWil:SM21}. The factor graph representing factorization \eqref{eq:factorGraph} is shown in Fig.~\ref{fig:FactorGraph}.  \rd{An interesting observation is that this factor graph has the same structure as a conventional multi-scan tracking problem \cite{MeyKroWilLauHlaBraWin:J18,WilLau:J18} with $M$ scans. (Here, every measurement is considered an individual scan.)} In what follows, we consider a single time step and remove the time index $n$ for the sake of readability.
 
\section{The Proposed Sum-Product Algorithm}
\label{sec:proposedSPA}

Since our factor graph in Fig.~\ref{fig:FactorGraph} has cycles, we have to decide on a specific order of message computation \cite{KscFreLoe:01}. We choose this order according to the following rules: 
(i) messages are not sent backward in 
time\footnote{This 
is equivalent to a joint filtering approach with the assumption that the \ac{po} states are conditionally independent given the past measurements.} 
\cite{MeyBraWilHla:J17,MeyKroWilLauHlaBraWin:J18}; and (ii) at each time step messages are computed and processed in parallel. With these rules, the generic message passing equations of the \ac{spa} \cite{KscFreLoe:01}
yield the following operations at each time step. The corresponding messages are shown in \vspace{-2mm}Fig.~\ref{fig:FactorGraph}.

\subsection{Prediction Step}
\label{sec:predictionStep}

First, a \emph{prediction} step is performed for all legacy POs $k \in \underline{K}$. Based on \ac{spa} rule \cite[Eq.~(6)]{KscFreLoe:01}, we\vspace{.5mm} obtain
\begin{align}
\underline{\alpha}(\underline{\V{x}}_{k}, \underline{\V{e}}_{k}, \underline{r}_{k}) &= \rrmv \sum_{r^{-}_{k} \in \{0,1\}}\int \rmv\rmv\rmv \int \rmv \underline{q}( \underline{\V{x}}_{k}, \underline{\V{e}}_{k}, \underline{r}_{k} \ist|\ist \V{x}^{-}_{k}, \V{e}^{-}_{k}, r^{-}_{k})\nn\\[-.5mm]
&\hspace{15mm} \times \tilde{f}(\V{x}^{-}_{k}, \V{e}^{-}_{k}, r^{-}_{k}) \ist \mathrm{d}\V{x}^{-}_{k} \ist \mathrm{d}\V{e}^{-}_{k} \label{eq:prediction} \\[-4.5mm] \nn
\end{align}
where $\tilde{f}(\V{x}^{-}_{k}, \V{e}^{-}_{k}, r^{-}_{k})$ is the belief that was calculated at the previous time step. Recall that the integration $\int \mathrm{d}\V{e}^{-}_{k}$ in \eqref{eq:prediction} is performed over the support of $\V{e}^{-}_{k},$ which corresponds to all positive-semidefinite matrices $\M{E}^{-}_{k}$. 

Next, we use the expression for $\underline{q}( \underline{\V{x}}_{k}, \underline{\V{e}}_{k},  \underline{r}_{k} \ist |$ $\V{x}^{-}_{k}, \V{e}^{-}_{k}, r^{-}_{k})$ as introduced in Section \ref{sec:factorGraph} and in turn \eqref{eq:singleTargetStateTrans_0} and \eqref{eq:singleTargetStateTrans_1} for $f( \underline{\V{x}}_{k}, \underline{\V{e}}_{k},  \underline{r}_{k} \ist | \V{x}^{-}_{k}, \V{e}^{-}_{k}, r^{-}_{k})$. In this way, we obtain the following expressions for \eqref{eq:prediction}
\begin{align}
\underline{\alpha}(\underline{\V{x}}_{k},\underline{\V{e}}_{k}, \underline{r}_{k} \rmv=\rmv 1)&\hspace{-1mm}= p_{\ist\mathrm{s}} \ist\ist \mathpzc{e}^{-\mu_{\mathrm{m}}(\underline{\V{x}}_{k,n},\underline{\V{e}}_{k,n})} \rmv\rmv\rmv  \int \rmv\rmv\rmv\rmv \int \rmv\rmv f( \underline{\V{x}}_{k}, \underline{\V{e}}_{k}  | \V{x}^{-}_{k}, \V{e}^{-}_{k}) \nn\\[.8mm]
&\hspace{10mm}\times\tilde{f}(\V{x}^{-}_{k}, \V{e}^{-}_{k}, 1) \ist \mathrm{d}\V{x}^{-}_{k} \ist \mathrm{d}\V{e}^{-}_{k}  \rmv \label{eq:alpha1}  \\[0mm]
\nn\\[-10mm]
\nn
\end{align}
and $\underline{\alpha}(\underline{\V{x}}_{k}, \underline{\V{e}}_{k}, \underline{r}_{k} \rmv=\rmv 0) = \underline{\alpha}^{\mathrm{n}}_k f_{\mathrm{d}}(\underline{\V{x}}_{k}, \underline{\V{e}}_{k})$ with
\begin{align}
\underline{\alpha}^{\mathrm{n}}_k &\triangleq\ist \tilde{f}^{-}_{k}  + \big(1\!-\rmv p_{\ist\mathrm{s}}\big) \rmv \int \rmv\rmv\rmv\rmv \int \! \tilde{f}(\V{x}^{-}_{k},\V{e}^{-}_{k},1) 
\,\mathrm{d}\V{x}^{-}_{k} \mathrm{d}\V{e}^{-}_{k}\nn\\[1mm]
&=\ist \tilde{f}^{-}_{k}  + \big(1\!-\rmv p_{\ist\mathrm{s}}\big) \rmv \big( 1- \tilde{f}^{-}_{k} \big).
\label{eq:alpha0}
\end{align}
We note that $\tilde{f}^{-}_{k} \rmv=\rmv \int \rmv\rmv\rmv\rmv \int \rmv\tilde{f}(\V{x}^{-}_{k}, \V{e}^{-}_{k}, 0) \ist \mathrm{d} \V{x}^{-}_{k} \mathrm{d}\V{e}^{-}_{k}$ approximates the probability of non-existence of legacy \ac{po} $k$ at the previous time step. After the prediction step, the iterative message passing is performed. For future reference, we also introduce $\underline{\alpha}_{k} \triangleq \int \rmv\rmv\rmv\rmv \int \rmv \underline{\alpha}_{k}(\underline{\V{x}}_{k}, \underline{\V{e}}_{k}, \underline{r}_{k} \rmv= 1) \ist \mathrm{d} \underline{\V{x}}_{k} \ist \mathrm{d} \underline{\V{e}}_{k} \rmv+\rmv \underline{\alpha}^{\mathrm{n}}_{k}$.

\subsection{Iterative Message Passing}

At iteration $p \in \{1,\dots,P\}$, the following operations are computed for all legacy and new \acp{po}.

\subsubsection{Measurement Evaluation}
\label{sec:measEval}

The messages $\beta^{(p)}_{kl}(b_{l})$, $k \rmv\in \{1,\dots,\underline{K}\}$, $l \rmv\in\rmv \{1,\dots,M\}$ sent from factor nodes $h_k\big( \underline{\V{y}}_{k}, b_{l}; \V{z}_l \big) \rmv=\rmv h_k\big( \underline{\V{x}}_{k},  \underline{\V{e}}_{k}, \underline{r}_{k}, b_{l}; \V{z}_l \big)$ to variable nodes $b_{l}$ can be calculated \vspace{0mm} as discussed next. First, by using the \ac{spa} rule \cite[Eq.~(6)]{KscFreLoe:01}, we obtain
\begin{align}
\underline{B}_{kl}^{(p)}(b_{l}) &= \rmv\rmv\rmv\rmv \sum_{\underline{r}_{k} \in \{0,1\}} \rmv \int \rmv\rmv\rmv\rmv \int h_k\big( \underline{\V{x}}_{k},  \underline{\V{e}}_{k}, \underline{r}_{k}, b_{l}; \V{z}_l \big) \nn\\[1mm]
&\hspace{11.5mm}\times \underline{\alpha}_{kl}^{(p)}(\underline{\V{x}}_{k},\underline{\V{e}}_{k},\underline{r}_{k}) \ist \mathrm{d} \underline{\V{x}}_{k}\ist \mathrm{d} \underline{\V{e}}_{k}.  \label{eq:beta1}\\[-3.5mm]
\nn
\end{align}
Note that for $p \rmv=\rmv 1$, we set  $\underline{\alpha}_{kl}^{(1)}(\underline{\V{x}}_{k},\underline{\V{e}}_{k},\underline{r}_{k})  \triangleq \underline{\alpha}(\underline{\V{x}}_{k}, \underline{\V{e}}_{k},\underline{r}_{k})$, and for $p\rmv>\rmv1$ we use $\underline{\alpha}_{kl}^{(p)}(\underline{\V{x}}_{k},\underline{\V{e}}_{k},\underline{r}_{k})$ (represented by $\underline{\alpha}_{kl}^{(p)}$ $(\underline{\V{x}}_{k},\underline{\V{e}}_{k},\underline{r}_{k} \rmv=\rmv 1)$ and $\underline{\alpha}^{\mathrm{n}(p)}_{kl}$) calculated in Section \ref{sec:exInfo}. By using the expressions for $h_k\big( \underline{\V{x}}_{k},\underline{\V{e}}_{k}, \underline{r}_{k}, b_{l}; \V{z}_l \big)$ introduced in Section \ref{sec:factorGraph}, \eqref{eq:beta1} can be further simplified\vspace{.5mm}, i.e.,
\begin{align}
\underline{B}_{kl}^{(p)}(b_{l} \rmv=\rmv k)  &= \frac{1}{\mu_{\mathrm{fa}} \ist f_{\mathrm{fa}}(\V{z}_{l})}  \int \rmv\rmv\rmv\rmv \int \mu_{\mathrm{m}}(\underline{\V{x}}_k,\underline{\V{e}}_k)  f(\V{z}_{l}| \underline{\V{x}}_k, \underline{\V{e}}_{k})\nn\\[1.8mm]
&\hspace{4mm}\times  \underline{\alpha}_{kl}^{(p)}(\underline{\V{x}}_{k}, \underline{\V{e}}_{k}, \underline{r}_{k} \rmv=\rmv 1) \ist \mathrm{d} \underline{\V{x}}_{k} \ist \mathrm{d} \underline{\V{e}}_{k}\label{eq:Beta1c} \end{align}
and $\underline{B}_{kl}^{(p)}(b_{l} \rmv\neq\rmv k) \rmv = \rmv \underline{\alpha}_{kl}^{(p)}$ with $\underline{\alpha}_{kl}^{(p)} \triangleq \int \rmv\rmv\rmv\rmv \int \rmv \underline{\alpha}_{kl}^{(p)}(\underline{\V{x}}_{k}, \underline{\V{e}}_{k}, \underline{r}_{k} \rmv= 1) \ist \mathrm{d} \underline{\V{x}}_{k} \ist \mathrm{d} \underline{\V{e}}_{k} \rmv+\rmv \underline{\alpha}^{\mathrm{n}(p)}_{kl}\rmv$.
After multiplying these two expressions by $1/\underline{\alpha}_{kl}^{(p)} $, the message $\underline{\beta}_{kl}^{(p)}(b_{l})$ is given by\footnote{Multiplying \ac{spa} messages by a constant factor does not alter the resulting approximate marginal posterior \acp{pdf} \cite{KscFreLoe:01}.}
\begin{align}
\underline{\beta}_{kl}^{(p)}(b_{l} \rmv=\rmv k)  &= \frac{1}{\mu_{\mathrm{fa}} \ist f_{\mathrm{fa}}(\V{z}_{l}) \ist \underline{\alpha}_{kl}^{(p)} }  \int \rmv\rmv\rmv\rmv \int \mu_{\mathrm{m}}(\underline{\V{x}}_k,\underline{\V{e}}_k)  f(\V{z}_{l}| \underline{\V{x}}_k, \underline{\V{e}}_{k})\nn\\[2.5mm]
&\hspace{7.5mm}\times  \underline{\alpha}_{kl}^{(p)}(\underline{\V{x}}_{k}, \underline{\V{e}}_{k}, \underline{r}_{k} \rmv=\rmv 1) \ist \mathrm{d} \underline{\V{x}}_{k} \ist \mathrm{d} \underline{\V{e}}_{k}\label{eq:Beta1b}
\end{align}
and $\underline{\beta}_{kl}^{(p)}(b_{l} \rmv\neq\rmv k) \rmv=\rmv 1$. This final normalization step makes it possible to perform data association and measurement update discussed in the next two sections more efficiently. For future reference, we introduce the short notation $\underline{\beta}_{kl}^{(p)} \triangleq \underline{\beta}_{kl}^{(p)}(b_{l} \rmv=\rmv k) $ for all $k \rmv\in\rmv \{1,\dots,\underline{K}\}$, $l \rmv\in\rmv \{1,\dots,M\}$.

The messages $\overline{\beta}_{kl}^{(p)}(b_{l})$, $k \rmv\in\rmv \{1,\dots,M-1\}$, $l \rmv\in\rmv \{k+1,$ $\dots,M\}$ sent from factor nodes $h_k\big( \overline{\V{y}}_{k}, b_{l}; \V{z}_l \big)$ to variable nodes $b_{l}$ can be obtained similarly. In particular, by replacing $\underline{\alpha}_{kl}^{(p)}(\underline{\V{x}}_{k},\underline{\V{e}}_{k},\underline{r}_{k} \rmv=\rmv 1)$ and $\underline{\alpha}_{kl}^{(p)}$  in \vspace{.5mm}\eqref{eq:Beta1b} by $\overline{\alpha}_{kl}^{(p)}(\overline{\V{x}}_{k},\overline{\V{e}}_{k},\overline{r}_{k} \rmv=\rmv 1)$ and  $\overline{\alpha}_{kl}^{(p)}$, respectively, we obtain $\overline{\beta}_{kl}^{(p)}(b_{l} \rmv=\rmv \underline{K} \rmv+\rmv k)$. Furthermore, we have $\overline{\beta}_{kl}^{(p)}(b_{l} \rmv \neq\rmv \underline{K} \rmv+\rmv k) = 1$. Note that for $p \rmv=\rmv 1$ we set  $\overline{\alpha}_{kl}^{(1)}(\overline{\V{x}}_{k},\overline{\V{e}}_{k},\overline{r}_{k})  \triangleq q(\overline{\V{x}}_{k},\overline{\V{e}}_{k},\overline{r}_{k})$ and for $p\rmv>\rmv1$ we again calculate $\overline{\alpha}_{kl}^{(p)}(\overline{\V{x}}_{k},\overline{\V{e}}_{k},\overline{r}_{k})$\vspace{1mm} as discussed in Section \ref{sec:exInfo}.

Finally, the messages $\overline{\beta}_{kk}^{(p)}(b_{k})$, $k \rmv\in\rmv \{1,\dots,M\}$ sent from factor nodes $g_k\big( \overline{\V{y}}_{k}, b_{k}; \V{z}_{k} \big) \rmv=\rmv g_k\big( \overline{\V{x}}_{k}, \overline{\V{e}}_{k}, \overline{\V{r}}_{k}, b_{k}; \V{z}_{k} \big)$ to variable nodes $b_k$ are calculated by also replacing $h_k\big( \underline{\V{x}}_{k},  \underline{\V{e}}_{k}, \underline{r}_{k}, b_{l}; \V{z}_l \big)$ in \eqref{eq:beta1} by $g_k\big( \overline{\V{x}}_{k}, \overline{\V{e}}_{k}, \overline{r}_k, b_{k}; \V{z}_{k} \big)$ and  performing \vspace{-.2mm}similar simplification steps. In particular, we get $\overline{\alpha}_{kk}^{\ist\mathrm{n}(p)} \rmv\rmv\triangleq\rmv\rmv  \int \rmv\rmv\rmv\rmv \int \overline{\alpha}_{kk}^{(p)}(\overline{\V{x}}_{k},\overline{\V{e}}_{k},\overline{r}_{k} \rmv= 0) \ist \mathrm{d} \overline{\V{x}}_{k} \ist \mathrm{d} \overline{\V{e}}_{k}$ for $b_{k} \rmv\neq\rmv \underline{K}+k$ and a result equal to \eqref{eq:Beta1c} (with $l$ replaced by $k$ and $\underline{\alpha}_{kl}^{(p)}(\underline{\V{x}}_{k},\underline{\V{e}}_{k},\underline{r}_{k})$ replaced by  \vspace{1mm} $\overline{\alpha}_{kk}^{(p)}(\overline{\V{x}}_{k},\overline{\V{e}}_{k},\overline{r}_{k})$) for $b_{k} \rmv\neq\rmv k$. After \vspace{-1mm} multiplying both expressions by $1/\overline{\alpha}_{kk}^{\ist \mathrm{n}(p)}\rmv$, the message $\overline{\beta}_{kk}^{(p)}(b_{k})$ finally reads
\begin{align}
\overline{\beta}_{kk}^{(p)}(b_{k} \rmv=\rmv \underline{K} \rmv+\rmv k)  &= \frac{1}{\mu_{\mathrm{fa}} \ist f_{\mathrm{fa}}(\V{z}_{k}) \ist \overline{\alpha}_{kk}^{ \ist \mathrm{n}(p)}}  \int \rmv\rmv\rmv\rmv \int \rmv f(\V{z}_{k}| \overline{\V{x}}_k, \overline{\V{e}}_{k})   \nn\\[2mm]
&\hspace{0mm}\times\mu_{\mathrm{m}}(\overline{\V{x}}_k,\overline{\V{e}}_k) \ist \overline{\alpha}_{kk}^{(p)}(\overline{\V{x}}_{k}, \overline{\V{e}}_{k}, \overline{r}_{k} \rmv=\rmv 1) \ist \mathrm{d} \overline{\V{x}}_{k} \ist \mathrm{d} \overline{\V{e}}_{k} \nn\\[1mm]
\label{eq:Beta1d} \\[-9mm]
\nn
\end{align}
and $\overline{\beta}_{kk}^{(p)}(b_{k} \rmv\neq\rmv \underline{K} \rmv+\rmv k) \rmv=\rmv 1$. For future reference, we again introduce the short notation $\overline{\beta}_{kl}^{(p)} \triangleq \overline{\beta}_{kl}^{(p)}(b_{l} \rmv=\rmv \underline{K} \rmv+\rmv k)$ for all $k \rmv\in\rmv \{1,\dots,M\}$, $l \rmv\in\rmv \{k,\dots,M\}$

\subsubsection{Data Association} \label{sec:dataAssociation}
The messages $\underline{\nu}^{(p)}_{l k}(b_l)$ sent from variable nodes $b_{l}$, $l \rmv\in \{1,\dots,M\}$ to factor nodes $h_k\big( \V{y}_{k}, b_{l}; \V{z}_l \big)$, $k \rmv\in\rmv \{1,\dots,\underline{K}\}$, can be \vspace{0mm}expressed as \cite[Eq.~(5)]{KscFreLoe:01}
\begin{align}
&\underline{\nu}^{(p)}_{l k}(b_l) \rmv=\rmv \Bigg( \ist \prod^{\underline{K}}_{\substack{k'=1\\k' \neq k}}  \underline{\beta}^{(p)}_{k' l}\big(b_{l}\big) \rmv \Bigg) \ist \prod^{l}_{\ell=1} \ist\ist \overline{\beta}^{(p)}_{\ell l}\big(b_{l}\big). 
  \label{eq:nu1}\\[-6mm]
&\nn
\end{align}
By using \eqref{eq:Beta1b} and \eqref{eq:Beta1d} in \eqref{eq:nu1}, we obtain $\underline{\nu}^{(p)}_{l k}(b_l \rmv=\rmv 0) = \underline{\nu}^{(p)}_{l k}(b_l \rmv=\rmv k) \rmv= 1$, $\underline{\nu}^{(p)}_{l k}(b_l \rmv=\rmv k') = \underline{\beta}^{(p)}_{k' l}$, $k' \rmv\in \{1,\dots,\underline{K}\} \backslash \{k\}$, and $\underline{\nu}^{(p)}_{l k}(b_l \rmv=\rmv \underline{K} \rmv+\rmv \ell) = \overline{\beta}^{(p)}_{\ell l}\rmv\rmv\rmv$, $\ell \rmv\in\rmv \{1,\dots,l\}$.

A similar expression is obtained for the messages $\overline{\nu}^{(p)}_{l k}(b_l)$ sent from variable nodes $b_{l}$, $l \rmv\in \{1,\dots,M\}$ to factor nodes $h_{\underline{K}+k}\big( \V{y}_{k}, b_{l}; \V{z}_l \big)$, $k \rmv\in \big\{1,\dots, l \rmv-\rmv1\big\}$ and factor node $g_{k}\big( \underline{\V{y}}_{k},$ $b_{l}; \V{z}_k \big)$, $k = l$, i.e.,\vspace{.4mm}
\begin{align}
&\overline{\nu}^{(p)}_{l k}(b_l) =\rmv \Bigg( \ist \prod^{\underline{K}}_{\substack{\ell=1}}  \ist\ist \underline{\beta}^{(p)}_{\ell l}\big(b_{l}\big) \rmv \Bigg) \ist \prod^{l}_{\substack{k'=1\\k' \neq k}} \ist\ist \overline{\beta}^{(p)}_{k' l}\big(b_{l}\big). \label{eq:nu2} \\[-6mm]
&\nn
\end{align}
By again using \eqref{eq:Beta1b} and \eqref{eq:Beta1d} in \eqref{eq:nu2}, we obtain $\overline{\nu}^{(p)}_{l k}(b_l \rmv= 0) = \overline{\nu}^{(p)}_{l k}(b_l \rmv=\rmv \underline{K} + k) \rmv= 1$, $\overline{\nu}^{(p)}_{l k}(b_l \rmv=\rmv k') = \underline{\beta}^{(p)}_{k' l}$, $k' \rmv\in \{1,\dots,\underline{K}\}$, and $\overline{\nu}^{(p)}_{l k}(b_l \rmv=\rmv \underline{K} \rmv+\rmv \ell) = \overline{\beta}^{(p)}_{\ell l}\rmv\rmv\rmv$, $\ell \rmv\in\rmv \{1,\dots,l\} \backslash \{k\}$.

\subsubsection{Measurement Update} Next, the messages sent from factor nodes $h\big( \underline{\V{y}}_{k}, b_{l}; \V{z}_l \big)$, $k \in \{1,\dots,\underline{K}\}$, $l \in \{1,$ $\dots,M\}$ to variable nodes $\underline{\V{y}}_{k}$ are calculated \vspace{1mm} as \cite[Eq.~(6)]{KscFreLoe:01}
\begin{align}
\underline{\gamma}^{(p)}_{lk}(\underline{\V{y}}_{k})  &= \rmv\rmv\rmv\rmv\rmv\sum^{\underline{K} + l}_{b_{l} = 0} h\big( \underline{\V{y}}_{k}, b_{l}; \V{z}_l \big)\ist\ist\underline{\nu}_{lk}^{(p)}(b_{l}). \label{eq:gammaLegacy}
\end{align}
\rd{By using again the expression for $h\big( \underline{\V{y}}_{k}, b_{l}; \V{z}_l \big)$ introduced in Section \ref{sec:factorGraph}, these messages $\underline{\gamma}^{(p)}_{lk}(\underline{\V{y}}_{k})\triangleq \underline{\gamma}^{(p)}_{lk}(\underline{\V{x}}_{k}, \underline{\V{e}}_{k}, \underline{r}_k)$ can be further simplified \vspace{1mm}as}
\begin{align}
\underline{\gamma}^{(p)}_{lk}(\underline{\V{x}}_{k}, \underline{\V{e}}_{k}, \underline{r}_k \rmv=\rmv 0)  &= \sum^{\underline{K}+ l}_{\substack{b_l=0\\b_l \neq k}} \ist \underline{\nu}_{l k}^{(p)}(b_l)  \nn\\[2mm]
\underline{\gamma}^{(p)}_{lk}(\underline{\V{x}}_{k}, \underline{\V{e}}_{k}, \underline{r}_k \rmv=\rmv 1) &= \frac{\mu_{\mathrm{m}}(\underline{\V{x}}_{k},\underline{\V{e}}_{k}) f(\V{z}_{l}| \underline{\V{x}}_k, \underline{\V{e}}_{k})}{\mu_{\mathrm{fa}}  f_{\mathrm{fa}}(\V{z}_{l})} \ist \underline{\nu}_{l k}^{(p)}(b_{l} \rmv=\rmv k) \nn\\[.8mm]
&\hspace{5mm}+ \sum^{\underline{K}+ l}_{\substack{b_l=0\\b_l \neq k}} \underline{\nu}_{l k}^{(p)}(b_l). \label{eq:gamma1}
\end{align}

By using the simplification of \eqref{eq:nu1} discussed in the previous Section~\ref{sec:dataAssociation}, we can \vspace{1.5mm} rewrite \eqref{eq:gamma1} as
\begin{align}
\underline{\gamma}^{(p)}_{lk}(\underline{\V{x}}_{k}, \underline{\V{e}}_{k}, \underline{r}_k \rmv=\rmv 0)  &= \underline{\xi}_{kl}^{(p)} \nn\\[2.5mm]
\underline{\gamma}^{(p)}_{lk}(\underline{\V{x}}_{k}, \underline{\V{e}}_{k}, \underline{r}_k \rmv=\rmv 1) &= \frac{\mu_{\mathrm{m}}(\underline{\V{x}}_{k},\underline{\V{e}}_{k}) f(\V{z}_{l}| \underline{\V{x}}_k, \underline{\V{e}}_{k})}{\mu_{\mathrm{fa}}  f_{\mathrm{fa}}(\V{z}_{l})} + \underline{\xi}_{kl}^{(p)} \label{eq:measUpdate} \\[-3mm]
\nn
\end{align}
where we introduced the \vspace{1.5mm} sum
\begin{equation}
\underline{\xi}_{kl}^{(p)} \triangleq \sum^{\underline{K}}_{\substack{k'=1\\k' \neq k}} \ist \underline{\beta}_{k'l}^{(p)} + \sum^{l}_{\substack{\ell=1}} \ist \overline{\beta}_{\ell l}^{(p)} + 1. \label{eq:sumMessages}
\end{equation}
\rd{A similar result can be obtained for the messages $\overline{\gamma}^{(p)}_{lk}(\overline{\V{y}}_{k}) $ sent from factor \vspace{.3mm} nodes $h\big( \overline{\V{y}}_{k}, b_l; \V{z}_l \big)$, $k \in \{1,\dots,M-1\}$, $l \in \{k+1,\dots,M\}$ to variable nodes $\overline{\V{y}}_{k}$. In particular, these messages are obtained by replacing \vspace{-.5mm} in \eqref{eq:gammaLegacy} the functions $h_k\big( \underline{\V{y}}_{k}, b_{l}; \V{z}_l \big)$ and $\underline{\nu}_{l k}^{(p)}(b_l)$ by $h_k\big( \overline{\V{y}}_{k}, b_l; \V{z}_l \big)$ and $\overline{\nu}_{l k}^{(p)}(b_l)$ as well as by performing the same simplification step discussed above.} 

By performing similar steps, we can then also obtain the messages $\varsigma_{kk}^{(p)}(\overline{\V{y}}_{k}) \rmv\triangleq\rmv \varsigma_{kk}^{(p)}(\overline{\V{x}}_{k},  \overline{\V{e}}_{k},\overline{r}_{k})$, $k \in \{1,\dots,M\}$ sent from $g_k\big( \overline{\V{x}}_{k},  \overline{\V{e}}_{k},$  $\overline{r}_{k}, b_k; \V{z}_k \big)$ to new \ac{po} state $\overline{\V{y}}_{k}$ \vspace{.3mm} as
\begin{align}
\varsigma_{kk}^{(p)}(\overline{\V{x}}_{k}, \overline{\V{e}}_{k},  \overline{r}_k \rmv=\rmv 0)  &= \overline{\xi}_{kk}^{(p)} \nn\\[2.5mm]
\varsigma_{kk}^{(p)}(\overline{\V{x}}_{k}, \overline{\V{e}}_{k},  \overline{r}_k \rmv=\rmv 1) &= \frac{\mu_{\mathrm{m}}(\overline{\V{x}}_{k},\overline{\V{e}}_{k}) f(\V{z}_{k} \ist| \ist \overline{\V{x}}_k, \overline{\V{e}}_{k})}{\mu_{\mathrm{fa}}  f_{\mathrm{fa}}(\V{z}_{k})}.\label{eq:varSigma}\\[-5.5mm]
\nn
\end{align}
\rd{The calculation of $\overline{\gamma}^{(p)}_{lk}(\overline{\V{y}}_{k}) $, $k \in \{1,\dots,M-1\}$, $l \in \{k+1,\dots,M\}$ and $\varsigma_{kk}^{(p)}(\overline{\V{y}}_{k})$, $k \in \{1,\dots,M\}$ is based on the sum \vspace{0mm}}
\begin{equation}
\overline{\xi}_{kl}^{(p)} \triangleq \sum^{\underline{K}}_{\substack{k'=1}} \ist \underline{\beta}_{k'l}^{(p)} + \sum^{l}_{\substack{\ell=1\\[.5mm] \ell \neq k}} \ist \overline{\beta}_{\ell l}^{(p)} + 1. \nn
\vspace{1mm}
\end{equation}

\subsubsection{Extrinsic Information} \label{sec:exInfo} Finally, updated messages used in the next message passing iteration $p+1$ for legacy \acp{po} $k \rmv\in\rmv \{1,\dots,\underline{K}\}$  are obtained \vspace{0mm} as \cite[Eq.~(5)]{KscFreLoe:01}
\begin{equation}
\underline{\alpha}_{kl}^{(p+1)}\big(\underline{\V{y}}_{k}\big) \ist=\ist \underline{\alpha}\big(\underline{\V{y}}_{k}\big) \hspace{-.2mm} \prod^{M}_{\substack{l'=1\\l' \neq l}} \hspace{-.2mm} \underline{\gamma}^{(p)}_{l'k}(\underline{\V{y}}_{k}).
\label{eq:alphaUpdated}
\end{equation}
\rd{Note that for non-existent objects, i.e., $ \underline{r}_{k} \rmv=\rmv 0$, the messages $\underline{\alpha}_{kl}^{(p+1)}(\underline{\V{x}}_{k},\underline{\V{e}}_{k}, \underline{r}_{k}) \rmv\triangleq\rmv  \underline{\alpha}_{kl}^{(p+1)}\big(\underline{\V{y}}_{k}\big)$ have the form $\underline{\alpha}_{kl}^{(p+1)} (\underline{\V{x}}_{k},$ $\underline{\V{e}}_{k}, \underline{r}_{k} \rmv=\rmv 0) = \underline{\alpha}^{\mathrm{n}(p+1)}_{kl} f_{\mathrm{d}}(\underline{\V{x}}_{k}, \underline{\V{e}}_{k})$.}

Similarly for new \acp{po} $k \rmv\in\rmv \{1,\dots,M\}$, we \vspace{.2mm} obtain
\begin{equation}
\overline{\alpha}_{kl}^{(p+1)}\big(\overline{\V{y}}_{k}\big) \ist=\ist \overline{q}\big(\overline{\V{y}}_{k}\big) \ist \varsigma_{kk}^{(p)}(\overline{\V{y}}_{k}) \hspace{-.2mm} \prod^{M}_{\substack{l'=k+1\\l' \neq l}} \hspace{-.2mm} \overline{\gamma}^{(p)}_{l'k}(\overline{\V{y}}_{k}) 
\label{eq:alphaUpdated2}
\vspace{.1mm}
\end{equation}
for $l \in \{k+1,\dots,M\}$ and $\overline{\alpha}_{kl}^{(p+1)}\big(\overline{\V{y}}_{k}\big) \ist=\ist \overline{q}\big(\overline{\V{y}}_{k}\big) \prod^{M}_{l'=k+1}$ $\overline{\gamma}^{(p)}_{l'k}(\overline{\V{y}}_{k})$ for $l = k$. \rd{Again, for $ \overline{r}_{k} \rmv=\rmv 0$, the messages $\overline{\alpha}_{kl}^{(p+1)}($ $\overline{\V{x}}_{k},\overline{\V{e}}_{k}, \overline{r}_{k}) \rmv\triangleq\rmv  \overline{\alpha}_{kl}^{(p+1)}\big(\overline{\V{y}}_{k}\big)$ have the form $\overline{\alpha}_{kl}^{(p+1)} (\overline{\V{x}}_{k}, \overline{\V{e}}_{k}, \overline{r}_{k} \rmv=\rmv 0) = \overline{\alpha}^{\hspace{.15mm}\mathrm{n}(p+1)}_{kl} f_{\mathrm{d}}(\overline{\V{x}}_{k}, \overline{\V{e}}_{k})$.}

\subsection{Belief Calculation}
After the last iteration $p \rmv=\rmv P$, the belief $\tilde{f}(\underline{\V{y}}_{k}) \triangleq \tilde{f}(\underline{\V{x}}_{k},$ $\underline{\V{e}}_{k},\underline{r}_{k})$ of legacy \ac{po} state $k \rmv\in \{1,\dots,\underline{K}\}$ can be calculated as the normalized product of all incoming messages \cite{KscFreLoe:01}, i.e.,
\begin{equation}
\tilde{f}(\underline{\V{y}}_{k}) = \underline{C}_k \ist\ist  \underline{\alpha}(\underline{\V{y}}_{k}) \ist\ist\ist \prod^{M}_{l=1} \ist\ist\ist \underline{\gamma}^{(P)}_{lk}(\underline{\V{y}}_{k}) \label{eq:beliefLegacy}
\vspace{0.2mm}
\end{equation}
where the normalization \vspace{-.3mm} constant (cf.~\eqref{eq:alpha0} and \eqref{eq:measUpdate}) reads
\begin{align}
\hspace{1mm}\underline{C}_k &= \bigg(\rmv\int \underline{\alpha}(\underline{\V{x}}_{k},\underline{\V{e}}_{k}, \underline{r}_{k} \rmv=\rmv 1) \prod^{M}_{l=1} \underline{\gamma}^{(P)}_{lk}(\underline{\V{x}}_{k},\underline{\V{e}}_{k}, \underline{r}_{k} \rmv=\rmv 1) \mathrm{d} \underline{\V{x}}_{k} \ist \mathrm{d} \underline{\V{e}}_{k} \nn\\[-1mm]
&+  \underline{\alpha}^{\mathrm{n}}_{k} \ist \prod^{M}_{l=1} \ist\ist  \underline{\xi}_{kl}^{(P)} \bigg)^{-1} \rmv\rmv\rmv\rmv\rmv.
\vspace{.8mm}
\label{eq:normConstant}
\end{align}

Similarly, the belief $b(\overline{\V{y}}_{k}) \triangleq b(\overline{\V{x}}_{k},\overline{\V{e}}_{k},\overline{r}_{k})$ of augmented new \ac{po} state $k \rmv=\rmv \{1,\dots,M\}$, is given \vspace{-1.2mm} by
\begin{equation}
\tilde{f}(\overline{\V{y}}_{k}) = \overline{C}_k \ist\ist \overline{q}(\overline{\V{y}}_{k}) \ist\ist \varsigma_{kk}^{(P)}(\overline{\V{y}}_{k}) \ist \prod^{M}_{l=k+1} \ist \overline{\gamma}^{(P)}_{lk}(\overline{\V{y}}_{k}).
\label{eq:beliefNew}
\end{equation}
Here, $\overline{C}_k$ is again the normalization constant that guarantees that \eqref{eq:beliefNew} is a valid probability distribution.

Note that a message passing order where messages are calculated sequentially and for each measurement individually is discussed in \cite{MeyWil:C20}. As demonstrated in \cite[Section V]{MeyWil:C20}, parallel processing leads to improved performance compared to sequential processing. 

\subsection{Computational Complexity and Scalability}
In the prediction step, \eqref{eq:Beta1c} has to be performed $\underline{K}$ times. Thus, its computational complexity scales as $\Set{O}(\underline{K})$. The computational complexity related to each message passing iteration $p \rmv\in\rmv \{1,\dots,P\}$ is discussed next. In the measurement evaluation step, for legacy PO $k \rmv\in\rmv \{1,\dots,\underline{K}\}$, a total of $M$ messages $\underline{\beta}_{kl}^{(p)}(b_{l})$ have to be calculated. Similarly, for every new PO $k \rmv\in\rmv \{1,\dots,M\}$, a total of $l \rmv\in\rmv \{k,\dots,M\}$ messages $\overline{\beta}_{kl}^{(p)}(b_{l})$ has to be obtained. Thus, the total number of messages is $\underline{K} M + 1/2 M^2$. The computational complexity related to calculating each individual message is constant in $\underline{K}$ and $M$. Also in the data association, measurement update, and extrinsic information steps, a total of $\underline{K} M + 1/2 \ist M^2$ messages have to be calculated at each of the three steps. \rd{The computational complexity related to the calculation of the individual messages is again constant in $\underline{K}$ and $M$.} For the data association and measurement update steps, this constant computational complexity is obtained by precomputing the sums  $\sum^{\underline{K}}_{k=1} \ist \underline{\beta}_{kl}^{(p)}(k) + \sum^{l}_{\substack{\ell=1}} \ist \overline{\beta}_{\ell l}^{(p)}(\underline{K} \rmv+\rmv \ell) + 1$ for each $l \rmv\in\rmv \{1,\dots,M\}$. Similarly, for the extrinsic information step, this constant computational complexity is obtained by precomputing the products $\prod^{M}_{l=1} \ist\ist\ist \underline{\gamma}^{(p)}_{lk}(\underline{\V{y}}_{k})$, $k \rmv\in\rmv \{1,\dots,\underline{K}\}$ and $\prod^{M}_{l=k+1} \ist\ist \overline{\gamma}^{(p)}_{lk}(\overline{\V{y}}_{k})$, $k \rmv\in\rmv \{1,\dots,M\}$. 

\rd{For $P$ fixed, it thus has been verified that the overall computational complexity only scales as $\Set{O}(\underline{K} M + M^2)$. (Here we have omitted $1/2$ since the $\Set{O}(\cdot)$ notation does not track constants.) Recalling that $K=\underline{K}+M$ is the total number of legacy and new POs, we can equivalently write $\Set{O}(KM)$. Conservatively, we consider this to be quadratic in the number of measurements and objects, since existing objects are also expected to contribute measurements.} We observed that increasing the number of message passing iterations beyond $P \rmv = \rmv 3$, does not significantly improve performance in typical \ac{eot} scenarios. Note that the computational complexity can be further reduced by preclustering of measurements (see, e.g., \cite[Section~IV]{GraLunOrg:J12}) and censoring of messages (see, e.g., \cite[Section~IV]{MeyWil:C20}). \rd{Preclustering combines the $M$ measurements  to a smaller number $M'\rmv<\rmv M$ of joint ``hyper measurements'' and replaces the single measurement ratios in \eqref{eq:pseudoLikelihood1} and \eqref{eq:pseudoLikelihood} by the corresponding product of \vspace{0mm} ratios. Censoring aims to omit messages related to new POs that are unlikely to represent an actual object.}

\section{Particle-Based Implementation}
\label{sec:particleBased}

\rd{For general state evolution and measurement models, the integrals in \eqref{eq:alpha1}--\eqref{eq:Beta1c} as well as the message products in \eqref{eq:alphaUpdated}--\eqref{eq:beliefNew} typically cannot be evaluated in closed form.} \rd{Therefore, we next present an approximate particle-based implementation of these operations that can be seen as a generalization of the particle-based implementation introduced in \cite{MeyBraWilHla:J17} to EOT. Pseudocode for the presented particle-based implementation is provided in \cite[Section~3]{MeyWil:SM21}.} Each belief $\tilde{f}(\V{y}_{k}) \triangleq \tilde{f}(\V{x}_{k}, \V{e}_{k}, r_{k})$ is represented by a set of particles and corresponding weights $\big\{ \big( \V{x}^{(j)}_{k} \ist, \V{e}^{(j)}_{k} \ist, w^{(j)}_{k} \big) \big\}_{j=1}^{J}$. 
More specifically, $\tilde{f}(\V{x}_{k},\V{e}_{k},1)$ is represented by $\big\{ \big( \V{x}^{(j)}_{k} \ist, \V{e}^{(j)}_{k} \ist, w^{(j)}_{k} \big) \big\}_{j=1}^{J}$, and $\tilde{f}(\V{x}_{k},\V{e}_{k},0)$ is 
given implicitly by the normalization property of $\tilde{f}(\V{x}_{k},\V{e}_{k},r_{k})$, i.e., $\tilde{f}(\V{x}_{k},\V{e}_{k},0) = 1 - \int \rmv\rmv\rmv\rmv \int \tilde{f}(\V{x}_{k},\V{e}_{k},1) \ist \mathrm{d} \V{x}_{k} \mathrm{d} \V{e}_{k}$. 
Contrary to conventional particle filtering \cite{AruMasGorCla:02,DouFreGor:01} and as in \cite{MeyBraWilHla:J17}, the particle weights $w^{(j)}_{k}$, $j \in \{1,\dots,J\}$ do not sum to one; instead, 
\begin{equation}
p^{\sist\text{e}}_{k} \ist\triangleq\ist \sum^{J}_{j=1} w^{(j)}_{k} \ist\approx\rmv \int \rmv\rmv\rmv\rmv \int \tilde{f}(\V{x}_{k}, \V{e}_{k}, 1) \ist \mathrm{d} \V{x}_{k} \mathrm{d} \V{e}_{k} \ist.
\label{eq:approxExistProb}
\end{equation}
Note that since $\int \rmv\rmv \int \tilde{f}(\V{x}_{k}, \V{e}_{k}, 1) \ist \mathrm{d} \V{x}_{k} \mathrm{d} \V{e}_{k}$ approximates the posterior probability of existence for the object, it follows that the sum of weights $p^{\sist\text{e}}_{k}$ is approximately equal to the posterior probability of existence.

\vspace{-1mm}

\subsection{Prediction}
\label{sec:pred}

The particle operations discussed in this section are performed for all legacy \acp{po} $k \!\in\! \{1,\dots,\underline{K}\}$ in parallel. For each legacy \ac{po} $k$, $J$ particles and weights 
$\big\{ \big( \V{x}^{-(j)}_{k}, \V{e}^{-(j)}_{k}, w^{-(j)}_{k} \big) \big\}_{j=1}^{J}$ representing the previous belief $\tilde{f}( \V{x}^{-}_{k}, \V{e}^{-}_{k}, r^{-}_{k})$ were calculated at the previous time $n \rmv-\! 1$ as described further below. Weighted particles $\big\{ \big( \underline{\V{x}}^{(j)}_{k}, \underline{\V{e}}^{(j)}_{k}, \underline{w}^{(1,j)}_{k} \big) \big\}_{j=1}^{J}$ 
representing the message $\underline{\alpha}(\underline{\V{x}}_{k},\underline{\V{e}}_{k},1)$ in \eqref{eq:alpha1} are now obtained as 
follows. First, for each particle $\big(\V{x}^{-(j)}_{k},\V{e}^{-(j)}_{k}\big)\ist$, $j \in \{1,\dots,J\}$, one particle $(\underline{\V{x}}^{(j)}_{k},\underline{\V{e}}^{(j)}_{k})$ is drawn from $f\big( \V{x}_{k}, \V{e}_{k} \big| \V{x}^{-(j)}_{k}, \V{e}^{-(j)}_{k}\big)$. 
Next, corresponding weights $w^{(1,j)}_{k} \rmv\rmv$, $j \in \{1,\dots,J\}$ are obtained \vspace{.5mm} as
\begin{align}
\hspace{-1mm} \underline{w}^{(1,j)}_{k} \rmv= p_{\ist\mathrm{s}} \ist\ist \mathpzc{e}^{-\mu_{\mathrm{m}}\big(\underline{\V{x}}^{(j)}_{k}\rmv, \ist\underline{\V{e}}^{(j)}_{k}\big)} \ist w^{-(j)}_{k}, \iist j \in \{1, \dots, J \}. \label{eq:alphaweights}\\[-3mm]
\nn
\end{align}
Note that the proposal distribution \cite{AruMasGorCla:02, DouFreGor:01} underlying \eqref{eq:alphaweights} is $f\big( \V{x}_{k}, \V{e}_{k} \big| \V{x}^{-(j)}_{k}\rmv\rmv\rmv, \V{e}^{-(j)}_{k})$ for $j \in \{1,\dots,J\}$.
\rd{A particle-based approximation $\tilde{\underline{\alpha}}^{\mathrm{n}}_k$ of $\underline{\alpha}^{\mathrm{n}}_k$ in \eqref{eq:alpha0} is now obtained \vspace{1mm} as
\begin{align}
\tilde{\underline{\alpha}}^{\mathrm{n}}_k = \big(1-p^{-\text{e}}_{k}\big) + \big(1\!-\rmv p_{\ist\mathrm{s}}\big) p^{-\text{e}}_{k}\rmv\label{eq:alpha1a} \\[-4mm]
\nn
\end{align}
where $p^{-\text{e}}_{k} \triangleq \sum^{J}_{j=1} w^{-(j)}_{k}$. Finally, a particle approximation $\tilde{\underline{\alpha}}_k$ of $\underline{\alpha}_k$ introduced in Section \ref{sec:predictionStep} is given \vspace{-1mm} by
\begin{align}
\tilde{\underline{\alpha}}_k = \sum^{J}_{j\rmv=\rmv1} \underline{w}^{(1,j)}_{k} + \tilde{\underline{\alpha}}^{\mathrm{n}}_k. 
\nn\\[-13mm]
\nn
\end{align}}

\subsection{Measurement Evaluation}
\label{sec:meaEval}

Let the weighted \vspace{-.6mm} particles $\big\{ \big( \underline{\V{x}}^{(j)}_{k}, \underline{\V{e}}^{(j)}_{k}, \underline{w}^{(p,j)}_{kl} \big) \big\}_{j=1}^{J}$ and the scalar $\tilde{\underline{\alpha}}^{(p)}_{kl} $ be a particle-based representation of $\underline{\alpha}_{kl}^{(p)}(\underline{\V{x}}_{k}, \underline{e}_{k}, \underline{r}_{k})$. For $p \rmv=\rmv 1$, we set  $\big\{ \big( \underline{\V{x}}^{(j)}_{k}, \underline{\V{e}}^{(j)}_{k},$ $\underline{w}^{(1,j)}_{kl} \big) \big\}_{j=1}^{J} \triangleq \big\{ \big( \underline{\V{x}}^{(j)}_{k}, \underline{\V{e}}^{(j)}_{k},$ $\underline{w}^{(1,j)}_{k} \big) \big\}_{j=1}^{J}$ and $\tilde{\underline{\alpha}}^{(p)}_{kl} \triangleq \tilde{\underline{\alpha}}_k$; for $p\rmv>\rmv1$\vspace{.5mm} this representation is calculated as discussed in Section \ref{sec:measurementUpdate}. An approximation $\tilde{\underline{\beta}}_{kl}^{(p)}$ of \vspace{.2mm} the message value $\underline{\beta}_{kl}^{(p)} \rmv\triangleq\rmv \underline{\beta}_{kl}^{(p)}(b_{l} \rmv=\rmv k)$ in \eqref{eq:Beta1b}, can now be obtained \vspace{.5mm} as
\begin{align}
\tilde{\underline{\beta}}_{kl}^{(p)} \rmv\rmv=\rmv\rmv \frac{1}{\mu_{\mathrm{fa}} \ist f_{\mathrm{fa}}(\V{z}_{l}) \ist \tilde{\underline{\alpha}}_{kl}^{(p)} } \rmv\rmv \sum^{J}_{j=1}  \rmv \underline{w}^{(p,j)}_{kl} \rmv \mu_{\mathrm{m}}(\underline{\V{x}}^{(j)}_k,\underline{\V{e}}^{(j)}_k) \rmv f(\V{z}_{l}| \underline{\V{x}}^{(j)}_k, \underline{\V{e}}^{(j)}_{k}). \nn\\[-3.5mm]
\nn
\end{align}
\rd{Here, $\sum^{J}_{j=1} \ist \underline{w}^{(p,j)}_{kl} \ist \mu_{\mathrm{m}}(\underline{\V{x}}^{(j)}_k,\underline{\V{e}}^{(j)}_k) \ist f(\V{z}_{l}| \underline{\V{x}}^{(j)}_k, \underline{\V{e}}^{(j)}_{k})$ is the Monte Carlo integration \cite{DouFreGor:01} of 
$ \int \rmv\rmv\rmv\rmv \int \mu_{\mathrm{m}}(\underline{\V{x}}_k,\underline{\V{e}}_k)  f(\V{z}_{l}| \underline{\V{x}}_k, \underline{\V{e}}_{k})$ $\underline{\alpha}_{kl}^{(p)}(\underline{\V{x}}_{k}, \underline{\V{e}}_{k}, \underline{r}_{k} \rmv\rmv\rmv=\rmv\rmv\rmv1) \ist \mathrm{d} \underline{\V{x}}_{k} \ist \mathrm{d} \underline{\V{e}}_{k}$ in \eqref{eq:beta1}. This Monte Carlo integration is based on the proposal distribution  $\underline{\alpha}_{kl}^{(p)}(\underline{\V{x}}_{k}, \underline{\V{e}}_{k},$ $\underline{r}_{k}\rmv\rmv=\rmv\rmv1)$.} Similarly, an approximation $\tilde{\overline{\beta}}_{kl}^{(p)}$ of the message values $\overline{\beta}_{kl}^{(p)}$  related to new \acp{po} can be obtained. \rd{Here, for Monte Carlo integration and further particle-based processing, it was found useful to use a proposal distribution $f_{\mathrm{p}} \rmv\big(\overline{\boldsymbol{x}}_k,\overline{\boldsymbol{e}}_k;\V{z}_k\big)$ that is calculated from the measurement $\V{z}_k$ and its uncertainty characterization, e.g., by means of the unscented transformation \cite{JulUhl:04}.}
 
Note that calculation of these messages relies on the likelihood function $ f(\V{z}_{l} \ist | \ist \V{x}_k, \V{e}_{k})$ introduced in \eqref{eq:Likelihood1}, which involves the integration $\int \rmv \mathrm{d} \V{v}^{(l)}_{k,n}$. For general nonlinear and non-Gaussian measurement models, evaluation of the likelihood function $f(\V{z}_l|\V{x}_k,\V{e}_k)$ can potentially also be performed by means of Monte Carlo integration \cite{DouFreGor:01}. Alternatively, if the measurement model $\V{d}(\cdot)$ is invertible in the sense that we can reformulate \eqref{eq:measModel} as
\begin{equation}
\V{d}^{-1}\big(\RV{z}_l + \RV{u}_l)  = \RV{p}_{k} \rmv+\rmv \RV{v}^{(l)}_{k} \label{eq:measModel2}
\vspace{.4mm}
\end{equation}
\rd{then an approximated linear-Gaussian measurement model \eqref{eq:measModelLin} can be obtained.} In particular, the \ac{pdf} of $\V{d}^{-1}\big(\V{z}_l + \RV{u}_l)$ (for observed $\V{z}_l$) is approximated by a Gaussian with mean $\tilde{\V{z}}_l = \V{d}^{-1}\big(\V{z}_l + \V{\mu}_{\V{u}_l})$ and covariance matrix $\M{\Sigma}_{\tilde{\V{z}}_l}$. This approximation can be obtained, e.g., by linearizing $\V{d}^{-1}\big(\cdot)$ or by applying the unscented transformation \cite{JulUhl:04}. \rd{As a result, closed-form expressions for $f(\V{z}_l|\V{x}_k,\V{e}_k)$ discussed in Section \ref{sec:MeasurementModel} and \cite[Section.~2]{MeyWil:SM21} can be used.}

\subsection{Measurement Update, Belief Calculation, and Extrinsic Informations}
\label{sec:measurementUpdate}

\vspace{.4mm}

\rd{The approximate measurement \vspace{.3mm} evaluation messages discussed in Section \ref{sec:meaEval} are used for the \vspace{-.2mm} subsequent approximation of $\underline{\xi}_{kl}^{(p)}$ and $\overline{\xi}_{kl}^{(p)}$ (cf.~\eqref{eq:sumMessages}) required in the measurement \vspace{.1mm} update step (cf.~\eqref{eq:measUpdate} and \eqref{eq:varSigma}).}
The calculation of the weighted particles  $\big\{ \big( \underline{\V{x}}^{(j)}_{k} \ist, \underline{\V{e}}^{(j)}_{k} \ist, \underline{w}^{(j)}_{k} \big) \big\}_{j=1}^{J}$ that represent the legacy PO belief in \eqref{eq:beliefLegacy} is discussed next. Weighted particles representing new PO beliefs  \eqref{eq:beliefNew} and extrinsic information messages in \eqref{eq:alphaUpdated} and \eqref{eq:alphaUpdated2} can be obtained by performing similar steps. 

The measurement update step \eqref{eq:measUpdate} and the belief calculation step \eqref{eq:beliefLegacy} are implemented by means of importance sampling \cite{AruMasGorCla:02, DouFreGor:01}. To that end, we first rewrite the \vspace{-.4mm} belief $\tilde{f}(\underline{\V{y}}_{k}) = \tilde{f}(\underline{\V{x}}_{k}, \underline{\V{e}}_{k}, \underline{r}_{k})$ 
in \eqref{eq:beliefLegacy} by \vspace{1mm} using \eqref{eq:alpha1}, \eqref{eq:alpha0}, and \eqref{eq:measUpdate}, i.e., 
\begin{align}
\underline{\tilde{f}}_{k} &\ist\propto\ist \underline{\alpha}^{\mathrm{n}}_k  \ist\ist \prod^{M}_{l=1}  \ist\ist\ist  \tilde{\underline{\xi}}_{kl}^{(P)}  \nn \\[3mm]
\hspace{-.1mm}\tilde{f}(\underline{\V{x}}_{k},\underline{\V{e}}_{k},1) &\ist\propto\ist \underline{\alpha}(\underline{\V{x}}_{k},\underline{\V{e}}_{k},1)   \label{eq:approxbelief_1}\\[2.5mm]
&\hspace{3.5mm}\times\prod^{M}_{l=1} \rmv \bigg(  \frac{\mu_{\mathrm{m}}(\underline{\V{x}}_{k},\underline{\V{e}}_{k}) f(\V{z}_{l}| \underline{\V{x}}_k, \underline{\V{e}}_{k})}{\mu_{\mathrm{fa}}  f_{\mathrm{fa}}(\V{z}_{l})} \ist + \ist \tilde{\underline{\xi}}_{kl}^{(P)} \rmv \bigg) \nn.\\[-4mm]
\nn
\end{align}
Here, we also \vspace{-1mm} replaced $\underline{\xi}_{kl}^{(P)}$ introduced in \eqref{eq:sumMessages} by its\vspace{1.5mm} particle-based approximation $\tilde{\underline{\xi}}_{kl}^{(P)}\rmv\rmv$ (cf.~Section \ref{sec:meaEval}), even\vspace{.5mm} though we do not indicate this additional 
approximation in our notation $\tilde{f}(\underline{\V{x}}_{k},\underline{\V{e}}_{k},\underline{r}_{k}) $. 

We now calculate nonnormalized weights \vspace{-1.5mm} corresponding \vspace{1.5mm} to \eqref{eq:approxbelief_1} for each particle $j \in \{1,\dots,J\}$ \vspace{1mm} as
\begin{align}
\hspace{0mm}\underline{w}^{\text{A}(j)}_{k} &=\ist \underline{w}^{(1,j)}_{k} \prod^{M}_{l=1} \rmv \bigg(  \frac{\mu_{\mathrm{m}}(\underline{\V{x}}^{(j)}_{k},\underline{\V{e}}^{(j)}_{k}) f(\V{z}_{l}| \underline{\V{x}}^{(j)}_k, \underline{\V{e}}^{(j)}_{k})}{\mu_{\mathrm{fa}}  f_{\mathrm{fa}}(\V{z}_{l})} \ist + \ist \tilde{\underline{\xi}}_{kl}^{(P)} \rmv \bigg). \label{eq:weightsA}
\end{align}
Note that here we perform importance sampling with proposal density $\underline{\alpha}(\underline{\V{x}}_{k},\underline{\V{e}}_{k},1)$. This proposal density is represented \vspace{.3mm} by the weighted particles 
$\big\{ \big( \underline{\V{x}}^{(j)}_{k}\rmv\rmv, \underline{\V{e}}^{(j)}_{k}\rmv\rmv, \underline{w}^{(1,j)}_{k} \big) \big\}_{j=1}^{J}$. Similarly, we calculate a single nonnormalized weight corresponding to \eqref{eq:approxbelief_1} 
\vspace{-2.5mm}
as
\begin{equation}
\underline{w}^{\text{B}}_{k} =\ist \tilde{\underline{\alpha}}^{\mathrm{n}}_k \ist\ist\ist \prod^{M}_{l=1}  \ist\ist\ist  \tilde{\underline{\xi}}_{kl}^{(P)}
\label{eq:wConstant}
\vspace{-1.5mm}
\end{equation}
in which $ \tilde{\underline{\alpha}}^{\mathrm{n}}_k$ has been calculated in \eqref{eq:alpha1a}.


Next, weighted \vspace{-.5mm} particles $\big\{ \big( \underline{\V{x}}^{(j)}_{k} \rmv, \underline{\V{e}}^{(j)}_{k} \rmv, \underline{w}^{(j)}_{k} \big) \big\}_{j=1}^{J}$ representing the\vspace{.15mm} belief $\tilde{f}(\underline{\V{x}}_{k},\underline{\V{e}}_{k},r_{k})$ 
are obtained by reusing the particles $\big\{ \big( \underline{\V{x}}^{(j)}_{k}\rmv\rmv, \underline{\V{e}}^{(j)}_{k} \big) \big\}_{j=1}^{J}$\vspace{.5mm} and calculating the corresponding new weights 
\vspace{-2mm}
as
\begin{equation}
\underline{w}^{(j)}_{k} \ist=\ist \frac{\underline{w}^{\text{A}(j)}_{k}}{\underline{w}^{\text{B}}_{k}+\sum^{J}_{j'=1} \ist \underline{w}^{\text{A}(j')}_{k}} \ist.
\label{eq:normlization}
\vspace{0mm}
\end{equation}
Here, $\underline{w}^{\text{B}}_{k}+\sum^{J}_{j=1} \underline{w}^{\text{A}(j)}_{k}$ is a particle-based approximation of the normalization constant $\underline{C}_{k}$ in \eqref{eq:normConstant}. 
We recall that $p^{\sist\text{e}}_{k} \ist= \sum^{J}_{j=1} \underline{w}^{(j)}_{k}$. 

\rd{Next, we discuss the calculation of \vspace{0mm} the particle representations of the extrinsic \vspace{.3mm} information messages in \eqref{eq:alphaUpdated} and \eqref{eq:alphaUpdated2}. For example, weighted \vspace{-.7mm} particles $\big\{ \big( \underline{\V{x}}^{(j)}_{k} \rmv, \underline{\V{e}}^{(j)}_{k} \rmv,  \underline{w}^{(p+1,j)}_{kl'} \big) \big\}_{j=1}^{J}$ \vspace{0mm} representing $\underline{\alpha}_{kl'}^{(p+1)}\big(\underline{\V{y}}_{k}\big)$ can be obtained as follows. The particle locations and extents $\big\{ \big( \underline{\V{x}}^{(j)}_{k}\rmv\rmv, \underline{\V{e}}^{(j)}_{k} \big) \big\}_{j=1}^{J}$ are again reused and the corresponding new weights $\underline{w}^{(p+1,j)}_{kl'}$ are obtained by \vspace{.8mm} replacing in \eqref{eq:weightsA} $\prod^{M}_{l=1}$ with $\prod^{M}_{\substack{l=1\\l \neq l'}}$ and $\tilde{\underline{\xi}}_{kl}^{(P)}$ with $\tilde{\underline{\xi}}_{kl}^{(p)} \rmv$\vspace{-.3mm}. This is equal to plugging \eqref{eq:measUpdate} into \eqref{eq:alphaUpdated} and evaluating \eqref{eq:alphaUpdated} at\vspace{.4mm} the particles $\big\{ \big( \underline{\V{x}}^{(j)}_{k}\rmv\rmv, \underline{\V{e}}^{(j)}_{k} \big) \big\}_{j=1}^{J}$. Here\vspace{-.5mm}, normalization of the weights in \eqref{eq:normlization} can be avoided and the constant $\tilde{\underline{\alpha}}_{kl'}^{(p+1)}$ \vspace{-.2mm} is obtained as $\tilde{\underline{\alpha}}_{kl'}^{(p+1)} \rmv=\rmv \tilde{\underline{\alpha}}_k\prod^{M}_{\substack{l=1\\l \neq l'}}  \ist  \tilde{\underline{\xi}}_{kl}^{(p)}$ (cf.~\eqref{eq:wConstant}).}

\subsection{Detection, Estimation, Pruning, and Resampling}
\label{sec:resampling}

\vspace{.7mm}

The weighted particles $\big\{ \big( \V{x}^{(j)}_{k} \ist, \V{e}^{(j)}_{k} \ist, w^{(j)}_{k} \big) \big\}_{j=1}^{J}$ can now be used for object detection and estimation. First, for each (legacy or new) \ac{po} $k$, 
an approximation $p^{\sist\text{e}}_{k}$ of the existence probability $p(r_{k} \!=\! 1|\V{z})$ is calculated from the particle weights $\big\{ w^{(j)}_{k} \big\}_{j=1}^{J}$ 
as in \eqref{eq:approxExistProb}. \ac{po} $k$ is then detected (i.e., considered to exist) if $p^{\sist\text{e}}_{k}$ is above a threshold $P_{\text{th}}$ (cf.\ Section \ref{sec:prob}).
For the detected objects $k$, an approximation of the MMSE estimate $\hat{\V{x}}^\text{MMSE}_{k}$ of the kinematic state in \eqref{eq:mmse} is calculated according 
\vspace{0mm}
to
\begin{equation}
\hat{\V{x}}_{k} \ist=\ist \frac{1}{p^{\sist\text{e}}_{k}} \sum_{j=1}^{J} \rmv w_{k}^{(j)} \ist \V{x}_{k}^{(j)} \ist. 
\label{eq:approxStateEstimation}
\vspace{.5mm}
\end{equation}
Similarly, an MMSE estimate  $\hat{\V{e}}_{k}$ of the extent state can be obtained by replacing $\V{x}_{k}^{(j)}$ in \eqref{eq:approxStateEstimation} by $\V{e}_{k}^{(j)}$.

As discussed in Section \ref{sec:MeasNewPOs}, the number of \acp{po} would grow with time $k$. Therefore, legacy and new \acp{po} whose approximate existence probabilities $p^{\sist\text{e}}_{k}$ are below a  threshold $P_{\text{pr}}$ are pruned \cite{MeyKroWilLauHlaBraWin:J18,Wil:J15} from the state space. In addition, a resampling step may be performed to avoid particle degeneracy \cite{DouFreGor:01,AruMasGorCla:02}.
 \vspace{-3mm}

\section{Numerical Results}
\label{sec:numericalResults}

Next, we report simulation results evaluating the performance of our method and comparing it with that of the PMBM filter implementation in \cite{GraBauReu:J17}. Note that a performance comparison with other data association algorithms based on the \ac{spa} has been presented in \cite{MeyWil:C20}.
\vspace{-2mm}

\subsection{Simulation Scenario}
 \vspace{-.5mm}
\label{sec:simScenarion}

We simulated ten extended objects whose states consist of two-dimensional (2-D) position and velocity, i.e., $\RV{x}_{k,n} \rmv= [\rv{p}^{(1)}_{k,n} \;\ist \rv{p}^{(2)}_{k,n} \;\ist \dot{\rv{p}}^{(1)}_{k,n} \;\ist \dot{\rv{p}}^{(2)}_{n,k}]^{\text{T}}\rmv\rmv$. The objects move in a region of interest (ROI) defined as $[-150\ist\text{m}, \ist 150\ist\text{m} ]  \times [-150\ist\text{m}, \ist 150\ist\text{m}]$ and according to the nearly constant-velocity motion model, i.e., $\RV{x}_{k,n} = \M{A}\ist\RV{x}_{k,n-1} + \M{W}\RV{c}_{k,n}$, where $\M{A} \rmv\in\rmv \mathbb{R}^{4 \times 4}$ and $\M{W} \rmv\in\rmv \mathbb{R}^{4 \times 2}$ are chosen as in \cite[Section 6.3.2]{ShaKirLi:B02} with $T \!=\! 0.2\ist$s, and $\RV{c}_{k,n} \rmv\sim \Set{N}(\V{c}_{k,n};\V{0},\sigma^2_{\rv{c}} \ist \M{I}_2)$ with $\sigma_{\rv{c}} \!=\rmv 1 \ist\ist \text{m}/\text{s}^2$ is an \ac{iid} sequence of 2-D Gaussian random vectors. 

\begin{figure*}[t!]
\centering
\subfloat[\label{fig:topology1}]{\scalebox{0.3}{\includegraphics[scale=.88]{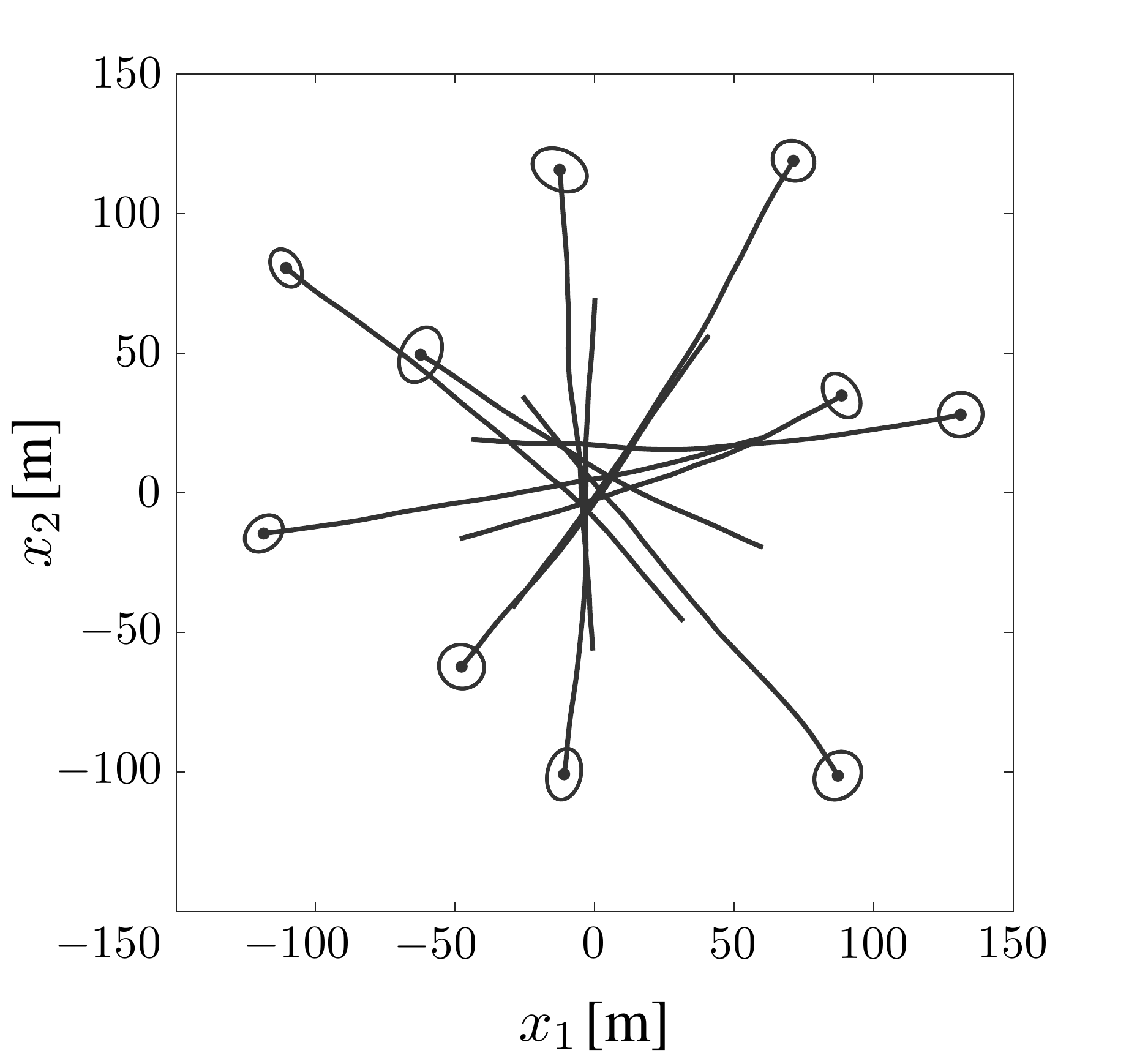}}}
\hspace{10mm} 
\subfloat[\label{fig:topology2}]{\scalebox{0.3}{\includegraphics[scale=.88]{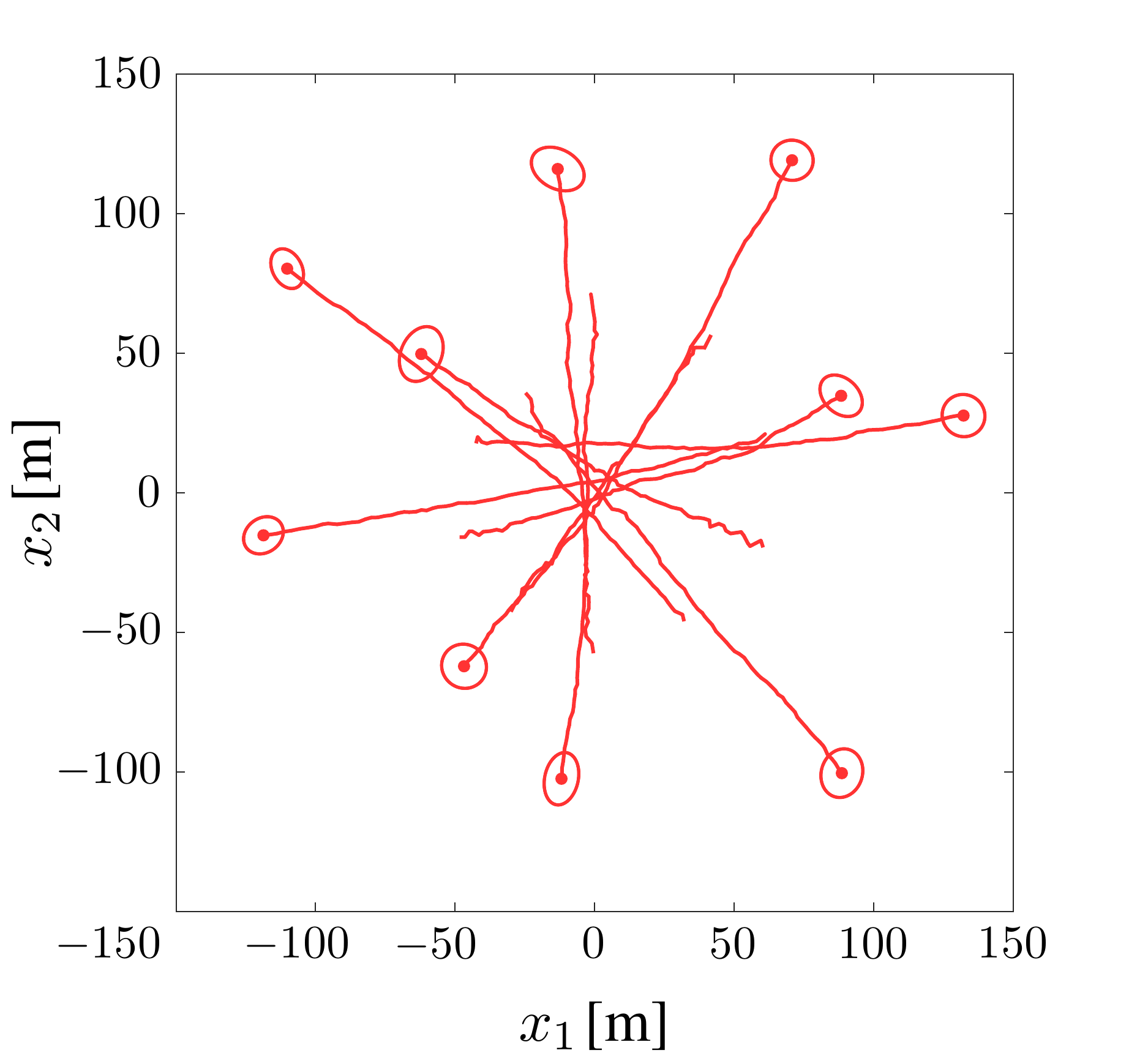}}}
\hspace{10mm}        
\subfloat[\label{fig:topology3}]{\scalebox{0.3}{\includegraphics[scale=.88]{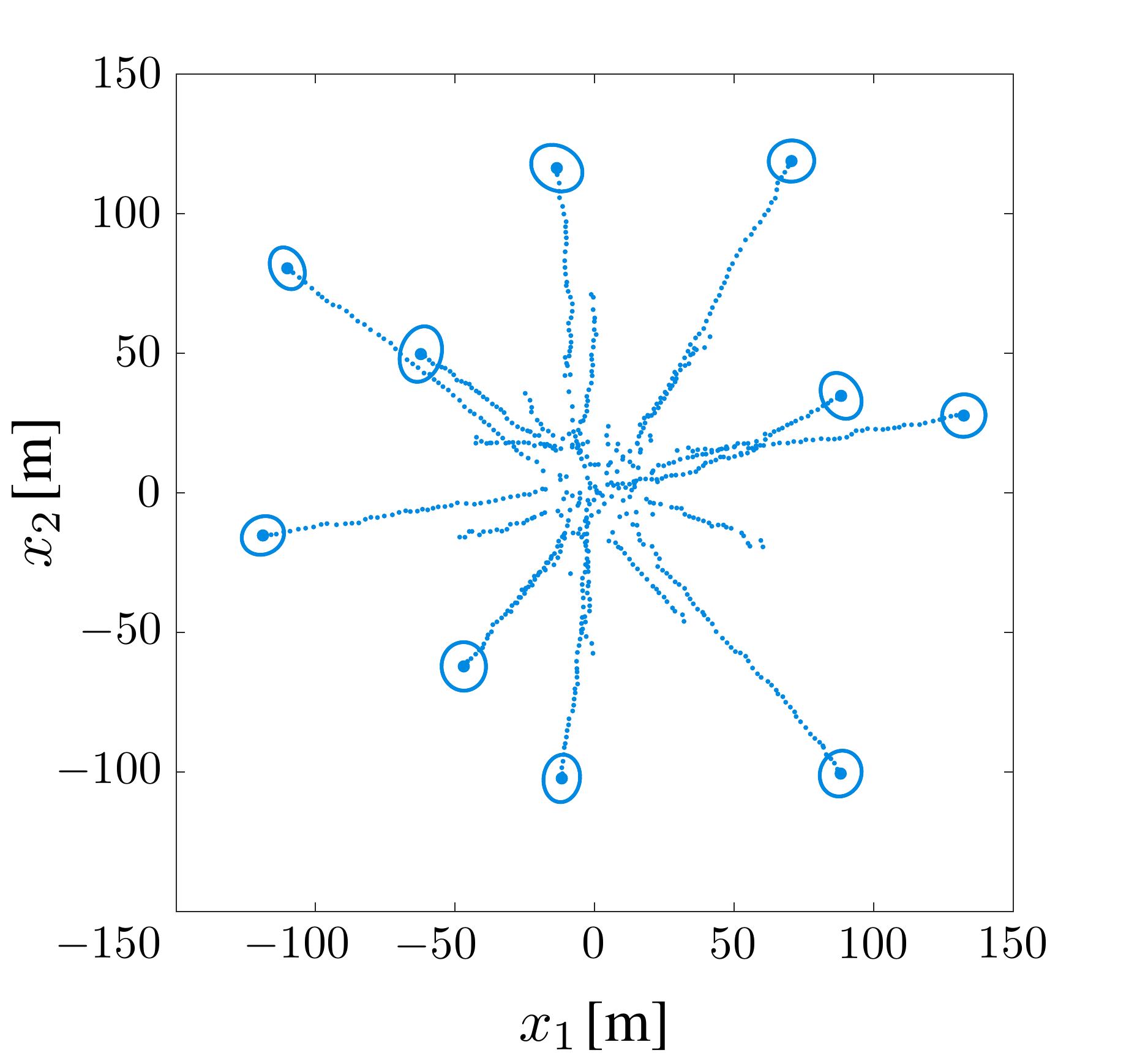}}}
\captionsetup{singlelinecheck = false, justification=justified}
\vspace{.5mm}
\caption{\rd{Example realization of true object tracks are depicted in (a). Estimation results of the proposed SPA and the PMBM-FR are depicted in (b) and (c), respectively. True and estimated object shapes at the last time step of an object's existence are also shown. In (a), track estimates provided by the proposed SPA method are shown as red line. In (b), state estimates provided by PMBM-FR are shown as blue dots.}}
\label{fig:topology}
\end{figure*}

We considered a challenging scenario where the ten object tracks intersect at the ROI center. The object tracks were generated by first assuming that the ten objects start moving toward the ROI center from positions uniformly placed on a circle of radius $75$ m about the ROI center, with an initial speed of $10$ m/s, and then letting the objects start to exist (i.e., generate measurements) in pairs at times $n \rmv\in\rmv \{ 3, 6, 9, 12, 15\}$. Object tracks intersect at the ROI center at around time 40 and disappear in pairs at times $n \rmv\in\rmv \{ 83, 86, 89, 92, 95\}$. The extent of each object is obtained by drawing a sample from the inverse Wishart distribution with mean matrix $3 \M{I}_2$ and $100$ degrees of freedom. The extent state of objects does not evolve in time, i.e., it remains unchanged for all time steps.  The survival probability is $p_{\mathrm{s}} \rmv=\rmv 0.99$. Example realizations of object tracks and extents are shown in Fig.~\ref{fig:topology}(a).

\begin{figure*}[t!]
\centering
\hspace{0mm} \subfloat[\label{fig:ospaError}]{\scalebox{0.3}{\includegraphics[scale=1]{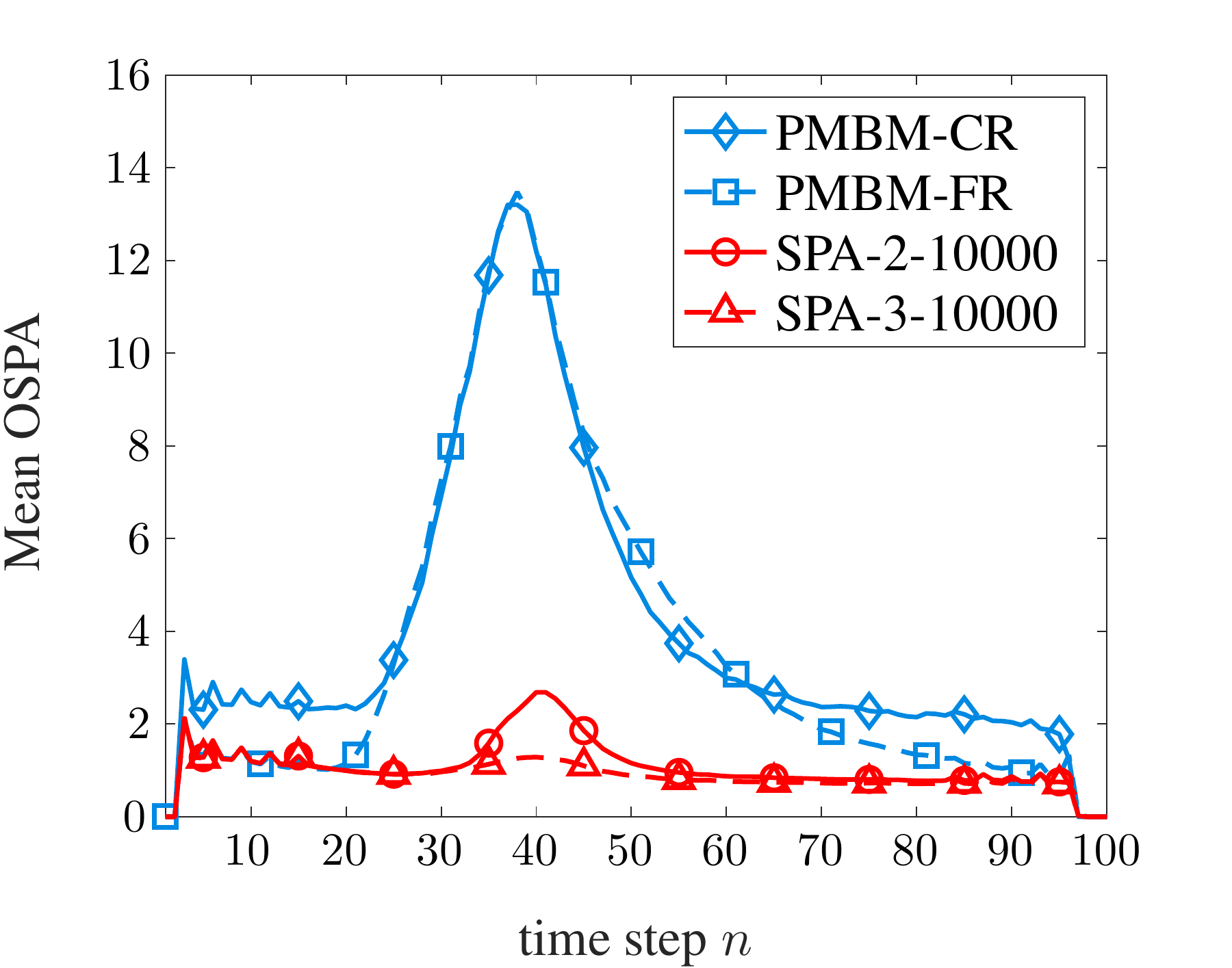}}}
\hspace{4mm} 
\subfloat[\label{fig:localizationError}]{\scalebox{0.3}{\includegraphics[scale=1]{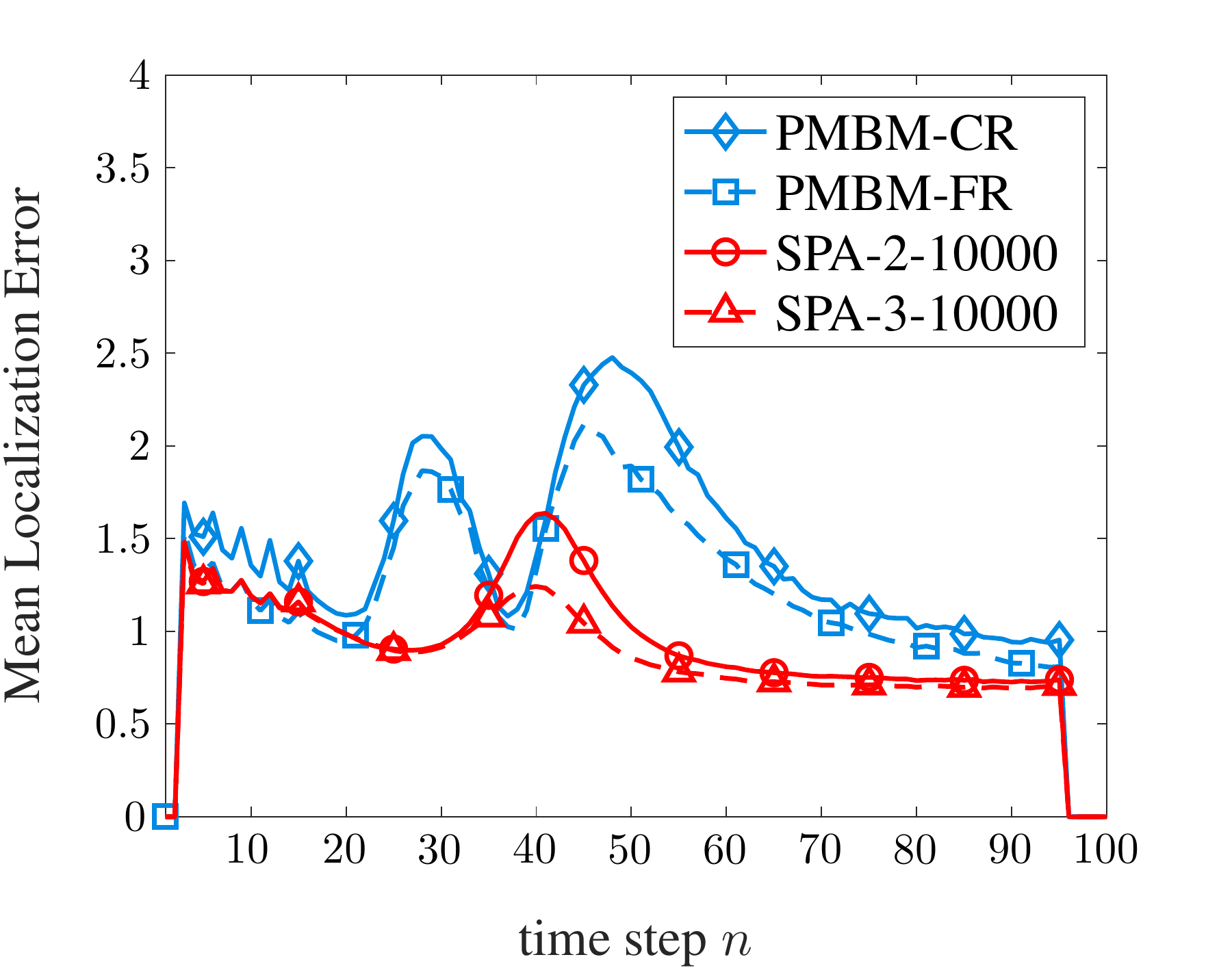}}}
\hspace{4mm}        
\subfloat[\label{fig:cardinalityError}]{\scalebox{0.3}{\includegraphics[scale=1]{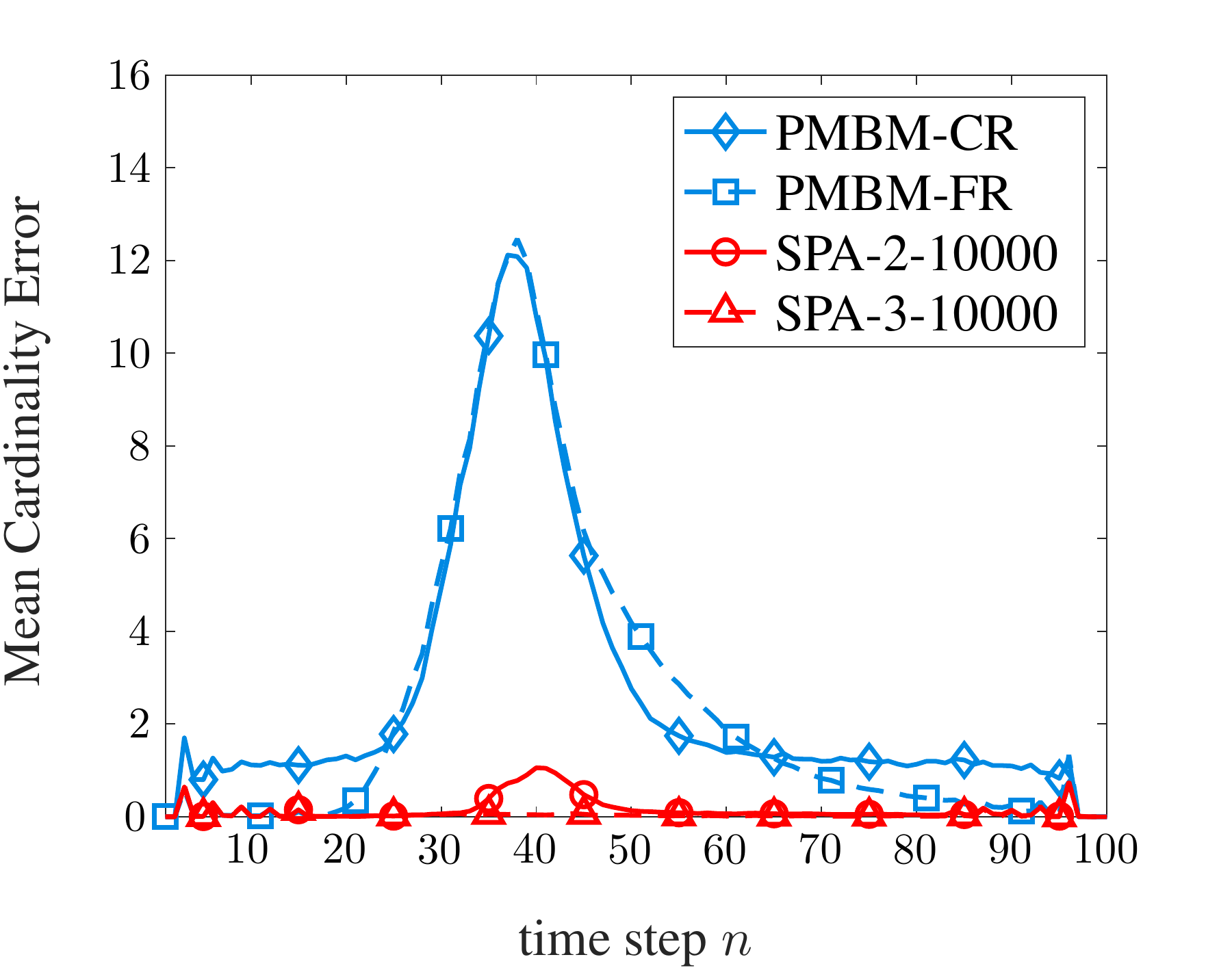}}}
\captionsetup{singlelinecheck = false, justification=justified}
\caption{Mean total OSPA (a) mean state error (b) and mean cardinality error (c) versus time of the proposed \ac{spa} method compared to the PMBM filter in a scenario with 10 closely-spaced extended objects.}
\label{fig:errors}
\end{figure*}

Since the PMBM filter is based on the Gaussian inverse Wishart model \cite{GraFatSve:J19}, we consider elliptical object shapes and a linear-Gaussian measurement model. \rd{In particular, a measurement $l$ that was originated by object $k$, is modeled as}  
\begin{equation}
\RV{z}_{l,n} = \RV{p}_{k,n} \rmv+\rmv \RV{v}^{(l)}_{k,n}  + \RV{u}_{l,n} \label{eq:measModelSim3}
\vspace{.8mm}
\end{equation}
\rd{where $\RV{u}_{l,n} \sim \mathcal{N}(\V{u}_{l,n} ;\V{0},\sigma^2_{\rv{u}}\M{I}_2)$ is the measurement noise with $\sigma_{\rv{u}} = 1$m. In addition, $\RV{v}_{k,n}^{(l)} \sim \mathcal{N}\big(\V{v}^{(l)}_{k,n};\V{0}, \M{\Sigma}_{\rv{v}} \big) $ is the random relative position of the reflection point with $\M{\Sigma}_{\rv{v}}$ determined by the extent state.} The mean of the number of measurements $\rv{L}_{k,n}$ is $\mu_{\mathrm{m}} \rmv=\rmv 8$ for all objects and the mean number of false alarm measurements is $\mu_{\mathrm{fa}} \!\rmv=\! 10$. The false alarm \ac{pdf} $f_{\mathrm{fa}}(\V{z}_{l,n})$ is uniform on the ROI. 

\rd{The performance of all simulated methods is measured by the \ac{ospa} \cite{SchVoVo:J08} and the generalized \ac{ospa} (GOSPA) \cite{RahGarSve:C17} metrics. Both metrics are based on the Gaussian-Wasserstein distance with parameters $p\rmv=\rmv1$ and $c\rmv=\rmv20$.} The threshold for object declaration is $P_{\text{th}} \rmv=\rmv 0.5$ and the threshold for pruning \acp{po} is $P_{\text{th}} \rmv=\rmv 10^{-3}$. \rd{These parameters are used for all simulated methods.}
\vspace{-2mm}

\subsection{Performance Comparison with the PMBM Filter}
\label{sec:compPMBM}

For the proposed \ac{spa}-based method we use the following settings. Newly detected objects are modeled as $f_{\mathrm{n}}(\overline{\V{x}}_{k,n},\overline{\V{e}}_{k,n})= f_{\mathrm{n}}(\overline{\V{p}}_{k,n}) f_{\mathrm{n}}(\overline{\V{m}}_{k,n})  f_{\mathrm{n}}(\overline{\V{e}}_{k,n}) $, with $f_{\mathrm{n}}(\overline{\V{p}}_{k,n})$ uniformly distributed on the ROI,  $f_{\mathrm{n}}(\overline{\V{m}}_{k,n})$ following a zero-mean Gaussian \ac{pdf} with covariance matrix $10^2 \ist \text{m}^2/\text{s}^2  \ist \M{I}_2 $, and $f_{\mathrm{n}}(\overline{\V{e}}_{k,n})$ distributed according to an inverse Wishart distribution with mean matrix $3 \ist \M{I}_2$ and $100$ degrees of freedom. The proposed method uses the state transition model in \eqref{eq:stateTransition} with $\M{V}(\V{m}_{k,n-1})$ given by $\M{I}_2$ and $q_{k,n} = 20000$. The mean number of newly detected objects is set to $\mu_{\mathrm{n}} = 10^{-2}\rmv$. To facilitate track initialization, we perform message censoring and ordering of measurements as discussed in \cite{MeyWil:C20}. The number of \ac{spa} iterations was set to $P \in \{2,3\}$ and the number of particles was set to $J\rmv=\rmv \{1000,10000\}$. The resulting parameter combinations are denoted as ``SPA-2-1000'', ``SPA-3-1000'', ``SPA-2-10000'',  and ``SPA-3-10000'', respectively. 

For the PMBM filter, we set the Poisson point process that represents object birth consistent with the newly detected object representation introduced above. Furthermore, the probability of detection is set to $1$. The Gamma distribution has an a priori mean of $\mu_{\mathrm{m}} \rmv=\rmv 8$ and a variance of $10^{-4}$. Its parameters remain unchanged at all time steps. The transformation matrix and maneuvering correlation constant (see \cite[Table III]{GraFatSve:J19}) used for extent prediction are set to $\M{I}_2$ and $10^5$, respectively. The PMBM implementation described in \cite{GraBauReu:J17} relies on measurement gating and clustering as well as pruning of global association events \cite{GraFatSve:J19}. The gate threshold is chosen such that the probability that an object-oriented measurement is in the gate is $0.999$. 

Clusters of measurements and likely association events are obtained by using the density-based spatial clustering (DBSCAN) and Murthy's algorithm, respectively. We simulated two different settings for measurement clustering and event pruning. Coarse clustering ``PMBM-C'' calculates measurement partitions by using the $50$ different distance values equally spaced between  $0.1$ and $5$ as well as a maximum number of $20$ assignments for each partition of measurements. Fine clustering ``PMBM-F'' clusters with $2000$ different distance values equally spaced between  $0.01$ and $20$ as well as uses a maximum number of $200$ assignments for each partition of measurements. \rd{Clustering is performed for each distance value individually. All resulting clusters are then combined into one joint set of clusters. In this way, a diverse set of overlapping clusters is obtained.} \rd{We also simulated variants of the PMBM that perform recycling of pruned Bernoulli components \cite{Wil:C12}. These variants are denoted as ``PMBM-CR'' and ``PMBM-FR''.}

\rd{Fig.\ \ref{fig:topology}(b) and (c) shows estimation results of SPA-10000 and PMBM-FR. By comparing Fig.\ \ref{fig:topology}(c) with Fig.\ \ref{fig:topology}(a), it can be seen that PMBM-FR is unable to accurately estimate the state of objects that are in close proximity.}
Fig.\ \ref{fig:errors} shows the mean OSPA error and its state and cardinality error contributions---averaged over 1200 simulation runs---of four simulated methods versus time. It can be seen that the proposed \ac{spa}-based methods outperform the PMBM at those time steps where objects are in close proximity. This can be explained by the fact that, due to its excellent scalability, the proposed \ac{spa}-based method can avoid clustering as performed by the PMBM filter implementations. In Fig.\ \ref{fig:errors}(c), it is shown that the main reason for the increased OSPA error of PMBM is an increased cardinality error. This is because large clusters that consist of measurements generated by multiple objects are associated to a single object. Thus, to certain other objects, no measurements are assigned, their probability of existence is reduced, and they are declared to be non-existent. The reduced state error of the PMBM methods compared to the proposed SPA method around time step 40 in Fig.\ \ref{fig:errors}(b) can be explained as follows. Since in this scenario PMBM methods tend to underestimate the number of objects, the optimum assignment step performed for OSPA calculation tends to find a solution with a lower state error. Table \ref{fig:table1} shows the mean GOSPA error and corresponding individual error contributions as well as runtimes per time step for MATLAB implementations on a single core of an Intel Xeon Gold 5222 CPU. Notably, despite not using gating, measurement clustering, and pruning of association events, the proposed SPA method has a runtime that is comparable with the PMBM\vspace{-1mm} filter.

\begin{table}
\vspace{3mm}
\centering
\begin{small}
\hspace{-.5mm}\begin{tabular}{  C{1.75cm} | C{.85cm} |  C{.85cm} |  C{.85cm} |  C{.85cm}  |  C{.85cm}  }
\hline\\[-3.4mm]
{\rule{0mm}{3.8mm} Method}&{Total}&{State}&{Missed}&{False}&{Runtime}\\[1mm]
\hline

\rowcolor{black!15!white} SPA-2-1000 & $13.3$ & $11.4$ &  $1.9$& $0.05$ & \underline{$1.05$} \rule{0mm}{3.3mm} \\[1mm]

SPA-2-10000 & $8.7$  & $8.0$ & $0.7$& $0.04$& $8.39$ \rule{0mm}{3.3mm} \\[1mm]

\rowcolor{black!15!white} SPA-3-1000 & $11.4$ & $10.3$ &  $1.1$& $0.07$ & $1.36$ \rule{0mm}{3.3mm} \\[1mm]

SPA-3-10000 & $\underline{7.5}$  & $\underline{7.3}$ & $\underline{0.1}$& $\underline{0.03}$& $11.67$ \rule{0mm}{3.3mm} \\[1mm]

\rowcolor{black!15!white} PMBM-FR & $22.5$ & $10.3$ & $11.8$ & $0.41$ & $50.59$  \rule{0mm}{3.3mm} \\[1mm]

 PMBM-F & $22.7$& $10.1$ &  $12.3$ &  $0.37$ & $52.71$  \rule{0mm}{3.3mm} \\[1mm]

\rowcolor{black!15!white} PMBM-CR & $25.3$  & $11.1$ & $10.2$& $4.01$& $3.75$\rule{0mm}{3.3mm} \\[1mm]

PMBM-C & $25.2$  & $10.8$ &  $10.8$&$3.55$& $4.29$\rule{0mm}{3.3mm} \\[1mm]

\hline
\end{tabular}
\end{small}
\vspace{2.5mm}
\caption{Mean GOSPA and runtime per time step in seconds for the considered simulation. The total GOSPA error as well as individual error contributions are shown.} 
\label{fig:table1}
\end{table}

\subsection{Size, Orientation, and Scalability}
\label{sec:moreResults}

\begin{figure}[t!]
\centering
\subfloat[\label{fig:ospaError}]{\scalebox{0.3}{\includegraphics[scale=1.25]{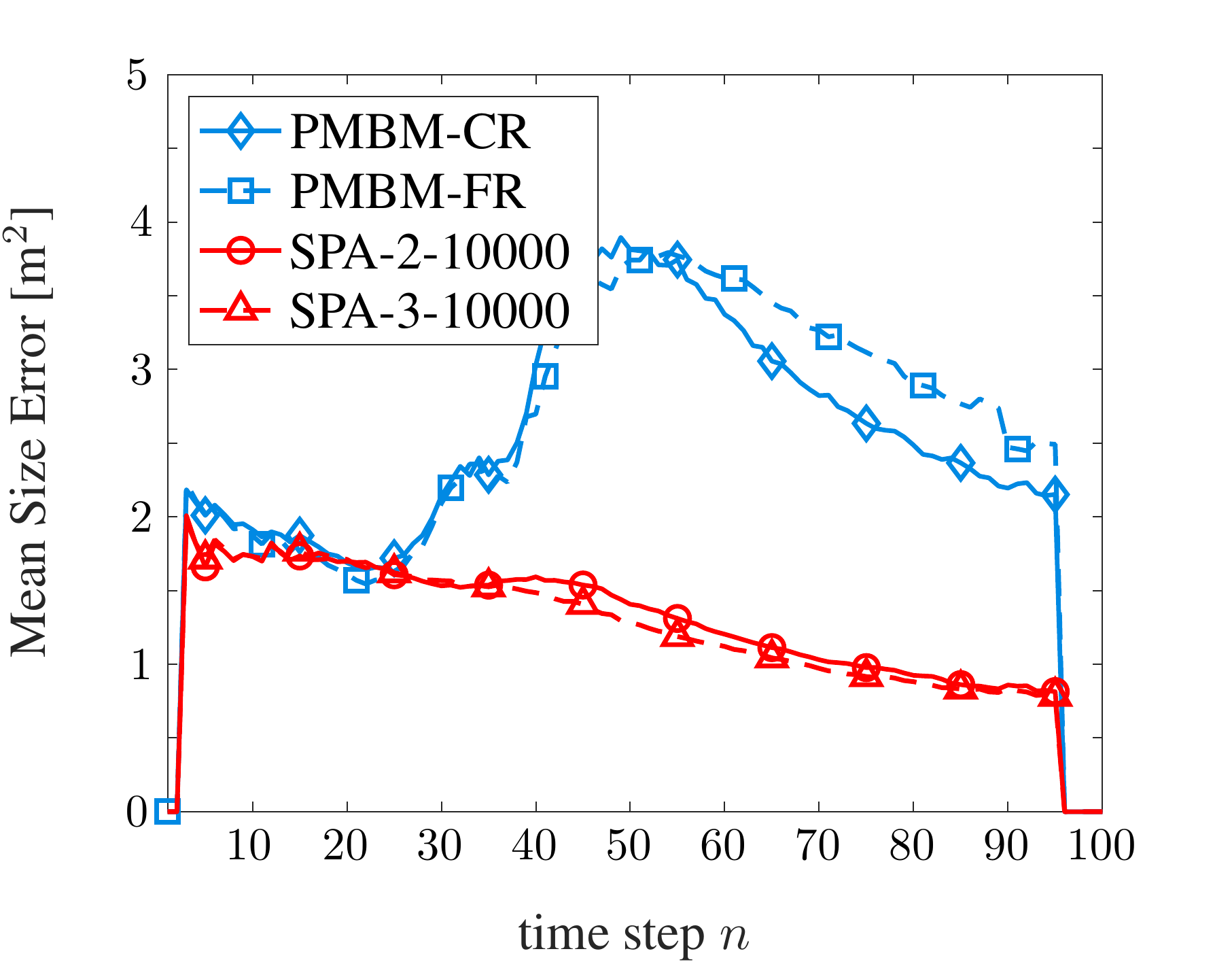}}} \\[0mm]
\subfloat[\label{fig:localizationError}]{\scalebox{0.3}{\includegraphics[scale=1.25]{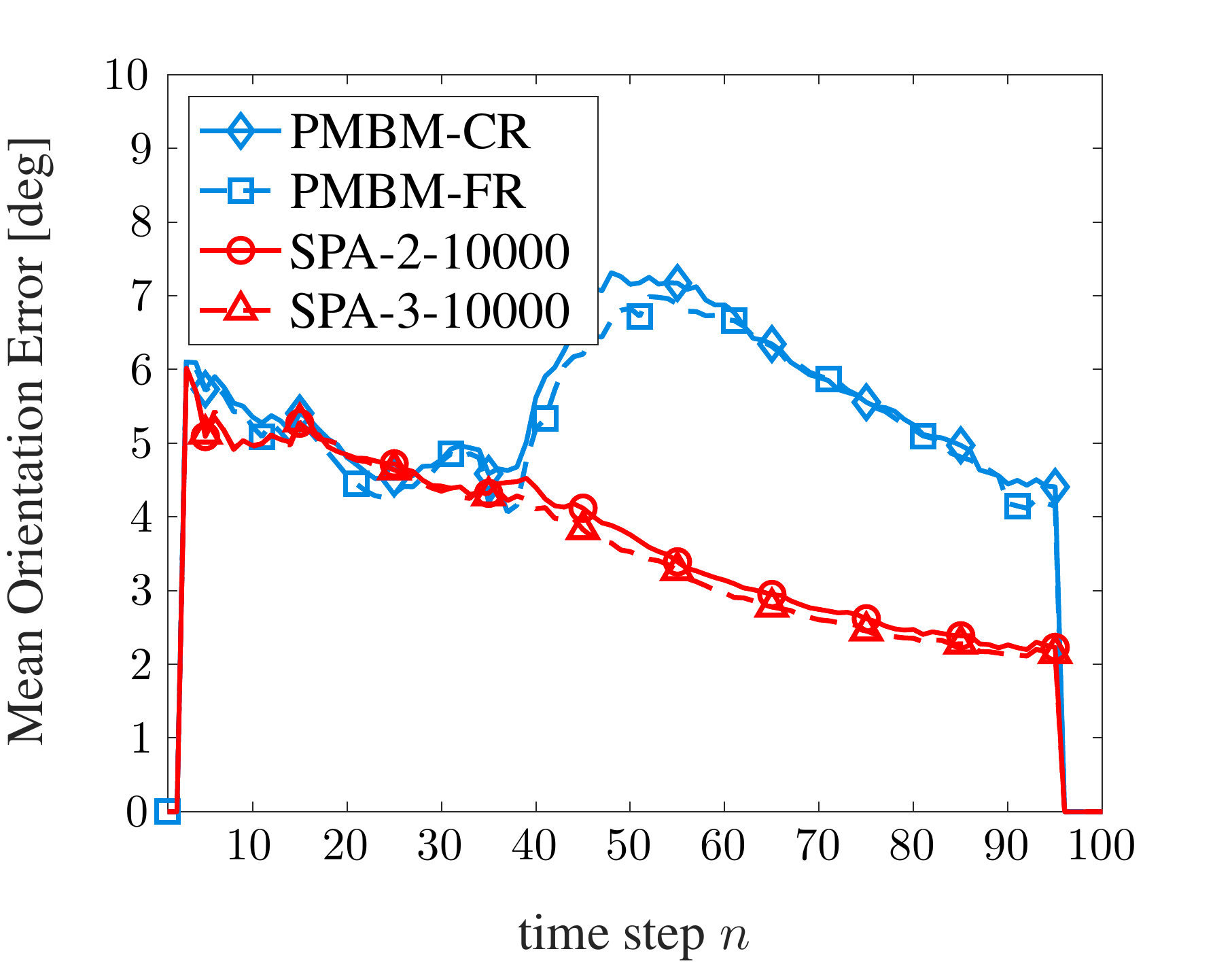}}}
\caption{Mean size error (a) and mean orientation error (b) of the proposed \ac{spa} method compared to the PMBM filter in a scenario with 10 closely-spaced extended objects.}
\label{fig:errorsOrientationExtent}
\end{figure}

\rd{To investigate the capability of the methods to estimate size and orientation, we considered a scenario similar to that discussed in Section \ref{sec:simScenarion}, changing the prior distribution of object extent to an inverse Wishart distribution with mean $\mathrm{diag}\big(3 \text{m}, 1.5 \text{m} \big)$. Estimated object sizes are calculated from estimated extent states $\hat{\M{E}}_{k,n}$ as the area of the represented ellipses, i.e., as $\pi \hat{\lambda}_{1,\M{E}_{k,n}} \rmv \hat{\lambda}_{2,\M{E}_{k,n}}$. Similarly, orientation is obtained by restricting the larger of the two eigenvectors $ \hat{\V{\lambda}}_{1,\M{E}_{k,n}}$ of $\hat{\M{E}}_{k,n}$ to the upper half plane and calculating its angle. To compute mean size errors and mean orientation errors, we use the optimum assignments of the \ac{ospa} \cite{SchVoVo:J08} metric  calculated as discussed in Section~\ref{sec:simScenarion} for each time steps and each of the 120 simulation runs. Then, we use these optimum assignments to calculate mean size errors and mean orientation errors. Fig.\ \ref{fig:errorsOrientationExtent} shows the resulting mean errors of the four simulated methods versus time. It can be seen that the proposed particle-based SPA method can estimate size and orientation more reliably than the PMBM filter when objects are in close proximity}

\begin{figure}[t!]
\centering
\scalebox{0.3}{\includegraphics[scale=1.25]{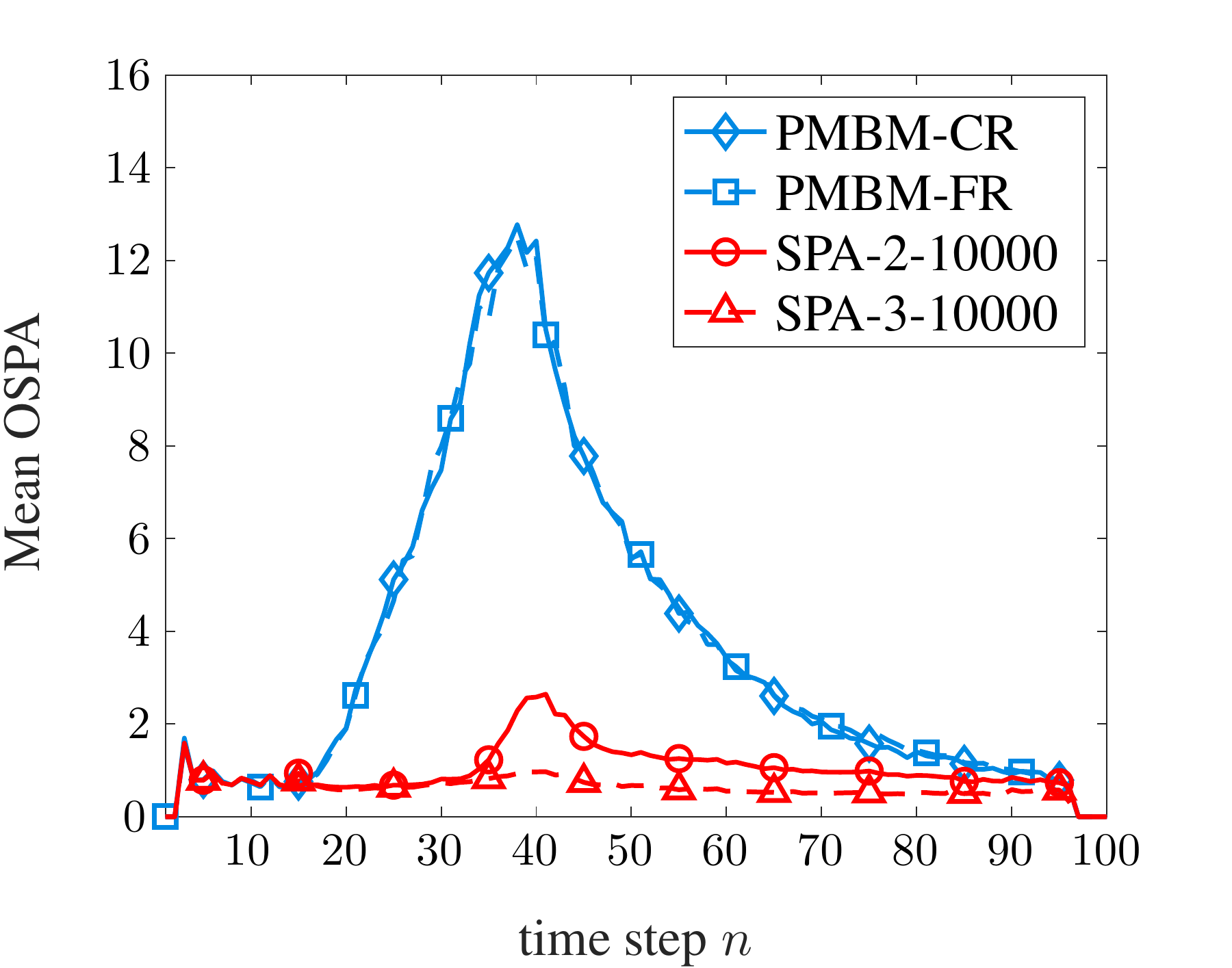}}
\caption{Mean total OSPA of the proposed \ac{spa} method compared to the PMBM filter in a scenario with 20 closely-spaced extended objects.}
\label{fig:errorsLarge}
\end{figure}

  
\rd{To demonstrate scalability, we furthermore simulated the same scenario discussed in Section \ref{sec:simScenarion} but with the number of objects increased to 20. Fig.\ \ref{fig:errorsLarge} shows the mean OSPA error---averaged over 120 simulation runs---of four simulated methods versus time. Also in this larger scenario the performance advantages of the SPA methods over the PMBM methods are comparable to the smaller scenario with 10 objects. Plots for individual error contributions are similar to the ones for the scenario with 10 objects and are thus omitted. The average runtimes per time step for MATLAB implementations on a single core of an Intel Xeon Gold 5222 CPU were measured as 48.1s for SPA-2-10000, 76.0s for SPA-3-10000, 6.7s for PMBM-CR, and 171.4s for PMBM-FR. We emphasize that this simulation is an extreme case, where all 20 objects come into close proximity at the mid-point in time. As previously discussed, additional, well-known  techniques can be utilized to limit the growth in complexity for well-spaced objects to linear by decoupling the joint EOT problem into smaller separate ones. The unique benefit of the proposed method is its ability to scale in problems where object proximity precludes trivial \vspace{-1mm} decoupling.}
   
\section{Real-Data Processing}

To validate our method, we present results in an urban autonomous driving scenario. The observations are part of the \textit{nuScenes} datasets \cite{nuScenes:W21} and were collected by a lidar sensor  mounted on the roof of an autonomous vehicle. The dataset consists of 850 scenes. Every scene has a duration of $20$s and consists of approximately 400 sets of lidar observations (or ``point clouds'') and corresponding ground truth annotations for vehicles and pedestrians. A single lidar measurement is given by the 3-D Cartesian coordinates of a reflecting obstacle in the environment.  Annotations are 3-D cubes defined by a position, dimensions, and heading. 

We focused on tracking of vehicles in the environment. To extract measurement points related to reflections on vehicles, we used the supervised learning method presented in \cite{ZhuJiaZhoLiYu:19}. In particular, we employed the point clouds and annotations of 700 scenes for the training of the deep neural network (see \cite{ZhuJiaZhoLiYu:19} for details). After training, measurement extraction provides cubes that indicate vehicle locations and corresponding confidence scores on $(0,1]$. We only used lidar observations inside vehicle-related detected cubes that have a confidence score larger than $0.25$ as measurements for the EOT methods. Measurements and ground truth annotations are converted to a global reference frame.  The ROI covers an area of $62,500$m$^2$. The four scenes S1--S4\footnote{ These four scenes can be identified by their tokens \texttt{8b31aa5cac3846a6}
\texttt{a197c507523926f9}, \texttt{9e79971671284ff482c63a1f658e703d}, \texttt{ee} 
\texttt{4979d9ef9e417f9337d09d62072692}, and \texttt{06b1e705b90641ce83}
\texttt{500b71c0dbf037}, respectively (see \cite{nuScenes:W21} for details). }
considered for performance evaluation, have not been used for the training.

We represent the extent state of vehicles by using the 2-D version of the cubical model introduced in Section \ref{sec:MeasurementModel} and \cite[Section 2]{MeyWil:SM21}. The kinematic states of vehicles is modeled by their  2-D position and velocity, as well as their turn rate $\rv{t}_{k,n}$ and their mean number of generated measurements $\rv{s}_{k,n}$, i.e., $\RV{x}_{k,n} \rmv= [\rv{p}^{(1)}_{k,n} \;\ist \rv{p}^{(2)}_{k,n} \;\ist \dot{\rv{p}}^{(1)}_{k,n} \;\ist \dot{\rv{p}}^{(2)}_{n,k} \;\ist \rv{t}_{k,n}  \;\ist \rv{s}_{k,n}  ]^{\text{T}}\rmv\rmv$. Since the number of measurements that a vehicle generates strongly depends on\vspace{0mm} its distance to the lidar sensor, the mean number of measurements is part of the random vehicle state, i.e., $\mu_{\mathrm{m}}(\RV{x}_{k,n},\RV{e}_{k,n}) = \rv{s}_{k,n}$. The mean number of measurements $\rv{s}_{k,n}$ follows the state transition model in \cite[Table III]{GraFatSve:J19} with parameter $\eta = 10$. The kinematic vehicle state and the extent state follow the state transition model in \eqref{eq:stateTransition} with $\mathpzc{f}(\cdot)$ and  $\M{V}(\cdot)$ as given in \cite{ GraOrg:J14a}. The parameters of the state transition model are  $q = 2000$, $\sigma_{\rv{c}} \!=\rmv 10 \ist\ist \text{m}/\text{s}^2$, and  $\sigma_{\rv{t}} \!=\rmv 0.003 \ist\ist \text{rad}/\text{s}^2$. The survival probability is $p_{\mathrm{s}} \rmv=\rmv 0.999$. In addition, the linear-Gaussian measurement model in \eqref{eq:measModelSim3} with $\sigma_{\rv{u}} = 5 \cdot 10^{-3}\ist$m was used. Extracted lidar observations are voxelized with a resolution of $0.5$m and their z-coordinate is ignored. We model newly detected vehicles as $f_{\mathrm{n}}(\overline{\V{x}}_{k,n},\overline{\V{e}}_{k,n})= f_{\mathrm{n}}(\overline{\V{p}}_{k,n}) f_{\mathrm{n}}(\overline{\V{e}}_{k,n})  f_{\mathrm{n}}(\overline{\V{m}}_{k,n})$, with $f_{\mathrm{n}}(\overline{\V{p}}_{k,n})$ uniformly distributed on the ROI and $f_{\mathrm{n}}(\overline{\V{e}}_{k,n})$ following an inverse Wishart distribution with mean matrix $3 \ist \M{I}_2$ and $30$ degrees of freedom. In addition, we set $f_{\mathrm{n}}(\overline{\V{m}}_{k,n}) = f_{\mathrm{n}}(\dot{\overline{\V{p}}}_{k,n}, \overline{t}_{k,n}) f_{\mathrm{n}}(\overline{s}_{k,n})$ where $f_{\mathrm{n}}(\dot{\overline{\V{p}}}_{k,n}, \overline{t}_{k,n})$ is zero-mean Gaussian distributed with covariance matrix $\mathrm{diag}\big(5^2 \text{m}^2/\text{s}^2\rmv, 5^2 \ist \text{m}^2/\text{s}^2\rmv, \ist\ist 0.01^2 \ist \text{rad}^2/\text{s}^2\big)$ and $f_{\mathrm{n}}(\overline{s}_{k,n})$ follows a Gamma distribution with mean $30$ and variance $5$. The number of \ac{spa} iterations is $P=3$ and the number of particles was again either set to $J\rmv=\rmv1000$ or to $J\rmv=\rmv10000$ (denoted as SPA-1000 and SPA-10000, respectively). All the other parameters are set as described in Section \ref{sec:compPMBM}.

 \begin{figure}[t!]
\centering
\includegraphics[scale=.18]{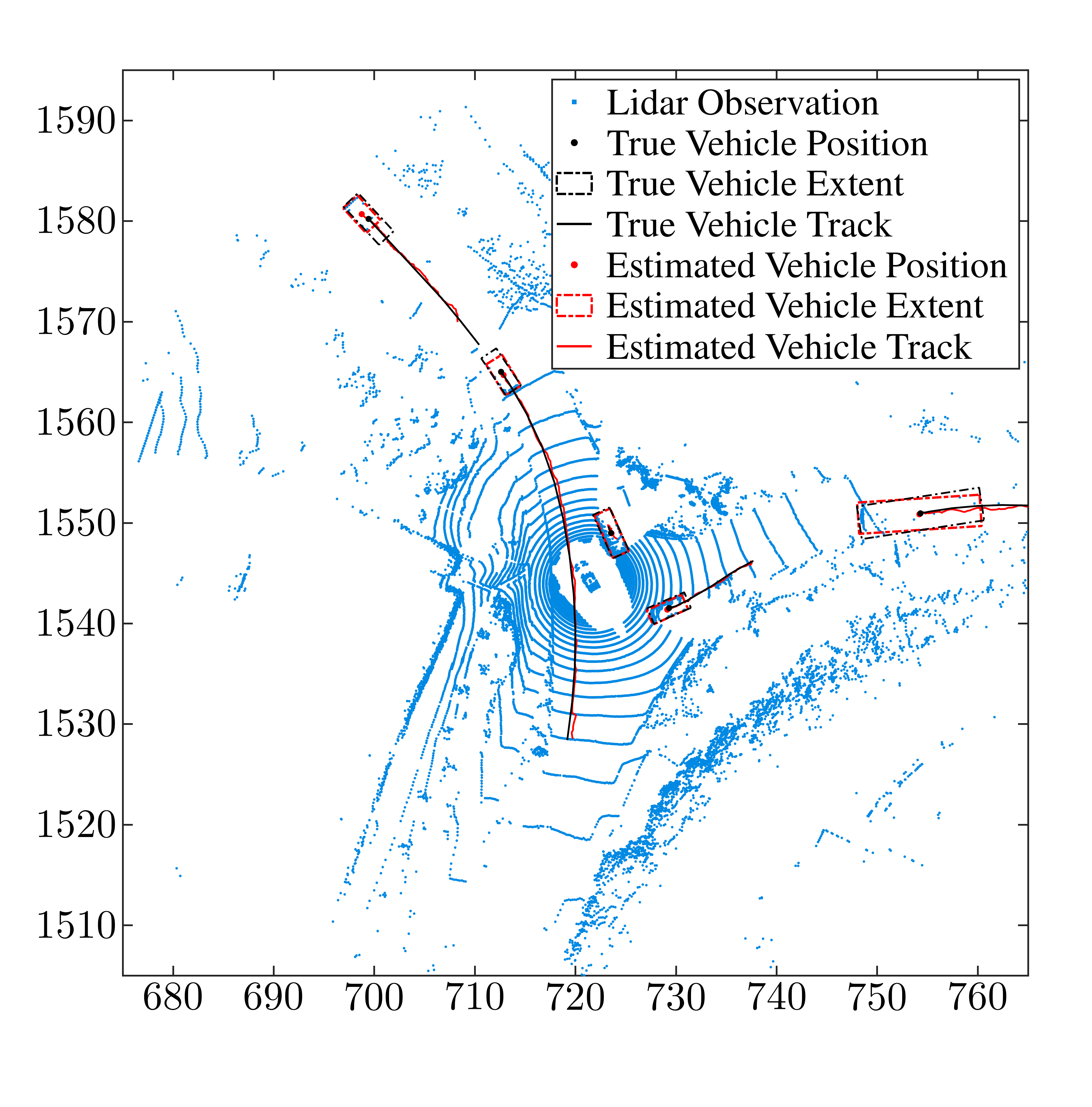}
\caption{Time step $n \rmv=\rmv 105$ of scene S4 in the \textit{nuScenes} urban autonomous driving dataset. Lidar observations, ground truth, and estimation results are shown. }
\label{fig:nuscenes}
\vspace{1mm}
\end{figure}

\rd{Fig.\ \ref{fig:nuscenes} shows the lidar observations, true vehicle positions, true vehicle extents, true vehicle tracks, estimated vehicle positions, estimated vehicle extents, and estimated vehicle tracks for time step $n\rmv=\rmv105$ of scene S4.} It can seen that all five actual vehicles are detected and tracked reliably. Videos of estimation results are available on \texttt{fmeyer.ucsd.edu}.

We use the PMBM filter based on the same system model described in Section \ref{sec:simScenarion} as a reference method. The driving noise standard deviation of the PMBM was set to $\sigma_{\rv{c}} \!=\rmv 30 \ist\ist \text{m}/\text{s}^2$. Furthermore, the transformation matrix and maneuvering correlation constant (see \cite[Table III]{GraFatSve:J19}) used for extent prediction were set to $\M{I}_2$ and $20$ respectively. All other model parameters were chosen as for the proposed method. The parameters for measurement clustering, event pruning, and recycling were set as discussed in Section \ref{sec:compPMBM} (denoted as PMBM-C, PMBM-F, PMBM-CR, and PMBM-FR).

As performance metric we use the generalized \ac{ospa} (GOSPA) \cite{RahGarSve:C17} based on the 2-norm with parameters $p\rmv=\rmv1$ and $c\rmv=\rmv5$. \rd{Only the kinematic state was used for the evaluation of the GOSPA.} The mean GOSPA for the individual methods---averaged over the four considered scenes and all time steps---as well as runtimes per time step on a single core of an Intel Xeon Gold 5222 CPU are summarized in Table \ref{fig:table2}. It can again be seen that the proposed \ac{spa}-based method outperforms the PMBM in terms of mean GOSPA and mean individual error contributions. These performance advantages of the proposed \ac{spa}-based method are related to its more accurate system model as well as its particle-based implementation. Interestingly, all PMBM variants perform very\vspace{0mm} similarly. 

\begin{table}
\vspace{1mm}
\centering
\begin{small}
\hspace{-.5mm}\begin{tabular}{  C{1.5cm} | C{.85cm} |  C{.85cm} |  C{.85cm} |  C{.85cm}  |  C{.85cm}  }
\hline\\[-3.4mm]
{\rule{0mm}{3.8mm} Method}&{Total}&{State}&{Missed}&{False}&{Runtime}\\[1mm]
\hline

\rowcolor{black!15!white} SPA-1000 & $7.63$ & $4.33$ &  $1.40$& $1.89$ & \underline{$0.07$} \rule{0mm}{3.3mm} \\[1mm]

SPA-10000 & $\underline{7.21}$  & $\underline{4.13}$ & $\underline{1.37}$& $\underline{1.71}$& $0.43$ \rule{0mm}{3.3mm} \\[1mm]

\rowcolor{black!15!white} PMBM-FR & $8.60$ & $4.99$ & $1.82$ & $1.80$ & $2.93$  \rule{0mm}{3.3mm} \\[1mm]

 PMBM-F & $8.60$& $4.98$ &  $1.82$ &  $1.81$ & $2.34$  \rule{0mm}{3.3mm} \\[1mm]

\rowcolor{black!15!white} PMBM-CR & $8.57$& $5.03$ &  $1.76$ &  $1.78$ & $0.16$ \rule{0mm}{3.3mm} \\[1mm]

PMBM-C & $8.61$  & $5.07$ &  $1.76$&$1.77$& $0.14$ \rule{0mm}{3.3mm} \\[1mm]

\hline
\end{tabular}
\end{small}
\vspace{2.5mm}
\caption{Mean GOSPA and runtime per time step in seconds for the considered vehicle tracking scenario. The total GOSPA error as well as individual error contributions are shown.} 
\label{fig:table2}
\end{table}

\section{Conclusion}

\rd{Detection, localization, and tracking of multiple objects is a key task in a variety of applications including autonomous navigation and applied ocean sciences. This paper introduced a scalable method for \ac{eot}. The proposed method is based on a factor graph formulation and the \ac{spa}. It dynamically introduces states of newly detected objects, and efficiently performs probabilistic data association, allowing for multiple measurements per object.  Scalable inference of object tracks and their geometric shape is enabled by modeling association uncertainty by measurement-oriented association variables and newly detected objects by a Poisson birth process. The fully particle-based approach can represent the extent of objects by different geometric shapes. In addition, it yields a computational complexity that only scales quadratically in the number of objects and the number of measurements. This excellent scalability translates to an improved EOT performance compared to existing methods because it makes it possible to (i)  generate and maintain a very large number of \acp{po} and (ii) avoid clustering of measurements, allowing problems with closely-spaced extended objects to be addressed.}

\rd{Simulation results in a challenging scenario with intersecting object tracks showed that the proposed method can outperform the recently introduced Poisson multi-Bernoulli (PMBM) filter despite yielding a reduced computational complexity. The application of the proposed method to real measurement data captured by a lidar sensor in an urban autonomous driving scenario also demonstrated performance advantages compared to the PMBM filter. For the extraction of vehicle-related lidar measurements, supervised learning based on a deep neural network was used. This motivates future research on multiobject tracking methods that closely integrate neural networks and iterative message passing. Other promising directions for future research are the development of highly parallelized variants of the proposed method that exploit particle flow \cite{LiCoates:17} and are suitable for real-time implementations on graphical processing units \vspace{0mm} (GPUs).}

\section*{Acknowledgment}
The authors would like to thank Dr.~E.~Leitinger and W.~Zhang for carefully reading the manuscript.

\begin{IEEEbiography}[{\includegraphics[width=1in,height=1.25in,clip,keepaspectratio]{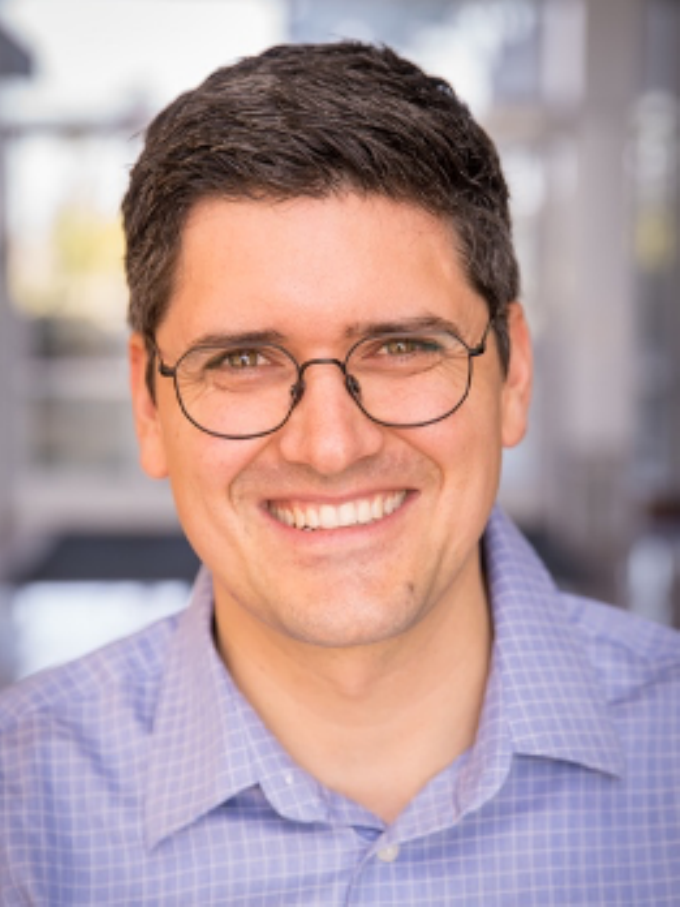}}]
{Florian~Meyer}
\input{Bios/FlorianMeyerBio} 
\end{IEEEbiography}
\begin{IEEEbiography}[{\includegraphics[width=1in,height=1.25in,clip,keepaspectratio]{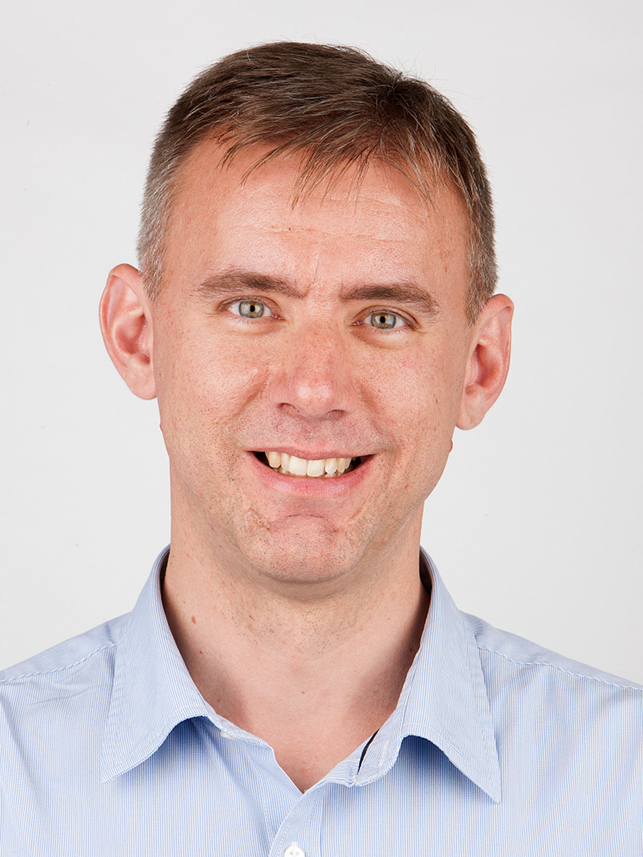}}]
{Jason~L.~Williams}
\input{Bios/JasonWilliams} 
\end{IEEEbiography}

\renewcommand{\baselinestretch}{1}
\selectfont
\bibliographystyle{IEEEtran}
\bibliography{IEEEabrv,StringDefinitions,Books,Papers,Temp}

\end{document}


\onehalfspacing

\title{\vspace{-5mm} Scalable Detection and Tracking\\[-1.5mm] of Geometric Extended Objects:\\[-1.5mm] Supplementary Material\vspace{0mm}}
\author{Florian Meyer and Jason L. Williams \vspace*{0mm}} 
\date{\today\vspace{-5mm}}

\maketitle

\renewcommand{\baselinestretch}{1.13}\small\normalsize

\noindent This manuscript provides derivations and pseudocode for the publication, ``Scalable Extended Object Tracking'' by the same authors \cite{MeyWil:J21}. \phantom{\ac{pdf}}

\section{The Joint Posterior \ac{pdf} $f \big( \V{y}_{0:n}, \V{b}_{1:n} \big |\V{z}_{1:n} \big)$}

In this section, we derive the expression of $f \big( \V{y}_{0:n}, \V{b}_{1:n} \big |\V{z}_{1:n} \big)$ which is represented by the factor graph in \cite[Fig.~2]{MeyWil:J21}  and provides the basis for the development of a scalable sum-product algorithm for extended object tracking.

\subsection{The Conditional Prior \ac{pdf} $p \big (\overline{\V{y}}_n, \V{b}_n, M_n \ist \big | \ist \underline{\V{y}}_n \big)$}

First, we derive the conditional \ac{pdf} $p(\overline{\V{y}}_n, b_n, M_n \ist | \ist \underline{\V{y}}_n)$, which will be used for the final expression of the joint posterior \ac{pdf} $f( \V{y}_{0:n}, \V{b}_{1:n}  \ist | \ist \V{z}_{1:n} )$ in Section~\ref{sec:jointPosterior}. For the sake of notational simplicity, we omit
the time index $n$ in the following.

We recall from \cite[Section~II]{MeyWil:J21}, that $\overline{\V{y}}$, $\underline{\V{y}}$, $\underline{K}$, and $M$ are the new \ac{po} states, legacy \ac{po} states, number of legacy \ac{po} states and number of measurements. For this derivation, we introduce the following additional auxiliary variables: (i) The number of newly detected objects $\overline{K}$ and their states $\tilde{\V{y}} = \big[\tilde{\V{y}}^{\T}_1 \rmv\cdots\ist \tilde{\V{y}}^{\T}_{\overline{K}} \big]^{\T}\rrmv$. (ii) The vector $\V{l} = [l_1 \cdots l_{\underline{K}+\overline{K}}]^{\T}$ that consists of the number of measurements $l_k$ generated by legacy \acp{po} $k \rmv\in\rmv \{1,\dots,\underline{K}\}$ and newly detected objects $k \rmv\in\rmv \{\underline{K}\rmv+\rmv1,\dots,\underline{K}\rmv+\rmv\overline{K}\}$. (iii) A pair of measurement-oriented association vectors $\tilde{\V{b}} \rmv= [\tilde{b}_{1},\dots,\tilde{b}_{M}]^{\mathrm{T}}$ and $\V{b} \rmv= [b_{1},\dots,b_{M}]^{\mathrm{T}}\rmv$, with elements 
\begin{equation}
\tilde{b}_{l} \ist\triangleq \begin{cases} 
    k \rmv\in\rmv \{1,\dots,\underline{K} + \overline{K}\} \ist , & \begin{minipage}[t]{110mm}if measurement $l$ is generated by legacy \ac{po} or newly detected object $k$\end{minipage}\\[-.5mm]
   0 \ist, & \begin{minipage}[t]{110mm}if measurement $l$ is not generated by an object (false alarm)\end{minipage}
  \end{cases}
\vspace{0mm}
\end{equation}
as well as $b_{l} = k' \rmv\in\rmv \{1,\dots,\underline{K} + M\}$ if measurement $l$ is generated by legacy \ac{po} or new \ac{po} $k'$ and $b_{l} = 0$ if measurement $l$ is not generated by an object (false alarm). (iv) A pair of binary object-oriented association vectors for newly detected objects $\tilde{\V{a}} \rmv= [\tilde{\V{a}}^{\mathrm{T}}_{1},\dots,\tilde{\V{a}}^{\mathrm{T}}_{\overline{K}}]^{\mathrm{T}}$ and new \acp{po} $\overline{\V{a}} \rmv= [\overline{\V{a}}^{\mathrm{T}}_{1},\dots,\overline{\V{a}}^{\mathrm{T}}_{M}]^{\mathrm{T}}\rmv$. In particular, the binary vector $\tilde{\V{a}}_k = [\tilde{a}_{k1} \cdots \tilde{a}_{kM} ]$ has elements $\tilde{a}_{kl}$, $l \in \{1,\dots,M\}$ given by $\tilde{a}_{kl} = 1$ if $\tilde{b}_l = \underline{K} \rmv+\rmv k$ and $\tilde{a}_{kl} = 0$ if $\tilde{b}_l \neq \underline{K} \rmv+\rmv k$. Similarly, the binary vector $\overline{\V{a}}_k = [\overline{a}_{k1} \cdots \overline{a}_{kM} ]$ has elements $\overline{a}_{kl}$, $l \in \{1,\dots,M\}$ given by $\overline{a}_{kl} = 1$ if $b_l = \underline{K} \rmv+\rmv k$ and $\overline{a}_{kl} = 0$ if $b_l \neq \underline{K} \rmv+\rmv k$. It is straightforward to see that there is a deterministic relationship between $\tilde{\V{b}}$ and  $\tilde{\V{a}}$ in the sense that $\tilde{\V{b}}$ implies $\tilde{\V{a}}$. Similarly, $\V{b}$ implies \vspace{2mm} $\V{a}$.

Before we can derive $p(\overline{\V{y}}, \V{b}, M \ist | \ist \underline{\V{y}}),$ we need to perform five preparatory \vspace{-1mm} steps.
\begin{enumerate}
\item In the first preparatory step, we will derive the conditional pdf $f(\overline{\V{y}}, \V{b} \ist | \ist \tilde{\V{b}}, \V{l}, \overline{K}, \tilde{\V{y}}, l_{\mathrm{fa}}, \underline{\V{y}}, \V{\pi})$. This \ac{pdf} establishes the relationship of the states of newly detected objects $\tilde{\V{y}}$ and association vector $\tilde{\V{b}}$ with new \ac{po} states $\overline{\V{y}}$ and the association vector $\V{b}$  \cite[Section~II-D]{MeyWil:J21}, based on a mapping $\V{\pi}$. The advantage of the new \ac{po} representation over a newly detected object representation is that after measurements $\V{z} \triangleq [\V{z}^{\T}_{1}$ $\dots \V{z}^{\T}_{{M}} ]^{\T}\rmv\rmv$ have been observed, the number of new \ac{po} states $M$ is fixed, while the number of newly detected objects $\overline{K}$ is still random (as some measurements may belong to legacy \acp{po} or may be false alarms).

Let us recall the number of measurements $M = \sum_{k=1}^{\underline{K} + \overline{K}} l_k + \lfa$ as well as the vectors of new \ac{po} states $\overline{\V{y}} \rmv=\rmv [\overline{\V{y}}^{\T}_1 \dots \overline{\V{y}}^{\T}_M]^{\T}\rmv\rmv$ and association variables $\overline{\V{a}} \rmv= [\overline{\V{a}}^{\T}_1 \dots \overline{\V{a}}^{\T}_M]^{\T}\rmv\rmv$. We also introduce a mapping $\V{\pi} : \{1,\dots,M\} \rightarrow \{0,\dots,\Knew\}$, which maps each element $\yt_{\V{\pi}(k)}$, $\V{\pi}(k) \in \{1,\dots,\overline{K}\}$ in $\yt$ to a different element $\ynew_k$ in $\ynew$; $\V{\pi}(k)=0$ denotes that $\ynew_k$ has no corresponding element in $\yt$. Mapping is always possible since $M \geq \overline{K}$.

New \ac{po} states $\ynew_k$ with $\V{\pi}(k) \in \{1,\dots,\overline{K}\}$ are equal to $\yt_{\V{\pi}(k)}$. New \ac{po} states $\ynew_k$ with $\V{\pi}(k) = 0$ have a zero existence probability, i.e., $\overline{r}_k \rmv=\rmv 0$, and their kinematic $\overline{\V{x}}_k$ and extent $\overline{\V{e}}_k$ state are distributed according to the arbitrary dummy pdf $\fd(\overline{\V{x}}_k,\overline{\V{e}}_k)$. Similarly, $\V{\pi}$ also maps elements of the new \ac{po} association vector $\overline{\V{a}}$ to elements in $\tilde{\V{a}}$, i.e., $\avnew_k = \avt_{\V{\pi}(k)}$ if $\V{\pi}(k)>0$, and $\avnew_k$ is equal to an all-zero vector if $\V{\pi}(k) = 0$. Thus, we obtain
\begin{align}
f(\overline{\V{y}}, \V{b} \ist | \ist \tilde{\V{b}},\tilde{\V{y}},\V{\pi}) = \Bigg(\prod_{k=1 \atop \V{\pi}(k)>0}^M \rmv\rmv\rmv  \rnew_k \ist\ist \delta\big(\ynew_k-\yt_{\V{\pi}(k)}\big) \ist \delta\big(\avnew_k-\avt_{\V{\pi}(k)}\big)\Bigg) 
\rmv\rmv \prod_{k=1 \atop \V{\pi}(k)=0}^M \rmv\rmv \fd(\xnew_k,\enew_k) \ist (1-\rnew_k) \hspace{.5mm} \delta\big(\ist \| \avnew_k \|_1\big)
\label{eq:mapping}
\end{align}
 where $\|\cdot\|_1$ is the one-norm. Based on $\V{\pi}$, every $\tilde{\V{b}}$ is also mapped to exactly one ${\V{b}}$. (We recall that $\tilde{\V{b}}$ implies $\tilde{\V{a}}$ and $\V{b}$ implies $\V{a}$.)

Next, we make the common assumption \cite{MeyKroWilLauHlaBraWin:J18,GraFatSve:J19}, that given the association vector $\tilde{\V{b}}$, new \ac{po} states $\overline{\V{y}}$ and the association vector $\V{b}$ are statistically independent of the number of false alarm measurements $l_{\mathrm{fa}}$, and the legacy  \ac{po} states $\underline{\V{y}}$. Furthermore, $\tilde{\V{b}}$ fully determines $\V{l}$. Thus, we \vspace{-1mm} obtain
\begin{equation}
f(\overline{\V{y}}, \V{b} \ist | \ist \tilde{\V{b}}, \V{l}, \overline{K}, \tilde{\V{y}}, l_{\mathrm{fa}}, \underline{\V{y}},\V{\pi}) = f(\overline{\V{y}}, \V{b} \ist | \ist \tilde{\V{b}},\tilde{\V{y}},\V{\pi}) .  \label{eq:condPDF3b}
\vspace{-1mm}
\end{equation}
In what follows and if not noted otherwise, it is assumed that the random variables of probability density or mass functions are consistent. This also holds for random variables in the condition. For example, the probability density function in \eqref{eq:condPDF3b} is only defined for valid mappings $\V{\pi}$, i.e, $\V{\pi}$ that map every newly detected object state in $\yt$ to exactly one different new PO state in $\ynew$.

\item As a second preparatory step, we introduce our modeling choice for the conditional \ac{pmf} $p( \V{\pi} \ist |$ $\tilde{\V{b}}, \V{l}, \overline{K},\tilde{\V{y}},  l_{\mathrm{fa}}, \underline{\V{y}})$. Given $\tilde{\V{b}}$, we choose a single representative mapping $\V{\pi}_{\tilde{\V{b}}}$ according to the following \vspace{-2mm} rules:
\begin{itemize}
\item for each $k\in\{1,\dots,\Knew\}$, set $\V{\pi}_{\tilde{\V{b}}}(\kmin)=k$, where $\kmin = \min_{l \ist | \ist \tilde{b}_l = \underline{K}+k} l$; and
\item for remaining $k'\in\{1,\dots,M\}$, set $\V{\pi}_{\tilde{\V{b}}}(k')=0$.
\end{itemize}
Note that this mapping $\V{\pi}_{\tilde{\V{b}}}$ implies that (i) new \ac{po} $k \rmv\in\rmv \{1,\dots,M\}$ exists ($\overline{r}_k = 1$) if and only if $b_{k} = \underline{K} \rmv+\rmv k$, (ii) new \ac{po} $k$ can only generate the $M-k+1$ measurements with indexes $l \rmv\in\rmv \{k,\dots,M\}$ and, (iii) the support of the measurement-oriented association variable with index $l \rmv\in\rmv \{1,\dots,M\}$ can be limited to $b_{l} \in \{1,\dots \underline{K} + l\}$.

The \ac{pmf} of $\V{\pi}$ conditioned on the association vector $\tilde{\V{b}}$ can now be written \vspace{-1.5mm} as 
\begin{equation}
p(\V{\pi} | \tilde{\V{b}}) = \delta (\V{\pi} - \V{\pi}_{\tilde{\V{b}}}).
\label{eq:mapping1}
\vspace{.1mm}
\end{equation}
It is straightforward to see that given $\tilde{\V{b}}$, the mapping $\V{\pi}$ is statistically independent of legacy \ac{po} states $\tilde{\V{y}}$, numbers of measurements $\V{l}$, number of false alarms $l_{\mathrm{fa}}$, as well as number of newly detected objects $\overline{K}$ and their states $\tilde{\V{y}}$. We can thus \vspace{-1mm} write
\begin{equation}
p( \V{\pi} \ist | \ist \tilde{\V{b}}, \V{l}, \overline{K}, \tilde{\V{y}},  l_{\mathrm{fa}},  \underline{\V{y}}) = p(\V{\pi} | \tilde{\V{b}}).
\label{eq:mapping2}
\end{equation}
Note that other choices are possible for  $p( \V{\pi} \ist |$ $\tilde{\V{b}}, \V{l}, \overline{K},\tilde{\V{y}},  l_{\mathrm{fa}}, \underline{\V{y}})$. For example, good choices can be made to obtain a conditional \ac{pdf} $p \big (\overline{\V{y}}_n, \V{b}_n, M_n \ist \big | \ist \underline{\V{y}}_n \big)$ that is symmetric with respect to permutations of the elements of the new \ac{po} vector $\overline{\V{y}}_n$. The particular choice used here, selects a single permutation of $\overline{\V{y}}_n$ as a representative such that the symmetric \ac{pdf} $p \big (\overline{\V{y}}_n, \V{b}_n, M_n \ist \big | \ist \underline{\V{y}}_n \big)$  can be obtained if desired by summing over all permutations of $\overline{\V{y}}_n$.

\item In the third preparatory step, we will start by deriving the conditional pmf \vspace{0mm} $p( \tilde{\V{b}} \ist | \ist \V{l}, \overline{K}, \tilde{\V{y}}, l_{\mathrm{fa}},  \underline{\V{y}})$. The number of association vectors $\tilde{\V{b}}$ can be calculated sequentially as discussed next. First, we draw (without replacement and without drawing order) $l_1$ measurements out of $ l_{\mathrm{fa}} + \sum^{\underline{K} + \overline{K}}_{k=1} l_{k}$ measurements; second we draw $l_2$ measurements out of $ l_{\mathrm{fa}} +  \sum^{\underline{K} + \overline{K}}_{k=2} l_{k}$ measurements; and finally we draw $l_{\overline{K}}$ measurements out of $l_{\mathrm{fa}} + l_{\overline{K}}$ measurements. Thus, by using basic results of combinatorics, the number of \vspace{-.5mm} valid $\tilde{\V{b}}$ can be calculated as
\begin{align}
\prod^{\underline{K} + \overline{K}}_{k=1} \rmv\rmv \left(\!
    \begin{array}{c}
      l_{\mathrm{fa}}  + \sum^{\underline{K} + \overline{K}}_{k'=k} \ist l_{k'} \\[1mm]
      l_{k} 
    \end{array}
  \!\right) &= \prod^{\underline{K}}_{k=1} \ist \frac{\Big(  l_{\mathrm{fa}}  + \sum^{\underline{K} + \overline{K}}_{k'=k} \ist l_{k'} \Big)!}{l_k!  \ist \Big(l_{\mathrm{fa}}  + \sum^{\underline{K} + \overline{K}}_{k'=k+1} \ist l_{k'}   \Big)!} \nn\\
  &= \frac{\Big( l_{\mathrm{fa}}  + \sum^{\underline{K} + \overline{K}}_{k=1} \ist l_{k}  \Big)!}{l_{\mathrm{fa}}! \ist \Big(\prod^{\underline{K} + \overline{K}}_{k=1} l_{k} ! \Big) \ist }. \nn\\[-8mm]
\end{align}

Assuming that each draw is equally likely, we \vspace{-1mm} obtain

\begin{equation}
p(\tilde{\V{b}} \ist | \ist \V{l},  l_{\mathrm{fa}} ) = \frac{l_{\mathrm{fa}}! \ist \Big(\prod^{\underline{K} + \overline{K}}_{k=1} l_{k} ! \Big) \ist }{\Big( l_{\mathrm{fa}}  + \sum^{\underline{K} + \overline{K}}_{k=1} \ist l_{k}  \Big)!}. \label{eq:condPDF1a}
\vspace{1mm}
\end{equation}
Now, we make the common assumption \cite{MeyKroWilLauHlaBraWin:J18,GraFatSve:J19}, that given the numbers of measurements $\V{l}$ and $l_{\mathrm{fa}}$ and without knowledge of the measurement vector, the association vector $\tilde{\V{b}}$ is statistically independent of the legacy and new object states $\underline{\V{y}}$ and $\tilde{\V{y}}$. Furthermore, the number of newly detected objects $\overline{K}$ is implied by $\V{l}$. Thus, we can write
\begin{equation}
p(\tilde{\V{b}} \ist | \ist \V{l}, \overline{K},  \tilde{\V{y}}, l_{\mathrm{fa}},  \underline{\V{y}}) = p(\tilde{\V{b}} \ist | \ist \V{l},  l_{\mathrm{fa}}) .  \label{eq:condPDF1b}
\vspace{1mm}
\end{equation}

\item As a fourth preparatory step, we derive the conditional \ac{pmf} $p( \ist \V{l},  \overline{K}, \tilde{\V{y}} \ist | \ist l_{\mathrm{fa}},$ $\underline{\V{y}}  \ist)$. Assuming existence of object $k \rmv\in \{1,\dots,\underline{K} + \overline{K}\}$, i.e., $r_k \rmv=\rmv 1$, the number of measurements it generates is Poisson distributed with mean $\mu_{\mathrm{m}}(\V{x}_k,\V{e}_k)$. It is assumed \cite{MeyKroWilLauHlaBraWin:J18,GraFatSve:J19} that the measurement generation is statistically independent across objects. Let us introduce the notation $\V{l} = \big[\underline{\V{l}}^{\T} \ist\ist\ist\ist \tilde{\V{l}}^{\ist\T} \ist\big]^{\T}$\rmv\rmv\rmv, where $\underline{\V{l}}  = \big[l_1 \cdots l_{\underline{K}}  \big]^{\T}$ are the numbers of measurements related to legacy \acp{po} and  $\tilde{\V{l}}  = \big[ l_{\underline{K} + 1} \cdots  l_{\underline{K} + \overline{K}}  \big]^{\T}\rmv\rmv$ are the numbers of measurements related to newly detected objects. By definition, newly detected objects  must generate a measurement, i.e., $\tilde{l}_k = 0$ has zero probability. Thus, for all $\tilde{l}_k > 0$, $k \in \{1,\dots,\overline{K}\}$ the conditional \ac{pmf} $p( \ist \tilde{\V{l}} \ist | \ist \overline{K}, \tilde{\V{y}} \ist) $ can be expressed \vspace{-1mm} as
\begin{equation}
p( \ist \tilde{\V{l}} \ist | \ist \overline{K}, \tilde{\V{y}}\ist)= \ist\ist\prod^{ \overline{K}}_{k= 1 }\ist\ist \frac{\mu_{\mathrm{m}}(\tilde{\V{x}}_{k},\tilde{\V{e}}_{k})^{\ist \tilde{l}_{k}}}{\tilde{l}_{k}!} \frac{\mathpzc{e}^{-\mu_{\mathrm{m}}(\tilde{\V{x}}_{k},\tilde{\V{e}}_{k})}}{1-\mathpzc{e}^{-\mu_{\mathrm{m}(\tilde{\V{x}}_{k},\tilde{\V{e}}_{k})}}}. \label{eq:condPMF}
\vspace{-1mm}
\end{equation}
It is assumed \cite{MeyKroWilLauHlaBraWin:J18,GraFatSve:J19} that conditioned on $\overline{K}$ and $\tilde{\V{y}}$, the numbers of measurements $\tilde{\V{l}}$ are conditionally independent of the legacy \ac{po} states $\underline{\V{y}}$, i.e., $p( \ist \tilde{\V{l}} \ist | \ist \overline{K}, \tilde{\V{y}}\ist) = p( \ist \tilde{\V{l}} \ist | \ist \overline{K}, \tilde{\V{y}}, \underline{\V{y}}\ist)$.

Next, we derive the conditional \ac{pmf} $p(\ist \underline{\V{l}} \ist | \ist \underline{\V{y}} \ist)$. By definition, legacy \ac{po} $k \rmv\in \{1,\dots,\underline{K}\}$ can only generate measurements if it exists, i.e., $\underline{r}_k = 1$. Let us define $\Set{D}_{\underline{\V{y}}} \triangleq \big\{ k \in \{1,\dots, \underline{K}\} \ist \big| \ist \underline{r}_k = 1 \big\}$. We can now \vspace{.5mm} write
\begin{equation}
p(\ist \underline{\V{l}} \ist | \ist \underline{\V{y}} \ist) = \Bigg ( \ist \prod_{k \in \cl{D}_{\underline{\V{y}}}} \frac{\mu_{\mathrm{m}} (\underline{\V{x}}_k,\underline{\V{e}}_k)^{\underline{l}_k}}{\underline{l}_{k}!} \ist \mathpzc{e}^{-\mu_{\mathrm{m}}(\underline{\V{x}}_k,\underline{\V{e}}_k)} \Bigg) \rmv \rmv \prod_{k \notin \cl{D}_{\underline{\V{y}}}}  \rmv \delta \big(\underline{l}_{k}\big)  \label{eq:condPDF6}
\vspace{.5mm}
\end{equation}
where $k \notin\Set{D}_{\underline{\V{y}}}$ is short for $ k \in \{1,\dots,\underline{K}\} \backslash \Set{D}_{\underline{\V{y}}}$ and $\delta( \cdot )$ is the Kronecker delta. It is assumed \cite{MeyKroWilLauHlaBraWin:J18,GraFatSve:J19} that conditioned on $\overline{K}$ and $\tilde{\V{y}}$, the numbers of measurements $\underline{\V{l}}$ are conditionally independent of the number $\overline{K}$ of newly detected objects and their states $\tilde{\V{y}}$, i.e., $p(\ist \underline{\V{l}} \ist | \ist \underline{\V{y}} \ist)= p( \ist \underline{\V{l}} \ist | \ist \overline{K}, \tilde{\V{y}}, \underline{\V{y}} \ist)$. Recalling that measurement generation is statistically independent across objects, we have $p( \ist \V{l} \ist | \ist \overline{K}, \tilde{\V{y}}, \underline{\V{y}} \ist) = p( \ist \tilde{\V{l}} \ist | \ist \overline{K}, \tilde{\V{y}}, \underline{\V{y}}\ist) \ist p( \ist \underline{\V{l}} \ist | \ist \overline{K}, \tilde{\V{y}}, \underline{\V{y}}\ist)$. By plugging in  \eqref{eq:condPMF} for $p( \ist \tilde{\V{l}} \ist | \ist \overline{K}, \tilde{\V{y}}, \underline{\V{y}}\ist) = p( \ist \tilde{\V{l}} \ist | \ist \overline{K}, \tilde{\V{y}}\ist)$ and \eqref{eq:condPDF6} for $p( \ist \underline{\V{l}} \ist | \ist \overline{K}, \tilde{\V{y}}, \underline{\V{y}}\ist) = p( \ist \underline{\V{l}} \ist | \ist \underline{\V{y}} \ist)$, we obtain
\begin{equation}
p( \ist \V{l} \ist | \ist \overline{K}, \tilde{\V{y}}, \underline{\V{y}} \ist) = \Bigg(  \prod^{ \overline{K}}_{k= 1 }\ist\ist \frac{\mu_{\mathrm{m}}(\tilde{\V{x}}_{k},\tilde{\V{e}}_{k})^{\ist \tilde{l}_{k}}}{\tilde{l}_{k}!} \frac{\mathpzc{e}^{-\mu_{\mathrm{m}}(\tilde{\V{x}}_{k},\tilde{\V{e}}_{k})}}{1-\mathpzc{e}^{-\mu_{\mathrm{m}(\tilde{\V{x}}_{k},\tilde{\V{e}}_{k})}}} \Bigg) \Bigg ( \ist \prod_{k \in \cl{D}_{\underline{\V{y}}}} \frac{\mu_{\mathrm{m}} (\underline{\V{x}}_k,\underline{\V{e}}_k)^{l_k}}{l_{k}!} \ist \mathpzc{e}^{-\mu_{\mathrm{m}}(\underline{\V{x}}_k,\underline{\V{e}}_k)} \Bigg) \rmv\rmv  \prod_{k \notin \cl{D}_{\underline{\V{y}}}}  \rmv \delta \big(l_{k}\big). \label{eq:condPDF7}
\vspace{.5mm}
\end{equation}

Newly detected objects must exist by definition. Thus, conditioned on their number $\overline{K}$, their joint state $\tilde{\V{y}}$ is distributed according \vspace{-4.5mm} to
\begin{align}
f(\yt | \Knew)  &=  \prod_{k=1}^{\Knew} \rt_k \ist f_{\mathrm{n}}(\xt_k, \et_k). \label{eq:priorNewPOs} \\[-8mm]
\nn
\end{align}

Furthermore, the number of newly detected objects is Poisson distributed with mean $\mu_{\mathrm{n}}$, i.e., 
\begin{equation}
p ( \ist \overline{K} \ist )= \frac{\mu^{\overline{K}}_{\mathrm{n}}}{\overline{K}!} \mathpzc{e}^{-\mu_{\mathrm{n}}}.
 \label{eq:priorK}
\end{equation}

We make the assumption \cite{MeyKroWilLauHlaBraWin:J18,GraFatSve:J19} that the number of newly detected objects $\overline{K}$ and their states $\yt$ are statistically independent of 
legacy PO states $\underline{\V{y}}$ and can thus \vspace{0mm} write $p( \ist \V{l}, \overline{K}, \tilde{\V{y}} \ist | \ist  \underline{\V{y}} \ist) = p( \ist \V{l} \ist | \ist \overline{K}, \tilde{\V{y}}, \underline{\V{y}} \ist) f(\yt | \Knew)  \ist p ( \ist \overline{K} \ist )$. Finally, we assume \cite{MeyKroWilLauHlaBraWin:J18,GraFatSve:J19} that $\V{l}$, $\overline{K}$, and $\tilde{\V{y}}$ are statistically independent of false alarm measurements $l_{\mathrm{fa}}$ conditioned on $\underline{\V{y}}$, i.e.\vspace{-4mm},
\begin{align}
f( \ist \V{l}, \overline{K}, \tilde{\V{y}}  \ist | \ist l_{\mathrm{fa}}, \underline{\V{y}} )  &=p( \ist \V{l}, \overline{K}, \tilde{\V{y}} \ist | \ist  \underline{\V{y}} \ist) \nn\\[2mm]
&= p( \ist \V{l} \ist | \ist \overline{K}, \tilde{\V{y}},  \underline{\V{y}}\ist)\ist p(\ist \tilde{\V{y}} |  \overline{K} \ist) \ist p( \ist \overline{K} \ist). 
\label{eq:condPDF4b} \\[-7mm]
\nn
\end{align}

\item In the fifth preparatory step, we derive the conditional \ac{pmf} $p(l_{\mathrm{fa}} | \underline{\V{y}}  \ist)$. We recall that false alarm measurements are Poisson distributed, \vspace{-3mm} i.e.,
\begin{equation}
p(l_{\mathrm{fa}}) = \frac{\mu^{l_{\mathrm{fa}}}_{\mathrm{fa}}}{l_{\mathrm{fa}}!} \mathpzc{e}^{-\mu_{\mathrm{fa}}}. \label{eq:condPDF5a}
\vspace{1mm}
\end{equation}
Now, we make the common assumption \cite{MeyKroWilLauHlaBraWin:J18,GraFatSve:J19}, that the number of false alarm measurements $l_{\mathrm{fa}}$, is statistically independent of legacy \ac{po} states $\underline{\V{y}}$, i.e.,
\begin{equation}
p(l_{\mathrm{fa}} \ist | \ist \underline{\V{y}} )  = p(l_{\mathrm{fa}}).  \label{eq:condPDF5b}
\vspace{2mm}
\end{equation}

\end{enumerate}
Next, we develop the hybrid probability distribution of $\overline{\V{y}}$, $\tilde{\V{y}}$, $\V{b}$, $\tilde{\V{b}}$, $\V{\pi}$, $\V{l}$, $\overline{K}$, and $l_{\mathrm{fa}}$, conditioned on $\underline{\V{y}}$. First, we use the chain rule for probability distributions and get the following expression
\begin{align}
f(\overline{\V{y}}, \tilde{\V{y}}, \V{b}, \tilde{\V{b}}, \V{\pi}, \V{l}, \overline{K}, l_{\mathrm{fa}} \ist | \ist \underline{\V{y}}) &= f(\overline{\V{y}}, \V{b} \ist | \ist \tilde{\V{b}}, \V{l}, \overline{K}, \tilde{\V{y}}, l_{\mathrm{fa}}, \underline{\V{y}}, \V{\pi} ) \ist\ist p( \V{\pi} \ist | \ist \tilde{\V{b}}, \V{l}, \overline{K}, \tilde{\V{y}},  l_{\mathrm{fa}},  \underline{\V{y}}) \ist p(\tilde{\V{b}} \ist | \ist \V{l}, \overline{K},  \tilde{\V{y}}, l_{\mathrm{fa}},  \underline{\V{y}}) \nn\\[1.5mm]
&\hspace{66.3mm} \times f( \ist \V{l}, \overline{K}, \tilde{\V{y}}  \ist | \ist l_{\mathrm{fa}}, \underline{\V{y}} )  \ist p(l_{\mathrm{fa}} \ist | \ist \underline{\V{y}} ). \label{eq:chainRule}  \\[-6mm]
\nn
\end{align}
By using the simplifications for $f(\overline{\V{y}}, \V{b} \ist | \ist \tilde{\V{b}}, \V{l}, \overline{K}, \tilde{\V{y}}, l_{\mathrm{fa}}, \underline{\V{y}}, \V{\pi} )$, $p( \V{\pi} \ist | \ist \tilde{\V{b}}, \V{l}, \overline{K}, \tilde{\V{y}},  l_{\mathrm{fa}},  \underline{\V{y}})$, $p(\tilde{\V{b}} \ist | \ist \V{l}, \overline{K},  \tilde{\V{y}}, l_{\mathrm{fa}},  \underline{\V{y}}) $, $f( \ist \V{l}, \overline{K},$ $\tilde{\V{y}}  \ist | \ist l_{\mathrm{fa}}, \underline{\V{y}} )$, 
and $p(l_{\mathrm{fa}} \ist | \ist \underline{\V{y}} )$  in  \eqref{eq:condPDF3b}, \eqref{eq:mapping2}, \eqref{eq:condPDF1b}, \eqref{eq:condPDF4b}, and \eqref{eq:condPDF5b} in \eqref{eq:chainRule}, we obtain
\begin{align}
\hspace{-1mm} f(\overline{\V{y}}, \tilde{\V{y}}, \V{b}, \tilde{\V{b}}, \V{\pi}, \V{l}, \overline{K}, l_{\mathrm{fa}} \ist | \ist \underline{\V{y}})  &= f(\overline{\V{y}}, \V{b} \ist | \ist \tilde{\V{b}},\tilde{\V{y}},\V{\pi}) \ist p( \V{\pi} \ist | \ist \tilde{\V{b}})\ist p(\tilde{\V{b}} \ist | \ist \V{l},  l_{\mathrm{fa}}) \ist p( \ist \V{l} \ist | \ist \overline{K}, \tilde{\V{y}},  \underline{\V{y}} \ist) \ist f(\yt | \Knew)  \ist p( \ist \overline{K} \ist)  \ist p(l_{\mathrm{fa}} ). \label{eq:chainRule2}  \\[-5mm]
\nn
\end{align}

By exploiting the fact that $\tilde{\V{b}}$ implies $\V{l}$ and using expressions for $ f(\overline{\V{y}}, \V{b} \ist | \ist \tilde{\V{b}},\tilde{\V{y}},\V{\pi})$, $p(\tilde{\V{b}} \ist | \ist \V{l},  l_{\mathrm{fa}}) $, $p( \ist \V{l} \ist | \ist \overline{K}, \tilde{\V{y}}, \underline{\V{y}} \ist)$, and $f(\yt | \Knew)$, in \eqref{eq:mapping}, \eqref{eq:condPDF1a}, \eqref{eq:condPDF7}, and \eqref{eq:priorNewPOs}, we can rewrite \eqref{eq:chainRule2} \vspace{1mm} as
\begin{align}
f(\overline{\V{y}}, \tilde{\V{y}}, \V{b}, \tilde{\V{b}}, \V{\pi}, \overline{K}, l_{\mathrm{fa}} \ist | \ist \underline{\V{y}})  &=  p( \V{\pi} \ist | \ist \tilde{\V{b}})  \ist p( \ist \overline{K} \ist)     \ist\ist p(l_{\mathrm{fa}} )  \nn\\[1.5mm]
&\hspace{2mm}\times  \Bigg(\prod_{k=1 \atop \V{\pi}(k)>0}^M \rmv\rmv\rmv  \rnew_k \ist\ist \delta\big(\ynew_k-\yt_{\V{\pi}(k)}\big) \ist \delta\big(\avnew_k-\avt_{\V{\pi}(k)}\big)\Bigg) \Bigg( \prod_{k=1 \atop \V{\pi}(k)=0}^M \rmv\rmv \fd(\xnew_k,\enew_k) \ist (1-\rnew_k) \hspace{.5mm} \delta\big(\ist  \tilde{l}_{\V{\pi}(k)} \big) \Bigg) \nn\\
&\hspace{2mm}\times\Bigg( \prod^{\overline{K}}_{k= 1} \ist \frac{\mu_{\mathrm{m}}(\tilde{\V{x}}_{k},\tilde{\V{e}}_{k})^{\ist \tilde{l}_{k}}}{\tilde{l}_{k}!} \frac{\mathpzc{e}^{-\mu_{\mathrm{m}}(\tilde{\V{x}}_{k},\tilde{\V{e}}_{k})}}{1-\mathpzc{e}^{-\mu_{\mathrm{m}(\tilde{\V{x}}_{k},\tilde{\V{e}}_{k})}}}\Bigg)  \Bigg ( \ist \prod_{k \in \cl{D}_{\underline{\V{y}}}} \frac{\mu_{\mathrm{m}} (\underline{\V{x}}_k,\underline{\V{e}}_k)^{l_k}}{l_{k}!} \ist \mathpzc{e}^{-\mu_{\mathrm{m}}(\underline{\V{x}}_k,\underline{\V{e}}_k)}  \rmv  \Bigg) \ist \nn\\
&\hspace{2mm}\times \Bigg( \prod_{k \notin \cl{D}_{\underline{\V{y}}}}  \rmv \delta \big(l_{k}\big) \rmv \rmv   \Bigg) \frac{l_{\mathrm{fa}}! \ist \Big(\prod^{\underline{K} + \overline{K}}_{k=1} l_{k} ! \Big) \ist }{\Big( l_{\mathrm{fa}}  + \sum^{\underline{K} + \overline{K}}_{k=1} \ist l_{k}  \Big)!} \hspace{1.3mm} \prod_{k=1}^{\Knew} \rt_k \ist f_{\mathrm{n}}(\xt_k, \et_k). \\[-5mm]
\nn
\end{align}
Note that $l_k = \sum^M_{l=1} \delta(\tilde{b}_l - k)$ for $k \in \big\{1,\dots,\underline{K} \rmv+\rmv \overline{K}\big\}$ and $\tilde{l}_k = \sum^M_{l=1} \delta(\tilde{b}_l - \underline{K} + k)$ for $k \in \big\{1,\dots,\overline{K}\big\}$.

Next, we marginalize out $\tilde{\V{y}}$, i.e., $f(\overline{\V{y}}, \V{b}, \tilde{\V{b}}, \V{\pi}, \overline{K}, l_{\mathrm{fa}} \ist | \ist \underline{\V{y}}) =  \int f(\overline{\V{y}}, \tilde{\V{y}}, \V{b}, \tilde{\V{b}}, \V{\pi}, \overline{K}, l_{\mathrm{fa}} \ist | \ist \underline{\V{y}}) \ist \mathrm{d} \tilde{\V{y}}$. After performing the marginalization, we \vspace{1mm} obtain
\begin{align}
f(\overline{\V{y}}, \V{b}, \tilde{\V{b}}, \V{\pi}, \overline{K}, l_{\mathrm{fa}} \ist | \ist \underline{\V{y}})  =   f\big(\overline{\V{y}}, \V{b}, \tilde{\V{b}} \ist | \ist  \overline{K}, \underline{\V{y}}, l_{\mathrm{fa}}, \V{\pi}\big)   p( \V{\pi} \ist | \ist \tilde{\V{b}}) \ist p( \ist \overline{K} \ist)     \ist\ist p(l_{\mathrm{fa}} ) \label{eq:margInt3}\\[-5mm]
\nn
\end{align}
where the conditional \ac{pdf} $f\big(\overline{\V{y}}, \V{b},\tilde{\V{b}} \ist | \ist  \overline{K},  \underline{\V{y}}, l_{\mathrm{fa}}, \V{\pi}\big)$ is given by
\begin{align}
f\big(\overline{\V{y}}, \V{b},\tilde{\V{b}} \ist | \ist  \overline{K},  \underline{\V{y}}, l_{\mathrm{fa}}, \V{\pi}\big) &= \Bigg(\prod_{k=1 \atop \V{\pi}(k)>0}^M \rmv\rmv\rmv\rmv \rnew_k \ist\ist \frac{\mu_{\mathrm{m}}(\overline{\V{x}}_k,\overline{\V{e}}_k)^{\ist \tilde{l}_{\V{\pi}(k)}}}{\tilde{l}_{ \V{\pi}(k)}!} \frac{\mathpzc{e}^{-\mu_{\mathrm{m}}(\overline{\V{x}}_k,\overline{\V{e}}_k)}}{1-\mathpzc{e}^{-\mu_{\mathrm{m}(\overline{\V{x}}_k,\overline{\V{e}}_k)}}} \ist f_{\mathrm{n}}(\overline{\V{x}}_k, \overline{\V{e}}_k) \ist \delta\big(\V{a}_k-\avt_{\V{\pi}(k)}\big)\Bigg) \nn\\[2mm]
&\times \Bigg ( \ist \prod_{k \in \cl{D}_{\underline{\V{y}}}} \frac{\mu_{\mathrm{m}} (\underline{\V{x}}_k,\underline{\V{e}}_k)^{l_k}}{l_{k}!} \ist \mathpzc{e}^{-\mu_{\mathrm{m}}(\underline{\V{x}}_k,\underline{\V{e}}_k)}  \rmv  \Bigg) \Bigg( \prod_{k \notin \cl{D}_{\underline{\V{y}}}}  \rmv \delta \big(l_{k}\big) \rmv \rmv   \Bigg) \nn\\
&\times  \Bigg(\prod_{k=1 \atop \V{\pi}(k)=0}^M \rmv\rmv \fd(\xnew_k,\enew_k) \ist (1-\rnew_k) \hspace{.5mm} \delta\big(\ist \tilde{l}_{\V{\pi}(k)}  \big) \Bigg) \frac{l_{\mathrm{fa}}! \ist \Big(\prod^{\underline{K} + \overline{K}}_{k=1} l_{k} ! \Big) \ist }{\Big( l_{\mathrm{fa}}  + \sum^{\underline{K} + \overline{K}}_{k=1} \ist l_{k}  \Big)!} 
\label{eq:margInt2}
\end{align}
(We recall that $\tilde{\V{b}}$ implies $\tilde{\V{a}}$ and $\V{b}$ implies $\V{a}$.)

Next, we marginalize out $\V{\pi}$. In particular, we replace the \ac{pmf} $p( \V{\pi} \ist | \ist \tilde{\V{b}} )$ in \eqref{eq:margInt3} by $\delta (\V{\pi} - \V{\pi}_{\tilde{\V{b}}})$ (cf. \eqref{eq:mapping1}) and perform the summation $\sum_{\V{\pi}}$ over all possible mappings $\V{\pi}$. In this way, we obtain
\begin{align}
&f(\overline{\V{y}}, \V{b}, \tilde{\V{b}}, \overline{K}, l_{\mathrm{fa}} \ist | \ist \underline{\V{y}})  = p( \ist \overline{K} \ist)     \ist\ist p(l_{\mathrm{fa}} ) f\big(\overline{\V{y}},\V{b}, \tilde{\V{b}}  \ist | \ist  \overline{K}, \underline{\V{y}}, l_{\mathrm{fa}}, \V{\pi}_{\tilde{\V{b}}} \big) .  \nn
\end{align}
Now, we note that $\V{\pi}_{\tilde{\V{b}}}$ implies the total number of measurements  $M$ and marginalize \vspace{2mm} out $\tilde{\V{b}}$, i.e.,
\begin{align}
&f(\overline{\V{y}}, \V{b}, \overline{K}, l_{\mathrm{fa}} \ist | \ist \underline{\V{y}}) =  p\big( \overline{K} \big)    \ist\ist p(l_{\mathrm{fa}} ) \ist \sum_{\tilde{\V{b}}} \ist\ist f\big(\overline{\V{y}}, \V{b}, \tilde{\V{b}}  \ist | \ist  \overline{K},   \underline{\V{y}}, l_{\mathrm{fa}}, \V{\pi}_{\tilde{\V{b}}}\big)  \nn\\[-9mm]
\nn
\end{align}
where $\sum_{\tilde{\V{b}}}$ is short for $\sum^{\underline{K} + \overline{K}}_{\tilde{b}_1 \rmv=\rmv1} \cdots \sum^{\underline{K} + \overline{K}}_{\tilde{b}_M \rmv=\rmv1}$. By replacing $f\big(\overline{\V{y}}, \V{b}, \tilde{\V{b}}  \ist | \ist  \overline{K},   \underline{\V{y}}, l_{\mathrm{fa}}, \V{\pi}_{\tilde{\V{b}}}\big) $ by the expression provided in \eqref{eq:margInt2} and reorganizing certain factors, we can write
\begin{align}
f(\overline{\V{y}}, \V{b}, \overline{K}, l_{\mathrm{fa}} \ist | \ist \underline{\V{y}})    &\hspace{0mm}=  p\big( \overline{K} \big)    \ist\ist p(l_{\mathrm{fa}} )\frac{l_{\mathrm{fa}}! \ist \Big(\prod^{\underline{K}}_{k=1} l_{k} ! \Big) \ist \Big(\prod^{M}_{k=1} \overline{l}_{k} ! \Big)}{\Big( l_{\mathrm{fa}}  + \sum^{\underline{K}}_{k=1}  l_{k}  +  \sum^{M}_{k'=1} \ist \overline{l}_{k'}  \Big)!} \nn\\[2mm]
&\hspace{8mm}\times \rmv \Bigg ( \ist \prod_{k \in \cl{D}_{\underline{\V{y}}}} \frac{\mu_{\mathrm{m}} (\underline{\V{x}}_k,\underline{\V{e}}_k)^{l_k}}{l_{k}!} \ist \mathpzc{e}^{-\mu_{\mathrm{m}}(\underline{\V{x}}_k,\underline{\V{e}}_k)}  \rmv  \Bigg) \Bigg( \prod_{k \notin \cl{D}_{\underline{\V{y}}}}  \rmv \delta \big(l_{k}\big) \rmv \rmv   \Bigg) \nn\\[2mm]
&\hspace{8mm}\times  \ist\ist\ist \sum_{\tilde{\V{b}}}\Bigg( \rmv\rmv\rmv  \prod_{k=1 \atop \V{\pi}_{\tilde{\V{b}}}(k)>0}^M \rmv\rmv\rmv\rmv\rmv \rnew_k \ist\ist\ist \frac{\mu_{\mathrm{m}}(\overline{\V{x}}_k,\overline{\V{e}}_k)^{\ist \tilde{l}_{\V{\pi}_{\tilde{\V{b}}}(k) }}}{\tilde{l}_{ \V{\pi}_{\tilde{\V{b}}}(k) }!} \frac{\mathpzc{e}^{-\mu_{\mathrm{m}}(\overline{\V{x}}_k,\overline{\V{e}}_k)}}{1-\mathpzc{e}^{-\mu_{\mathrm{m}(\overline{\V{x}}_k,\overline{\V{e}}_k)}}} \ist f_{\mathrm{n}}(\overline{\V{x}}_k, \overline{\V{e}}_k) \ist \delta\big(\avnew_k-\avt_{\V{\pi}_{\tilde{\V{b}}}(k)}\big) \rmv\rmv\Bigg) \nn\\[2mm]
&\hspace{8mm}\times  \rmv\rmv\rmv \prod_{k=1 \atop \V{\pi}_{\tilde{\V{b}}}(k)=0}^M \rmv\rmv\rmv\rmv \fd(\xnew_k,\enew_k) \ist (1-\rnew_k) \hspace{.5mm} \delta\big(\ist \tilde{l}_{\V{\pi}_{\tilde{\V{b}}}(k)} \big) . \nn
\end{align}
Here, we exploited the fact that the numbers of measurements produced by the the objects are also implied by $\V{b} = [b_1 \cdots b_M]^{\T}\rmv$. In particular, for $k \in \big\{1,\dots,\underline{K}\big\}$ we have $l_k = \sum^M_{l=1} \delta(b_l - k)$. Furthermore, we used the equalities $\prod^{\underline{K} + \overline{K}}_{k=\underline{K}+1} l_{k} ! = \prod^{M}_{k=1} \overline{l}_k !$ and $\sum^{\underline{K} + \overline{K}}_{k=\underline{K}+1} l_{k} = \sum^{M}_{k=1} \overline{l}_k$ with $\overline{l}_k = \sum^M_{l=1} \delta(b_l - \underline{K} + k)$ for $k \in \big\{1,\dots,M\big\}$. 

We recall that $\tilde{\V{b}}$ implies $\tilde{\V{a}}$ and $\V{b}$ implies $\V{a}$. Since for all $\overline{K}!$  permutations of a vector $\tilde{\V{a}}$ we get the same $\overline{\V{a}}$, there are $\overline{K !}$ different $\tilde{\V{a}}$ that map to the same $\overline{\V{a}}$. Thus, as a result of the marginalization, we \vspace{0mm} obtain
\begin{align}
f(\overline{\V{y}}, \V{b}, \overline{K}, l_{\mathrm{fa}} \ist | \ist \underline{\V{y}})   &\hspace{0mm}=  p\big( \overline{K} \big)    \ist\ist p(l_{\mathrm{fa}} )\frac{\overline{K}! \ist l_{\mathrm{fa}}! \ist \Big(\prod^{\underline{K}}_{k=1} l_{k} ! \Big) \ist \Big(\prod^{M}_{k=1} \overline{l}_{k} ! \Big)}{\Big( l_{\mathrm{fa}}  + \sum^{\underline{K}}_{k=1}  l_{k}  +  \sum^{M}_{k'=1} \ist \overline{l}_{k'}  \Big)!}  \Bigg ( \ist \prod_{k \in \cl{D}_{\underline{\V{y}}}} \frac{\mu_{\mathrm{m}} (\underline{\V{x}}_k,\underline{\V{e}}_k)^{l_k}}{l_{k}!} \ist \mathpzc{e}^{-\mu_{\mathrm{m}}(\underline{\V{x}}_k,\underline{\V{e}}_k)}  \rmv  \Bigg) \nn\\[2mm]
&\hspace{0mm\times }\Bigg( \prod_{k \notin \cl{D}_{\underline{\V{y}}}}  \rmv \delta \big(l_{k}\big) \rmv \rmv   \Bigg) \rmv \Bigg(  \prod_{k \in \Set{D}_{\V{b}}} \rmv\rmv \rnew_k \ist\ist \frac{\mu_{\mathrm{m}}(\overline{\V{x}}_k,\overline{\V{e}}_k)^{ \overline{l}_{k}  }}{ \overline{l}_{k}  !} \frac{\mathpzc{e}^{-\mu_{\mathrm{m}}(\overline{\V{x}}_k,\overline{\V{e}}_k)}}{1-\mathpzc{e}^{-\mu_{\mathrm{m}(\overline{\V{x}}_k,\overline{\V{e}}_k)}}} \ist\ist f_{\mathrm{n}}(\overline{\V{x}}_k, \overline{\V{e}}_k) \rmv \Bigg) \nn\\[2mm]
&\hspace{0mm}\times  \prod_{k \notin \Set{D}_{\V{b}}}  \rmv\rmv\rmv \fd(\xnew_k,\enew_k) \ist (1-\rnew_k) \hspace{.5mm} \delta\big(\ist \overline{l}_k \big) \rmv \nn
\end{align}
where we introduced the set of indexes $\Set{D}_{\V{b}} \triangleq \big\{ k \in \{1,\dots, M\} \ist \big| \ist b_k = \underline{K} + k\big\}$, and used $k \notin\Set{D}_{\V{b}}$ short for $ k \in \{1,\dots,M\} \backslash \Set{D}_{\V{b}}$. 

We can now also replace $p\big( \ist \overline{K} \ist\big)$ and $p(l_{\mathrm{fa}} )$ by the expressions provided in \eqref{eq:priorK} and  \eqref{eq:condPDF5a}, respectively. In this way, we \vspace{-2mm} get
\begin{align}
f(\overline{\V{y}}, \V{b}, \overline{K}, l_{\mathrm{fa}} \ist | \ist \underline{\V{y}})&=  \frac{\ist \mu^{\overline{K}}_{\mathrm{n}} \ist\ist\ist \mu^{l_{\mathrm{fa}}}_{\mathrm{fa}} \ist\ist \mathpzc{e}^{- \mu_{\mathrm{n}} - \mu_{\mathrm{fa}}}}{\Big( l_{\mathrm{fa}}  + \sum^{\underline{K}}_{k=1}  l_{k}  +  \sum^{M}_{k'=1} \ist \overline{l}_{k'}  \Big)!} \ist \Bigg ( \ist \prod_{k \in \cl{D}_{\underline{\V{y}}}} \mu_{\mathrm{m}} (\underline{\V{x}}_k,\underline{\V{e}}_k)^{l_k} \ist \mathpzc{e}^{-\mu_{\mathrm{m}}(\underline{\V{x}}_k,\underline{\V{e}}_k)} \Bigg) \rmv \rmv \Bigg( \prod_{k \notin \cl{D}_{\underline{\V{y}}}}  \rmv \delta \big(l_{k}\big) \rmv\rmv  \Bigg) \nn \\[1.5mm]
&\hspace{5mm}\times \rmv \Bigg(  \prod_{k \in \Set{D}_{\V{b}}} \rmv\rmv \rnew_k \ist\ist \mu_{\mathrm{m}}(\overline{\V{x}}_k,\overline{\V{e}}_k)^{\overline{l}_{k} }\ist \frac{\mathpzc{e}^{-\mu_{\mathrm{m}}(\overline{\V{x}}_k,\overline{\V{e}}_k)}}{1-\mathpzc{e}^{-\mu_{\mathrm{m}}(\overline{\V{x}}_k,\overline{\V{e}}_k)}} \ist\ist f_{\mathrm{n}}(\overline{\V{x}}_k, \overline{\V{e}}_k) \rmv \Bigg) \nn\\[2.3mm]
&\hspace{5mm}\times \prod_{k \notin \Set{D}_{\V{b}}}  \rmv\rmv\rmv \fd(\xnew_k,\enew_k) \ist (1-\rnew_k) \ist \delta\big(\ist \overline{l}_{k} \big) \rmv \nn \\[-7.5mm]
\nn
\end{align}
where we used $\prod_{k \in \Set{D}_{\V{b}}}  \overline{l}_{k}! = \prod^{M}_{k = 1}  \overline{l}_{k}! $ and $\prod_{k \in \Set{D}_{\underline{\V{y}}}} l_{k}! = \prod^{\underline{K}}_{k=1} l_{k}!$. (This is possible because for $k \notin \Set{D}_{\V{b}}$, $\overline{l}_{k} \rmv>\rmv 0$ or $k \notin \Set{D}_{\underline{\V{y}}}$, $l_{k} \rmv>\rmv 0$, we have $f(\overline{\V{y}}, \V{b}, \V{\pi}, \tilde{\V{l}}, \overline{K}, l_{\mathrm{fa}} \ist | \ist \underline{\V{y}}) \rmv = \rmv 0$ anyway.)

It is now straightforward to see that $\overline{K}$ is fully determined by $\V{b}$ via the relation $\overline{K} = |\Set{D}_{\V{b}}|$. Therefore, we get
\begin{equation}
f(\overline{\V{y}}, \V{b}, l_{\mathrm{fa}} \ist | \ist \underline{\V{y}})  = f\big(\overline{\V{y}}, \V{b}, \overline{K} \rmv=\rmv |\Set{D}_{\V{b}}|, l_{\mathrm{fa}} \ist \big| \ist \underline{\V{y}}\big). \nn
\vspace{1mm}
\end{equation}
Furthermore, the total number of measurements is given by $M \rmv=\rmv l_{\mathrm{fa}} \rmv+ \sum^{\underline{K}}_{k=1} l_k + \sum^{M}_{k'=1} \overline{l}_{k'}$. Thus, it follows that there is a one-to-one relation between $(\overline{\V{y}}, \V{b}, l_{\mathrm{fa}} )$ and $(\overline{\V{y}}, \V{b}, M )$. After reorganizing certain factors, we can thus \vspace{0mm} write
\begin{align}
f\big(\overline{\V{y}}, \V{b}, M \ist | \ist \underline{\V{y}} \big) &= \Bigg( \prod_{k\in\Set{D}_{\V{b}}} \rmv\rmv \overline{r}_k \ist f_{\mathrm{n}}(\overline{\V{x}}_k, \overline{\V{e}}_k) \ist\ist \mu_{\mathrm{m}}(\overline{\V{x}}_k,\overline{\V{e}}_k)^{\overline{l}_k} \frac{\mathpzc{e}^{-\mu_{\mathrm{m}}(\overline{\V{x}}_k,\overline{\V{e}}_k) }}{1-\mathpzc{e}^{-\mu_{\mathrm{m}}(\overline{\V{x}}_k,\overline{\V{e}}_k) }}   \Bigg)  \Bigg(  \prod_{k \notin\Set{D}_{\V{b}}} (1 -\overline{r}_{k}) \ist f_{\mathrm{d}}(\overline{\V{x}}_{k}, \overline{\V{e}}_{k}) \ist\ist \delta\big(\ist \overline{l}_k \big) \rmv\rmv \Bigg)  \nn\\[2mm]
&\hspace{0mm} \times\Bigg( \prod_{k \in \cl{D}_{\underline{\V{y}}}} \mu_{\mathrm{m}} (\underline{\V{x}}_k,\underline{\V{e}}_k)^{\underline{l}_k} \ist \mathpzc{e}^{-\mu_{\mathrm{m}}(\underline{\V{x}}_k,\underline{\V{e}}_k)} \Bigg)  \Bigg( \prod_{k \notin \cl{D}_{\underline{\V{y}}}} \rmv\rmv  \delta \big( \hspace{.15mm} \underline{l}_k \big) \rmv\rmv \Bigg) \ist\ist \frac{ \mu^{|\Set{D}_{\V{b}}|}_{\mathrm{n}} \ist\ist\ist\ist \mu_{\mathrm{fa}}^{\big(M - \sum^{\underline{K}}_{k=1} \rmv\rmv \underline{l}_k - \sum^{M}_{k'=1} \rmv\rmv \overline{l}_{k'} \big)}\ist\ist\ist \mathpzc{e}^{-( \mu_{\mathrm{n}} + \mu_{\mathrm{fa}} )}}{M!}.  \nn \\[-3.5mm]
\label{eq:posterior2} \\[-8mm]
\nn
\end{align}
Note that  $f\big(\overline{\V{y}}, \V{b}, M \ist | \ist \underline{\V{y}} \big) $ is a valid probability distribution in the sense that $\sum^{\infty}_{M = 0} \sum_{\V{b}} \int f\big(\overline{\V{y}}, \V{b}, M \ist | \ist \underline{\V{y}} \big) \mathrm{d} \overline{\V{y}} \rmv=\rmv 1$.

\subsection{The Joint Prior \ac{pdf} $f \big (\V{y}_{0:n}, \V{b}_{1:n}, M_{1:n} \ist \big)$}
\label{sec:jointPrior}

As a preparatory step for the derivation of the joint prior \ac{pdf} $f \big (\V{y}_{0:n}, \V{b}_{1:n},  M_{1:n} \ist \big)$, we first derive the conditional \ac{pdf} $f\big(\V{y}_n, \V{b}_n,  M_n \ist | \ist \V{y}_{n-1}\big) \rmv=\rmv f\big(\overline{\V{y}}_n, \underline{\V{y}}_n, \V{b}_n, M_n \ist | \ist \V{y}_{n-1}\big) $. We start by assuming that legacy \acp{po} evolve independently across time \cite{MeyKroWilLauHlaBraWin:J18,GraFatSve:J19}, i.e.,
\begin{equation}
f\big(\underline{\V{y}}_n | \ist\V{y}_{n-1} \big) = 
    \prod^{\underline{K}_n}_{k=1} \ist f\big(\underline{\V{y}}_{k,n} | \ist\V{y}_{k,n-1} \big), \label{eq:stateTransitionAll}
\end{equation}
where $f\big(\underline{\V{y}}_{k,n} | \ist\V{y}_{k,n-1} \big) \triangleq f\big(\underline{\V{x}}_{k,n}, \underline{\V{e}}_{k,n}, \underline{r}_{k,n} | \ist \V{x}_{k,n-1}, \V{e}_{k,n-1}, r_{k,n-1}  \big)$ is the single \ac{po} state transition function provided in \cite[Eq.~(2) and (3)]{MeyWil:J21}. Next, we expand $f\big(\overline{\V{y}}_n, \underline{\V{y}}_n, \V{b}_n, M_n \ist | \ist \V{y}_{n-1}\big) $ as follows
\begin{align}
f\big(\overline{\V{y}}_n, \underline{\V{y}}_n, \V{b}_n, M_n \ist | \ist \V{y}_{n-1}\big) &= f\big(\overline{\V{y}}_n, \V{b}_n, M_n \ist | \underline{\V{y}}_n\rmv, \ist \V{y}_{n-1}\big) \ist f\big(\underline{\V{y}}_n | \ist\V{y}_{n-1} \big).
\label{eq:twoConditionalPDFs}
\end{align}
and make the common assumption \cite{MeyKroWilLauHlaBraWin:J18,GraFatSve:J19} that conditioned on the legacy \ac{po} states $\underline{\V{y}}_n$ a time $n$, new \ac{po} states $\overline{\V{y}}_n$, association vectors $\V{b}_n$, and number of measurements $M_n$, are statistically independent of all \ac{po} states $\V{y}_{n-1}$ at time $n \rmv-\rmv1$. Thus, \eqref{eq:twoConditionalPDFs} can be expressed as
\begin{align}
f\big(\overline{\V{y}}_n, \underline{\V{y}}_n, \V{b}_n, M_n \ist | \ist \V{y}_{n-1}\big) &= f\big(\overline{\V{y}}_n, \V{b}_n, M_n \ist | \underline{\V{y}}_n \big) \ist f\big(\underline{\V{y}}_n | \ist\V{y}_{n-1} \big).
\label{eq:otherTwoConditionalPDFs}
\end{align}

By using \eqref{eq:posterior2} and \eqref{eq:stateTransitionAll} in \eqref{eq:otherTwoConditionalPDFs} and rearranging certain factors, we obtain
\begin{align}
f\big(\V{y}_n, \V{b}_n,  M_n \ist | \ist \V{y}_{n-1}\big) &=f\big(\overline{\V{y}}_n, \underline{\V{y}}_n, \V{b}_n,  M_n \ist | \ist \V{y}_{n-1}\big) \nn\\[1mm]
&=  C(M_n)  \Bigg( \prod^{M_n}_{k = 1} \ist \ist  \overline{q}(\overline{\V{y}}_{k,n}) \ist\ist g_k\big( \overline{\V{y}}_{k,n}, b_{l,n} \big) \ist \prod^{M_n}_{l=k+1} \rmv h_{1,\underline{K}_n+k}\big( \overline{\V{y}}_{k,n}, b_{l,n} \big) \rmv\rmv\Bigg) \nn\\
&\hspace{30.8mm}\times\Bigg(\prod^{\underline{K}_n}_{k = 1} \ist \underline{q}(\underline{\V{y}}_{k,n} \ist | \ist \V{y}_{k,n-1} ) \ist\ist \prod^{M_n}_{l=1} \ist\ist h_{1,k}\big( \underline{\V{y}}_{k,n}, b_{l,n} \big) \rmv\rmv \Bigg)  \label{eq:priorSingle}  \\[-8.5mm]
\nn
\end{align}
where we introduced the constant $C(M_n) = \big(\mu_{\mathrm{fa}}^{M_n}\ist\ist\ist \mathpzc{e}^{-( \mu_{\mathrm{n}} + \mu_{\mathrm{fa}} )}\big)/M_n!$ and the functions $\overline{q}(\overline{\V{y}}_{k,n})$, $\underline{q}\big(\underline{\V{y}}_{k,n} | \V{y}_{k,n-1} \big)$, $h_{1,k}\big( \V{y}_{k,n}, b_{l,n} \big)$, and $g_k\big( \overline{\V{y}}_{k,n}, b_{l,n} \big)$ that will be presented in what follows. The \textit{factors containing prior distributions} for new \acp{po} $\overline{q}(\overline{\V{y}}_{k,n}) \triangleq \overline{q}(\overline{\V{x}}_{k,n},\overline{\V{e}}_{k,n},\overline{r}_{k,n})$ are given \vspace{0mm} by
\begin{align}
\overline{q}(\overline{\V{x}}_{k,n},\overline{\V{e}}_{k,n},\overline{r}_{k,n})  &\,\triangleq \begin{cases}
      \mu_{\mathrm{n}} f_{\mathrm{n}}(\overline{\V{x}}_{k,n}, \overline{\V{e}}_{k,n}) \ist \frac{\mathpzc{e}^{-\mu_{\mathrm{m}}(\overline{\V{x}}_{k,n},\overline{\V{e}}_{k,n}) }}{1 - \mathpzc{e}^{-\mu_{\mathrm{m}}}(\overline{\V{x}}_{k,n},\overline{\V{e}}_{k,n})}  \ist, 
       & \rmv\rmv \overline{r}_{k,n} = 1 \\[1mm]
     f_{\mathrm{d}}\big(\overline{\V{x}}_{k,n}, \overline{\V{e}}_{k,n}\big) \ist,  & \rmv\rmv \overline{r}_{k,n} = 0.
  \end{cases} \nn\\[-11.5mm]
    \nn\\
  \nn
\end{align}
Similarly, the \textit{pseudo transition functions} for legacy \acp{po} $\underline{q}\big(\underline{\V{y}}_{k,n} | \V{y}_{k,n-1} \big) \triangleq \underline{q}\big(\underline{\V{x}}_{k,n}, \underline{\V{e}}_{k,n}, \underline{r}_{k,n} | \V{x}_{k,n-1},$ $\V{e}_{k,n-1},$ $r_{k,n-1} \big)$ \vspace{-1.8mm} read
\begin{align}
&\underline{q}\big(\underline{\V{x}}_{k,n}, \underline{\V{e}}_{k,n}, \underline{r}_{k,n} | \V{x}_{k,n-1}, \V{e}_{k,n-1}, r_{k,n-1} \big) \nn\\
&\hspace{15mm} \triangleq \begin{cases}
      \mathpzc{e}^{-\mu_{\mathrm{m}} (\underline{\V{x}}_{k,n}, \underline{\V{e}}_{k,n}) } f\big(\underline{\V{x}}_{k,n}, \underline{\V{e}}_{k,n},\underline{r}_{k,n} \rmv\rmv\rmv=\rmv\rmv\rmv  1 |\V{x}_{k,n-1},\V{e}_{k,n-1},r_{k,n-1} \big)  \ist, 
       & \rmv\rmv \overline{r}_{k,n} = 1 \\[1mm]
      f\big(\underline{\V{x}}_{k,n}, \underline{\V{e}}_{k,n},$ $\underline{r}_{k,n} \rmv=\rmv  0 \ist | \ist \V{x}_{k,n-1}, \V{e}_{k,n-1}, r_{k,n-1} \big) \ist,  & \rmv\rmv \overline{r}_{k,n} = 0.
  \end{cases} \nn\\[-11mm]
    \nn\\
  \nn
\end{align}
Finally, functions $h_{1,k}\big( \V{y}_{k,n}, b_{l,n} \big) \rmv=\rmv h_{1,k}\big( \V{x}_{k,n},  \V{e}_{k,n},  r_{k,n},  b_{l,n} \big)$ and $g_k\big( \overline{\V{y}}_{k,n}, b_{k,n} \big) \rmv=\rmv g_k\big(\overline{\V{x}}_{k,n}, \overline{\V{e}}_{k,n},  \overline{r}_{k,n},  b_{k,n} \big)$ are given\vspace{-3mm} by
\begin{align}
&h_{1,k}\big(  \V{x}_{k,n}, \V{e}_{k,n}, r_{k,n} \rmv=\rmv 1, b_{l,n} \big) \triangleq \begin{cases}
       \frac{\mu_{\mathrm{m}}(\V{x}_{k,n},\V{e}_{k,n})}{\mu_{\mathrm{fa}}}   \ist, 
       & \rmv\rmv b_{l,n} = k \\[1mm]
     1 \ist,  & \rmv\rmv b_{l,n} \neq k 
  \end{cases} \nn\\[-8mm]
  \nn
\end{align}
and $h_{1,k}\big( \V{x}_{k,n}, \V{e}_{k,n}, r_{k,n} \rmv=\rmv 0,  b_{l,n}  \big) \rmv\triangleq\rmv 1 \rmv-\rmv \delta(b_{l,n} - k )$ as well \vspace{-1.5mm} as 
\begin{align}
&g_{1,k}\big(  \overline{\V{x}}_{k,n}, \overline{\V{e}}_{k,n}, \overline{r}_{k,n} \rmv=\rmv 1, b_{k,n}  \big) \triangleq \begin{cases}
       \frac{\mu_{\mathrm{m}} (\overline{\V{x}}_{k,n},\overline{\V{e}}_{k,n})}{\mu_{\mathrm{fa}}}   \ist, 
       & \rmv\rmv b_{k,n} = \underline{K}_n + k \\[1mm]
     0 \ist,  & \rmv\rmv b_{k,n} \neq \underline{K}_n + k 
  \end{cases} \nn\\[-8mm]
  \nn
\end{align}
and $g_{1,k}\big( \overline{\V{x}}_{k,n}, \overline{\V{e}}_{k,n}, \overline{r}_{k,n} \rmv=\rmv 0,  b_{k,n}\big) \rmv\rmv\triangleq\rmv\rmv 1 \rmv - \rmv \delta \big( b_{k,n} \rmv - \rmv ( \underline{K}_n + k ) \big)$, respectively. 
At time $n \rmv=\rmv0$, \ac{po} states are assumed statistically independent, \vspace{-1.5mm} i.e.,
\begin{equation}
 f(\V{y}_{0}) =  \prod^{K_0}_{k = 1} f(\V{y}_{k,0}).  \label{eq:priorTime0}
 \vspace{1mm}
\end{equation}
Note that there may be no prior information available and thus $K_0 = 0$. We now make the common assumption \cite{MeyKroWilLauHlaBraWin:J18,GraFatSve:J19} that given $\V{y}_{n-1}$, the random variables $\V{y}_{n}$, $\V{a}_{n}$, and $M_{n}$ are statistically independent of all $\V{y}_{0:n-2}$, $\V{a}_{1:n-1}$, and $M_{1:n-1}$. Thus, we \vspace{-.5mm} get
\begin{equation}
f\big(\V{y}_{0:n}, \V{a}_{1:n}, M_{1:n} \big) = f( \V{y}_{0}) \prod^{n}_{n' = 1} f\big(\V{y}_{n'}, \V{a}_{n'}, M_{n'} \ist | \ist \V{y}_{n'-1}\big). 
\end{equation}
Finally, we replace $f( \V{y}_{0})$ and $f\big(\V{y}_{n}, \V{a}_{n}, M_{n} \ist | \ist \V{y}_{n-1}\big)$ by the expression provided in \eqref{eq:priorTime0} and \eqref{eq:priorSingle}, respectively. In this way, we \vspace{1mm} finally obtain
\begin{align}
&f\big(\V{y}_{0:n}, \V{a}_{1:n}, M_{1:n} \big) \nn\\[1mm]
&\hspace{5mm}= \Bigg( \prod^{K_0}_{k = 1} f(\V{y}_{k,0})  \Bigg) \rmv \prod^{n}_{n' = 1}C(M_{n'})  \Bigg( \prod^{M_{n'}}_{k = 1} \ist \ist  \overline{q}(\overline{\V{y}}_{k,n'}) \ist\ist g_{1,k}\big( \overline{\V{y}}_{k,n'}, b_{l,n'} \big) \ist \prod^{M_{n'}}_{l=k+1} h_{1,\underline{K}_{n'} + k}\big( \overline{\V{y}}_{k,n'}, b_{l,n'} \big) \rmv\rmv\Bigg)  \nn\\[1.5mm]
&\hspace{73mm}\times \Bigg(\prod^{\underline{K}_{n'}}_{k = 1} \ist\ist \underline{q}(\underline{\V{y}}_{k,n'} \ist | \ist \V{y}_{k,n'-1} ) \ist\ist \prod^{M_{n'}}_{l=1} \ist\ist h_{1,k}\big( \underline{\V{y}}_{k,n'}, b_{l,n'} \big) \rmv\rmv \Bigg) . \label{eq:PriorFinal}
\end{align}

\subsection{The Joint Likelihood Function $f(\V{z}_{1:n} \ist | \ist \V{y}_{1:n}, \V{b}_{1:n}, M_{1:n})$}
\label{sec:jointLikelihoodFunction}

For \ac{po} $k \rmv\in\rmv \{1,\dots,\underline{K}_n \rmv+\rmv M_n\}$, we introduce the set of indexes $\Set{M}_{\V{b}_n,k} \triangleq \big\{ l \in \{1,\dots, M_n\} \ist \big| \ist b_{l,n} = k \big\}$. Now, we make the common assumption \cite{MeyKroWilLauHlaBraWin:J18,GraFatSve:J19}, that conditioned on $\overline{\V{y}}_n$, $\underline{\V{y}}_n$, $\V{b}_n$, and $M_n$ all measurements $\V{z}_{l,n}$, $l \in \{1,\dots,M_n\}$ are statistically independent. Thus, the joint likelihood function at time $n$, is given \vspace{-2mm} by
\begin{align}
&f(\V{z}_n \ist | \ist \overline{\V{y}}_n, \underline{\V{y}}_n, \V{b}_n, M_n) \nn\\
&\hspace{9mm} = \Bigg( \prod^{M_n}_{l \rmv=\rm1} \ist f_{\mathrm{fa}}(\V{z}_{l,n}) \Bigg) \Bigg( \prod^{M_n}_{k = 1} \ist\ist\ist \prod_{l \in \ist \Set{M}_{\V{b}_n,k+\underline{K}_n}} \rmv\rmv\rmv\rmv\rmv\rmv\rmv\rmv\rmv \frac{f(\V{z}_{l,n}| \overline{\V{x}}_{k,n},  \overline{\V{e}}_{k,n})}{f_{\mathrm{fa}}(\V{z}_{l,n})} \Bigg) \Bigg( \prod^{\underline{K}_n}_{k = 1} \ist \prod_{l \in \ist \Set{M}_{\V{b}_n,k}} \!\!\frac{f(\V{z}_{l,n}| \underline{\V{x}}_{k,n}, \underline{\V{e}}_{k,n})}{f_{\mathrm{fa}}(\V{z}_{l,n})}  \Bigg) \label{eq:likelihoodFunction}
\end{align}
for $\V{z}_n \in \mathbb{R}^{d_{\V{z}} M_n}$ and $f(\V{z}_n \ist | \ist \overline{\V{y}}_n, \underline{\V{y}}_n, \V{b}_n, M_n) = 0$ otherwise.

We can further reformulate \eqref{eq:likelihoodFunction} as \vspace{0mm} follows
\begin{align}
f(\V{z}_n \ist | \ist \overline{\V{y}}_n, \underline{\V{y}}_n, \V{b}_n, M_n) = C(\V{z}_n) \Bigg( \prod^{M_n}_{k = 1} \ist \prod^{M_n}_{l = k} h_{2,\underline{K}_n+k}\big( \overline{\V{y}}_{k,n}, b_{l,n}; \V{z}_{l,n} \big) \Bigg) \Bigg( \prod^{\underline{K}_n}_{k = 1} \ist \prod^{M_n}_{l=1} \ist\ist\ist h_{2,k}\big( \underline{\V{y}}_{k,n}, b_{l,n}; \V{z}_{l,n} \big) \Bigg)\label{eq:likelihoodFunction2} \\[-8.5mm]
\nn
\end{align}
where we introduced $C(\V{z}_n) \triangleq \prod^M_{l \rmv=\rm1} \ist f_{\mathrm{fa}}(\V{z}_{l,n})$ as well as $h_{2,k}\big( \V{y}_{k,n}, b_{l,n}; \V{z}_{l,n} \big) = h_{2,k}\big( \V{x}_{k,n}, \V{e}_{k,n}, r_{k,n}, b_{l,n}; \V{z}_{l,n} \big)$ given \vspace{-2.5mm} by
\begin{align}
&h_{2,k}\big( \V{x}_{k,n},  \V{e}_{k,n}, r_{k,n} \rmv=\rmv 1, b_{l,n}; \V{z}_{l,n} \big) \triangleq \begin{cases}
      \frac{ f(\V{z}_{l,n}| \V{x}_{k,n}, \V{e}_{k,n})}{  f_{\mathrm{fa}}(\V{z}_{l,n})}    \ist, 
       & \rmv\rmv b_{l,n} = k \\[1mm]
     1 \ist,  & \rmv\rmv b_{l,n} \neq k 
  \end{cases} \nn
\end{align}
and $h_{2,k}\big( \V{x}_{k,n}, \V{e}_{k,n}, r_{k,n} \rmv=\rmv 0,  b_{l,n}; \V{z}_{l,n} \big)  \triangleq 1 $. Assuming that measurement noise and false alarm measurements are statistically independent across time \cite{MeyKroWilLauHlaBraWin:J18,GraFatSve:J19}, the joint likelihood function can be expressed as
\begin{align}
f(\V{z}_{1:n} \ist | \ist \V{y}_{1:n}, \V{b}_{1:n}, M_{1:n})  &= f(\V{z}_{1:n} \ist | \ist \overline{\V{y}}_{1:n}, \underline{\V{y}}_{1:n},  \V{b}_{1:n}, M_{1:n}) \nn\\
&= \prod^{n}_{n' = 1} f(\V{z}_{n'} \ist | \ist \overline{\V{y}}_{n'}, \underline{\V{y}}_{n'}, \V{b}_{n'}, M_{n'}).
\label{eq:likelihoodFunctionFinal}
\end{align}

\subsection{Final Expression for the Joint Posterior \ac{pdf} $f \big( \V{y}_{0:n}, \V{b}_{1:n} \big |\V{z}_{1:n} \big)$}
\label{sec:jointPosterior}

We derive the factorization of $f \big( \V{y}_{0:n},\V{b}_{1:n} \big |\V{z}_{1:n} \big)$ considering that the measurements $\V{z}_{1:n}$ are observed and thus fixed (consequently $M_n$ is fixed as well). By using Bayes' rule and by exploiting the fact that $\V{z}_n$ implies $M_n$ according to \eqref{eq:likelihoodFunction}, we obtain
\begin{align}
f( \V{y}_{0:n}, \V{b}_{1:n}|\V{z}_{1:n})  &=\sum^{\infty}_{M_1' = 0} \sum^{\infty}_{M_2' = 0} \cdots \sum^{\infty}_{M_n' = 0} f( \V{y}_{0:n}, \V{b}_{1:n}, M'_{1:n}|\V{z}_{1:n})  \nn\\
&\propto \sum^{\infty}_{M_1' = 0} \sum^{\infty}_{M_2' = 0} \cdots \sum^{\infty}_{M_n' = 0} f(\V{z}_{1:n} \ist | \ist \V{y}_{1:n}, \V{b}_{1:n}, M'_{1:n}) \ist f \big (\V{y}_{0:n}, \V{b}_{1:n}, M'_{1:n} \ist \big) \nn\\[.5mm]
&=  f(\V{z}_{1:n} \ist | \ist \V{y}_{1:n}, \V{b}_{1:n}, M_{1:n}) \ist f \big (\V{y}_{0:n}, \V{b}_{1:n}, M_{1:n} \ist \big). \nn\\[-6mm]
\nn
\end{align}

Next, we replace the joint prior \ac{pdf} $f \big (\V{y}_{0:n}, \V{b}_{1:n}, M_{1:n} \ist \big)$ and the joint likelihood function $ f(\V{z}_{1:n} \ist | \ist \V{y}_{1:n}, \V{b}_{1:n},$ $M_{1:n})$ by the expressions provided in \eqref{eq:PriorFinal} and \eqref{eq:likelihoodFunctionFinal}, respectively. In this way, after rearrangement of certain factors, we \vspace{-6mm} obtain

\begin{align}
&f( \V{y}_{0:n}, \V{b}_{1:n}|\V{z}_{1:n})  \nn\\[1.3mm]
&\hspace{0mm}\propto  \Bigg( \prod^{K_0}_{k = 1} f(\V{y}_{k,0}) \rmv \Bigg) \rmv \prod^{n}_{n' = 1} \rmv \Bigg( \prod^{M_{n'}}_{k = 1} \ist \ist  \overline{q}(\overline{\V{y}}_{k,n'}) \ist\ist g_{1,k}\big( \overline{\V{y}}_{k,n'}, b_{k,n'} \big)  \ist    h_{2,\underline{K}_n+k}\big( \overline{\V{y}}_{k,n'}, b_{k,n'}; \V{z}_{k,n'} \big) \rmv\rmv \prod^{M_{n'}}_{l=k+1} \rmv\rmv h_{1,\underline{K}_n+k}\big( \overline{\V{y}}_{k,n'}, b_{l,n'} \big) \nn\\[1.3mm]
&\hspace{6mm}\times  h_{2,\underline{K}+k}\big( \overline{\V{y}}_{k,n'}, b_{l,n'}; \V{z}_{l,n'} \big)\rmv\Bigg) \ist \prod^{\underline{K}_{n'}}_{k' = 1} \ist\ist \underline{q}(\underline{\V{y}}_{k'\rmv,n'} \ist | \ist \V{y}_{k'\rmv,n'-1} ) \ist\ist \prod^{M_{n'}}_{l'=1} \ist\ist h_{1,k}\big( \underline{\V{y}}_{k'\rmv,n'}, b_{l'\rmv,n'} \big)  \ist   h_{2,k}\big( \underline{\V{y}}_{k'\rmv,n'}, b_{l'\rmv,n'}; \V{z}_{l'\rmv,n'} \big).
\nn
\end{align}
By introducing the combined factors $h_k\big( \V{y}_{k,n}, b_{l,n}; \V{z}_{l,n} \big) \triangleq h_{1,k}\big( \V{y}_{k,n}, b_{l,n} \big) \ist h_{2,k}\big( \V{y}_{k,n}, b_{l,n}; \V{z}_{l,n} \big)$ and  $g_k\big( \overline{\V{y}}_{k,n},$ $b_{k,n}; \V{z}_{k,n} \big) \triangleq g_{1,k}\big( \overline{\V{y}}_{k,n}, b_{k,n} \big) \ist h_{2,k}\big( \overline{\V{y}}_{k,n}, b_{k,n}; \V{z}_{k,n} \big)$, we obtained the final expression of the joint posterior \ac{pdf} in \cite[Eq.~(11)]{MeyWil:J21}.

\section{Evaluation of the Likelihood Function}
\label{sec:approxLikelihood}

We will next present an approximation that leads to a closed-form of the likelihood function $f(\V{z}_{l,n} | \V{x}_{k,n}, \V{e}_{k,n})$ in \cite[Eq.~6]{MeyWil:J21} for the case where measurements are uniformly distributed on the extent of objects. This resulting closed-form of $f(\V{z}_{l,n} | \V{x}_{k,n}, \V{e}_{k,n})$ makes it possible to develop efficient particle-based methods for EOT. To simplify the notation, we will again drop the time index $n$. 

It is assumed that any measurement $\V{z}_l$ that originates from object $k$ follows the linear\vspace{-1mm} model
\begin{align}
\RV{z}_{l} &= \RV{p}_{k} \rmv+\rmv \RV{v}^{(l)}_{k} + \RV{u}_{l} \label{eq:measModelLin}
\vspace{-1mm}
\end{align}
where $\RV{u}_{l} \sim \mathcal{N}(\V{u}_{l} ;\V{0},\M{\Sigma}_{\V{u}_{l}})$ and $\RV{v}^{(l)}_{k}$ is uniformly distributed with support $\Set{S}(\M{E}_{k})$. 

\begin{figure*}[t!]
\centering
\includegraphics[scale=1]{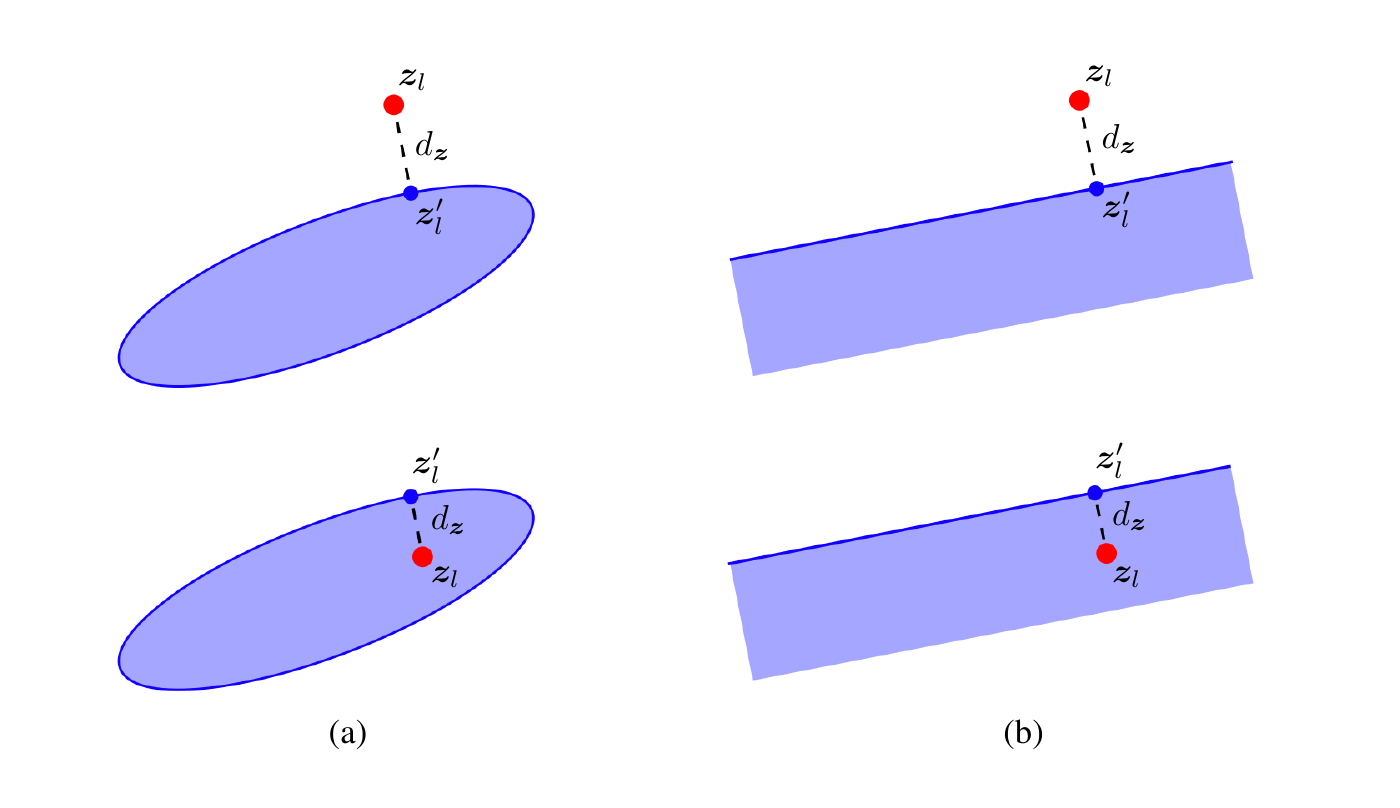}
\vspace{-5mm}
\caption{Graphical representation of the approximation used to obtain a closed form expression for $f(\V{z}_{l} | \V{x}_{k}, \V{e}_{k})$.}
\label{fig:likelihoodApproximation}
\end{figure*}

The likelihood function $f(\V{z}_{l} | \V{x}_{k}, \V{e}_{k})$ can be expressed as
\begin{align}
f(\V{z}_{l} | \V{x}_{k}, \V{e}_{k}) &= \int \rmv\rmv  f\big(\V{z}_{l} | \V{x}_{k}, \V{v}^{(l)}_k\big) \ist f\big(\V{v}^{(l)}_k| \V{e}_{k}\big) \ist  \mathrm{d} \V{v}^{(l)}_k \nn\\[.5mm]
&= \frac{1}{|\Set{S}(\M{E}_{k})|} \int_{\Set{S}(\M{E}_{k})} \rmv\rmv \rmv\rmv  \rmv\rmv  \mathcal{N}(\V{p}_{k} \rmv+\rmv \V{v}^{(l)}_{k};\V{z}_{l},\M{\Sigma}_{\V{u}_{l}} ) \ist \mathrm{d} \V{v}^{(l)}_k \label{eq:likelihoodInt}
\end{align}
where $|\Set{S}(\M{E}_{k})|$ is the surface or volume of $\Set{S}(\M{E}_{k})$. Let $\Set{S}_{\V{p}_{k}}\rmv\rmv(\M{E}_{k})$ be the area or surface $\Set{S}(\M{E}_{k})$ shifted by position $\V{p}_{k}$. With this notation, we can rewrite \eqref{eq:likelihoodInt} as
\begin{align}
f(\V{z}_{l} | \V{x}_{k}, \V{e}_{k}) &= \frac{1}{| \Set{S}(\M{E}_{k}) |} \int_{\Set{S}_{\V{p}_{k}}\rmv\rmv(\M{E}_{k})} \rmv\rmv \rmv\rmv  \rmv\rmv  \mathcal{N}(\V{v}^{(l)}_{k};\V{z}_{l},\M{\Sigma}_{\V{u}_{l}} ) \ist \mathrm{d} \V{v}^{(l)}_k. \label{eq:likelihoodInt2}
\end{align}
To obtain an approximate closed form expression for \eqref{eq:likelihoodInt}, we first introduce the point $\V{z}'_{l}$ on the boundary of $\Set{S}_{\V{p}_{k}}(\M{E}_{k})$ which is closest to the measurement $\V{z}_{l}$. This is shown in Fig.~\ref{fig:likelihoodApproximation}(a). Next, we also introduce the unit vector $\V{1}_{\V{z}_{l}} =  \frac{(\V{z}_{l} - \V{z}'_{l} )}{ \| \V{z}_{l} - \V{z}'_{l} \|}$ pointing from $\V{z}'_{l}$ to $\V{z}_{l}$ as well as the distance $d_{\V{z}_{l}} =  \| \V{z}_{l} - \V{z}'_{l} \|$  between $\V{z}'_{l}$ and $\V{z}_{l}$. By projecting $\RV{u}_{l}$ onto $\V{1}_{\V{z}_{l}}$, the scalar measurement noise $\rv{u}_{l}$ ``in the direction of $\V{z}_{l}$'' is obtained, i.e., $\rv{u}_{l} \sim \mathcal{N}\big(u_{l} ;0,\sigma^2_{u_{l}}\big)$ with $\sigma^2_{u_{l}} = \V{1}^{\mathrm{T}}_{\V{z}_{l}} \M{\Sigma}_{\V{u_l}} \V{1}_{\V{z}_{l}}$. 

We now assume that the extent of the object is much larger than the measurement noise standard deviation  $\sigma_{u_{l}}$ and approximate the extent of the object by an infinite plane or volume. This approximation is shown in Fig.~\ref{fig:likelihoodApproximation}(b). Based on this approximation it is now straightforward to see that the high-dimensional integration in \eqref{eq:likelihoodInt2} can be replaced by the following one-dimensional\vspace{1mm} integrations
\begin{align}
 &f(\V{z}_{l} | \V{x}_{k}, \V{e}_{k}) \approx  \begin{cases}
   \frac{1}{| \Set{S}(\M{E}_{k}) |}  \int_{d_{\V{z}_l}}^{\infty} \hspace{2mm} \mathcal{N}(v^{(l)}_{k};0,\sigma^2_{u_{l}}) \ist \mathrm{d} v^{(l)}_k  \ist, &  \V{z}_l \notin \Set{S}_{\V{p}_{k}}\rmv\rmv(\M{E}_{k}) \\[1.5mm]
    \frac{1}{| \Set{S}(\M{E}_{k}) |}  \int^{\infty}_{-d_{\V{z}_l}}  \mathcal{N}(v^{(l)}_{k};0,\sigma^2_{u_{l}} ) \ist \mathrm{d} v^{(l)}_k   \ist, & \V{z}_l \in \Set{S}_{\V{p}_{k}}\rmv\rmv(\M{E}_{k})  . 
   \end{cases} \label{eq:likelihoodApproximation1D} \\[-6mm]
   \nn
\end{align}
A closed-form expression for the integrations in \eqref{eq:likelihoodApproximation1D} can be obtained by means of the Q-function \cite{BorSun:J79}, i.e.,
\begin{align}
 &f(\V{z}_{l} | \V{x}_{k}, \V{e}_{k}) \approx  \begin{cases}
   \frac{1}{| \Set{S}(\M{E}_{k}) |} \ist\ist Q\Big(\frac{\phantom{-}d_{\V{z}_l}}{\sigma_{u_{l}}}\Big), &  \V{z}_l \notin \Set{S}_{\V{p}_{k}}\rmv\rmv(\M{E}_{k}) \\[1.5mm]
    \frac{1}{| \Set{S}(\M{E}_{k}) |} \ist\ist  Q\Big(\frac{-d_{\V{z}_l}}{\sigma_{u_{l}}}\Big), & \V{z}_l \in \Set{S}_{\V{p}_{k}}\rmv\rmv(\M{E}_{k}) . 
   \end{cases} \label{eq:likelihoodApproximation1DQ} \\[-6.5mm]
   \nn
\end{align}

In addition, we discuss an important special case where $f(\V{z}_{l} | \V{x}_{k}, \V{e}_{k})$ can be evaluated by avoiding any approximation. Let us assume the measurement noise is iid, i.e., $\RV{u}_{l} \sim \mathcal{N}\big(\V{u}_{l} ;0,\sigma^2_{u_{l}} \M{I}\big)$. Furthermore, $\V{v}_l$ is uniformly distributed on a support $\Set{S}_{\mathrm{c}}(\M{E}_{k})$ that has the shape of a rectangle or cube. For a 3-D scenario, we recall that the eigenvalues and  eigenvectors of $\M{E}_{k}$ that define dimension and orientation of the cube, are denoted as $\lambda_{1,\M{E}_{k,n}} > \lambda_{2,\M{E}_{k,n}} > \lambda_{3,\M{E}_{k,n}} \geq 0$ as well as  $\protect\V{\lambda}_{1,\M{E}_{k,n}}$, $\protect\V{\lambda}_{2,\M{E}_{k,n}}$, and $\protect\V{\lambda}_{3,\M{E}_{k,n}}$, respectively.  Let $\Set{S}_{\V{p}_{k}}\rmv\rmv(\M{E}_{k})$ be the cube $\Set{S}_{\mathrm{c}}(\M{E}_{k})$ shifted by position $\V{p}_{k}$. In addition, let $d^{(1)}_{1,\V{z}_l}$ and $d^{(2)}_{1,\V{z}_l}$ be the distances of the measurement $\V{z}_l$ to the two parallel edges of the cube $\Set{S}_{\V{p}_{k}}\rmv\rmv(\M{E}_{k})$ defined by dimension $\lambda_{1,\M{E}_{k,n}}$ and axis $\V{\lambda}_{1,\M{E}_{k,n}}\rmv$. The likelihood contribution $\mathpzc{f}_1(\V{z}_{l} | \V{x}_{k}, \V{e}_{k})$ ``along direction $\lambda_{1,\M{E}_{k,n}}\rmv\rmv$'' can now be written\vspace{0mm} as 
\begin{align}
 &\mathpzc{f}_1(\V{z}_{l} | \V{x}_{k}, \V{e}_{k}) =  \begin{cases}
   1- \int^{\infty}_{d^{(2)}_{1,\V{z}_l}}  \mathcal{N}(v^{(l)}_{k};0,\sigma^2_{u_{l}} ) \ist \mathrm{d} v^{(l)}_k - \int^{\infty}_{d^{(1)}_{1,\V{z}_l}}  \mathcal{N}(v^{(l)}_{k};0,\sigma^2_{u_{l}} ) \ist \mathrm{d} v^{(l)}_k, &  \quad \lambda_{1,\M{E}_{k,n}} > d^{(1)}_{1,\V{z}_l}, d^{(2)}_{1,\V{z}_l} \\[3mm]
  \hspace{6.5mm}  \int^{\infty}_{d^{(1)}_{1,\V{z}_l}}  \mathcal{N}(v^{(l)}_{k};0,\sigma^2_{u_{l}} ) \ist \mathrm{d} v^{(l)}_k - \int^{\infty}_{d^{(2)}_{1,\V{z}_l}}  \mathcal{N}(v^{(l)}_{k};0,\sigma^2_{u_{l}} ) \ist \mathrm{d} v^{(l)}_k, & \quad d^{(2)}_{1,\V{z}_l} \geq \lambda_{1,\M{E}_{k,n}}, d^{(1)}_{1,\V{z}_l} \\[3mm]
     \hspace{6.5mm} \int^{\infty}_{d^{(2)}_{1,\V{z}_l}}  \mathcal{N}(v^{(l)}_{k};0,\sigma^2_{u_{l}} ) \ist \mathrm{d} v^{(l)}_k - \int^{\infty}_{d^{(1)}_{1,\V{z}_l}}  \mathcal{N}(v^{(l)}_{k};0,\sigma^2_{u_{l}} ) \ist  \mathrm{d} v^{(l)}_k, & \quad d^{(1)}_{1,\V{z}_l} \geq  \lambda_{1,\M{E}_{k,n}}, d^{(2)}_{1,\V{z}_l}. \nn
   \end{cases} \\[-1mm]
   \label{eq:likelihood3D}  \\[-8mm]
   \nn
\end{align}
Here, the first line is the probability that $\V{z}_l$ is located between the two edges along direction $\lambda_{1,\M{E}_{k,n}}$ and lines 2 and 3 are the probabilities that $\V{z}_l$ is not located between the two edges and closer to edge one or closer to edge two, respectively. A closed-form expression for the integrations in \eqref{eq:likelihood3D} can again be obtained by means of the Q-function\vspace{-4mm}, i.e.,
\begin{align}
 &\mathpzc{f}_1(\V{z}_{l} | \V{x}_{k}, \V{e}_{k}) =  \begin{cases}
   1- Q\bigg(\frac{d^{(1)}_{1,\V{z}_l}}{\sigma_{u_{l}}}\bigg) - Q\bigg(\frac{d^{(2)}_{1,\V{z}_l}}{\sigma_{u_{l}}}\bigg), &  \lambda_{1,\M{E}_{k,n}} > d^{(1)}_{1,\V{z}_l}, d^{(2)}_{1,\V{z}_l} \\[3mm]
   Q\bigg(\frac{d^{(1)}_{1,\V{z}_l}}{\sigma_{u_{l}}}\bigg) - Q\bigg(\frac{d^{(2)}_{1,\V{z}_l}}{\sigma_{u_{l}}}\bigg), & d^{(2)}_{1,\V{z}_l} \geq \lambda_{1,\M{E}_{k,n}}, d^{(1)}_{1,\V{z}_l}\\[3mm]
   Q\bigg(\frac{d^{(2)}_{1,\V{z}_l}}{\sigma_{u_{l}}}\bigg) - Q\bigg(\frac{d^{(1)}_{1,\V{z}_l}}{\sigma_{u_{l}}}\bigg), & d^{(1)}_{1,\V{z}_l} \geq  \lambda_{1,\M{E}_{k,n}}, d^{(2)}_{1,\V{z}_l}. \nn
   \end{cases}\nn \\[-5mm]
   \nn
\end{align}

Now, we can introduce $\mathpzc{f}_2(\V{z}_{l} | \V{x}_{k}, \V{e}_{k})$ and $\mathpzc{f}_3(\V{z}_{l} | \V{x}_{k}, \V{e}_{k})$ based on remaining eigenvalues $\lambda_{2,\M{E}_{k,n}}$, $\lambda_{3,\M{E}_{k,n}}$ and eigenvectors $\V{\lambda}_{2,\M{E}_{k,n}}$, $\V{\lambda}_{3,\M{E}_{k,n}}$ of $\M{E}_{k,n}$. The final closed form of the likelihood function for the case where $\V{v}_l$ is uniformly distributed on cube $\Set{S}_{\mathrm{c}}(\M{E}_{k})$ and the measurement noise $\V{u}_l$ is iid Gaussian, then reads
\begin{align}
f(\V{z}_{l} | \V{x}_{k}, \V{e}_{k})  = \frac{1}{| \Set{S}_{\mathrm{c}}(\M{E}_{k}) |} \ist\ist \mathpzc{f}_1(\V{z}_{l} | \V{x}_{k}, \V{e}_{k}) \ist\ist \mathpzc{f}_2(\V{z}_{l} | \V{x}_{k}, \V{e}_{k}) \ist\ist \mathpzc{f}_3(\V{z}_{l} | \V{x}_{k}, \V{e}_{k}). \nn\\[-7mm]
\nn
 \end{align}

\section{Implementation of the Proposed EOT Method}
Pseudocode for one time step of the proposed EOT method is provided in Algorithm~\ref{al:totalEOT1}. This pseudocode closely follows the presentation of the particle-based implementation discussed in \cite[Section~V]{MeyWil:J21}. MATLAB code is available on \texttt{fmeyer.ucsd.edu}.

\vspace{-5mm}
\begin{algorithm}
\SetAlgoLined
\tiny
\vspace{.8mm}

$\bigg[\Big\{  \big\{ \big( \V{x}^{(j)}_{k}\rmv\rmv, \V{e}^{(j)}_{k}\rmv\rmv, w^{(j)}_{k} \big) \big\}_{j=1}^{J} \Big\}_{k=1}^{\underline{K} +M} \bigg]= \text{EOT}\Big[ \Big\{  \big\{ \big( \V{x}^{-(j)}_{k}\rmv\rmv, \V{e}^{-(j)}_{k}\rmv\rmv, w^{-(j)}_{k} \big) \big\}_{j=1}^{J} \Big\}_{k=1}^{\underline{K}}\rmv, \big\{\V{z}_l \big\}_{l=1}^{M} \Big]$\\
\vspace{1mm}

\For{$k = 1 : \underline{K}$}{ \hspace{120mm} \bl{\tcp{\tiny prediction for legacy POs}}
\vspace{-1mm}
  \For{$j = 1 : J$}{
  \vspace{-.8mm}
     Compute $\underline{w}^{(1,j)}_{k} \rmv= p_{\mathrm{s}} \ist\ist \mathpzc{e}^{-\mu_{\mathrm{m}}\big(\underline{\V{x}}^{(j)}_{k}\rmv, \ist\underline{\V{e}}^{(j)}_{k}\big)} \ist w^{-(j)}_{k}$ and set $\underline{w}^{(1,j)}_{kl} \rmv\triangleq \underline{w}^{(1,j)}_{k}\rmv, l \rmv\in\rmv \{1,\dots,M\}$ 
  } 
  Calculate $p^{-\text{e}}_{k} = \sum^{J}_{j=1} w^{-(j)}_{k}$; $\tilde{\underline{\alpha}}^{\mathrm{n}}_k = \big(1-p^{-\text{e}}_{k}\big) + \big(1\!-\rmv p_{\ist\mathrm{s}}\big) p^{-\text{e}}_{k}$; and $\tilde{\underline{\alpha}}_k = \sum^{J}_{j\rmv=\rmv1} \underline{w}^{(1,j)}_{k} + \tilde{\underline{\alpha}}^{\mathrm{n}}_k$ \\[.5mm]
  Set $\tilde{\underline{\alpha}}^{\mathrm{n}(p)}_{kl}  \triangleq \tilde{\underline{\alpha}}^{\mathrm{n}}_k $, $l \rmv\in\rmv \{1,\dots,M\}$ and $\tilde{\underline{\alpha}}^{(p)}_{kl}  \triangleq \tilde{\underline{\alpha}}_k $, $l \rmv\in\rmv \{1,\dots,M\}$
}
\vspace{1mm}

\For{$k = 1 : M$}{ \hspace{120mm} \bl{\tcp{\tiny initialization of new POs}}
\vspace{-1mm}
    \For{$j = 1 : J$}{ 
      Draw $\Big(\overline{\boldsymbol{x}}_k^{(j)}\rmv\rmv, \overline{\boldsymbol{e}}_k^{(j)} \Big) \sim f_{\mathrm{p}} \rmv\big(\overline{\boldsymbol{x}}_k,\overline{\boldsymbol{e}}_k;\V{z}_k\big)$,
      compute $\overline{w}^{(1,j)}_{k} \rmv= \frac{q \rmv\Big(\overline{\boldsymbol{x}}_k^{(j)}\rmv\rmv, \overline{\boldsymbol{e}}_k^{(j)}\rmv\rmv, \overline{r}_k = 1 \Big)}{f_{\mathrm{p}} \rmv\Big(\overline{\boldsymbol{x}}_k^{(j)}\rmv\rmv, \overline{\boldsymbol{e}}_k^{(j)};\V{z}_l\Big)}$, and set $\overline{w}^{(1,j)}_{kl} \rmv\triangleq \overline{w}^{(1,j)}_{k}\rmv, l \rmv\in\rmv \{k,\dots,M\}$ 
    } 
  Calculate $\tilde{\overline{\alpha}}_k = \sum^{J}_{j\rmv=\rmv1} \overline{w}^{(1,j)}_{k} + 1$ \\[.5mm]
    Set $\tilde{\overline{\alpha}}^{\hspace{.1mm}\mathrm{n}(1)}_{kl}  \triangleq 1$, $l \rmv\in\rmv \{k\dots M\}$ and $\tilde{\overline{\alpha}}^{(1)}_{kl}  \triangleq \tilde{\overline{\alpha}}_k $, $l \rmv\in\rmv \{k\dots M\}$
}
\vspace{1mm}

\For{$p = 1 : P$}{
\vspace{1mm}

\For{$k = 1 : \underline{K}$}{ \hspace{89mm} \bl{\tcp{\tiny perform measurement evaluation for legacy POs}}
\vspace{-2mm}
  \For{$l = 1 : M$}{
  \vspace{1mm}
    Calculate $\tilde{\underline{\beta}}_{kl}^{(p)} = \frac{1}{\mu_{\mathrm{fa}} \ist f_{\mathrm{fa}}(\V{z}_{l}) \ist \tilde{\underline{\alpha}}_{kl}^{(p)} } \hspace{.5mm} \sum^{J}_{j=1} \underline{w}^{(p,j)}_{kl} \mu_{\mathrm{m}}\Big(\underline{\V{x}}^{(j)}_k,\underline{\V{e}}^{(j)}_k\Big) \ist f\Big(\V{z}_{l}| \underline{\V{x}}^{(j)}_k, \underline{\V{e}}^{(j)}_{k}\Big)$ 
      } 
}
\vspace{1.5mm}

\For{$k = 1 : M$}{\hspace{93mm} \bl{\tcp{\tiny perform measurement evaluation for new POs}}
\vspace{-1.5mm}
Compute $\tilde{\overline{\beta}}_{kk}^{(p)} = \frac{1}{\mu_{\mathrm{fa}} \ist f_{\mathrm{fa}}(\V{z}_{k}) \ist \tilde{\overline{\alpha}}_{kk}^{\hspace{.15mm}\mathrm{n}(p)} } \hspace{.5mm} \sum^{J}_{j=1}  \overline{w}^{(p,j)}_{kk} \mu_{\mathrm{m}}\Big(\overline{\V{x}}^{(j)}_k\rmv,\overline{\V{e}}^{(j)}_k\Big) \ist f\Big(\V{z}_{l}| \overline{\V{x}}^{(j)}_k\rmv, \overline{\V{e}}^{(j)}_{k}\Big)$
 \vspace{1.5mm}
 
  \For{$l = k\rmv+\rmv1 : M$}{
  \vspace{1mm}
    Calculate $\tilde{\overline{\beta}}_{kl}^{(p)} = \frac{1}{\mu_{\mathrm{fa}} \ist f_{\mathrm{fa}}(\V{z}_{l}) \ist \tilde{\overline{\alpha}}_{kl}^{(p)} } \hspace{.5mm} \sum^{J}_{j=1}  \overline{w}^{(p,j)}_{kl} \mu_{\mathrm{m}}\Big(\overline{\V{x}}^{(j)}_k,\overline{\V{e}}^{(j)}_k \Big) \ist f\Big(\V{z}_{l}| \overline{\V{x}}^{(j)}_k, \overline{\V{e}}^{(j)}_{k}\Big)$  } 
}
\vspace{1mm}

\For{$k = 1 : \underline{K}$}{  \hspace{74mm} \bl{\tcp{\tiny combine measurement evaluation information for legacy POs}}
\vspace{-1mm}

    Compute $\tilde{\underline{\xi}}_{kl}^{(p)} \triangleq \sum^{\underline{K}}_{\substack{k'=1\\k' \neq k}} \ist \tilde{\underline{\beta}}_{k'l}^{(p)} + \sum^{l}_{\substack{\ell=1}} \ist \tilde{\overline{\beta}}_{\ell l}^{(p)} + 1$
}
\vspace{1mm}

\For{$k = 1 : M$}{  \hspace{78mm} \bl{\tcp{\tiny combine measurement evaluation information for new POs}}
\vspace{-1mm}

Calculate $\tilde{\overline{\xi}}_{kl}^{(p)} \triangleq \sum^{\underline{K}}_{\substack{k'=1}} \ist \tilde{\underline{\beta}}_{k'l}^{(p)} + \sum^{l}_{\substack{\ell=1\\[.5mm] \ell \neq k}} \ist \tilde{\overline{\beta}}_{\ell l}^{(p)} + 1$
}
\vspace{1mm}

\If{$p < P$}{ \hspace{66.5mm} \bl{\tcp{\tiny calculate extrinsic information messages for legacy and new POs}}
\vspace{-2mm}
\For{$k = 1 : \underline{K}$}{
\For{$l = 1 : M$}{
Compute $\underline{w}^{(p\hspace{.1mm}+\hspace{.1mm}1,j)}_{kl} =\ist \underline{w}^{(1,j)}_{k} \hspace{.5mm} \prod^{M}_{\substack{l'=1\\l' \neq l}} \rmv\rmv \Bigg(  \frac{\mu_{\mathrm{m}}\Big(\underline{\V{x}}^{(j)}_{k}\rmv\rmv,\underline{\V{e}}^{(j)}_{k}\Big) f\Big(\V{z}_{l'}| \underline{\V{x}}^{(j)}_k\rmv\rmv, \underline{\V{e}}^{(j)}_{k}\Big)}{\mu_{\mathrm{fa}}  f_{\mathrm{fa}}\big(\V{z}_{l'}\big)} \ist + \ist \tilde{\underline{\xi}}_{kl'}^{(p)} \rmv \Bigg)$ \\[1.2mm]

Calculate $\tilde{\underline{\alpha}}^{\mathrm{n}(p+1)}_{kl}  =\ist \tilde{\underline{\alpha}}^{\mathrm{n}}_k \hspace{.5mm} \prod^{M}_{\substack{l'=1\\l' \neq l}}  \tilde{\underline{\xi}}_{kl'}^{(p)}$ and $\tilde{\underline{\alpha}}^{(p+1)}_{kl} = \sum^{J}_{j\rmv=\rmv1} \underline{w}^{(p+1,j)}_{kl} + \tilde{\underline{\alpha}}^{\mathrm{n}(p+1)}_{kl} $
}
}

\For{$k = 1 : M$}{
\For{$l = k : M$}{
Compute $\overline{w}^{(p\hspace{.1mm}+\hspace{.1mm}1,j)}_{kl} =\ist \overline{w}^{(1,j)}_{k} \frac{\mu_{\mathrm{m}}\Big(\overline{\V{x}}^{(j)}_{k}\rmv\rmv,\overline{\V{e}}^{(j)}_{k}\Big) f\Big(\V{z}_{k}| \overline{\V{x}}^{(j)}_k\rmv\rmv, \overline{\V{e}}^{(j)}_{k}\Big)}{\mu_{\mathrm{fa}}  f_{\mathrm{fa}}\big(\V{z}_{k}\big)} \hspace{.5mm} \prod^{M}_{\substack{l'=k+1\\l' \neq l}} \rmv\rmv \Bigg(  \frac{\mu_{\mathrm{m}}\Big(\overline{\V{x}}^{(j)}_{k}\rmv\rmv,\overline{\V{e}}^{(j)}_{k}\Big) f\Big(\V{z}_{l'}| \overline{\V{x}}^{(j)}_k\rmv\rmv, \overline{\V{e}}^{(j)}_{k}\Big)}{\mu_{\mathrm{fa}}  f_{\mathrm{fa}}\big(\V{z}_{l'}\big)} \ist + \ist \tilde{\overline{\xi}}_{kl'}^{(p)} \rmv \Bigg)$ \\[1.2mm]

Calculate $\tilde{\overline{\alpha}}^{\hspace{.1mm}\mathrm{n}(p+1)}_{kl}  =\ist \tilde{\overline{\alpha}}^{\mathrm{n}}_k \hspace{.5mm} \prod^{M}_{\substack{l'=k\\l' \neq l}} \rmv \tilde{\overline{\xi}}_{kl'}^{(p)}$ and $\tilde{\overline{\alpha}}^{(p+1)}_{kl} = \sum^{J}_{j\rmv=\rmv1} \overline{w}^{(p+1,j)}_{kl} + \tilde{\overline{\alpha}}^{\hspace{.1mm}\mathrm{n}(p+1)}_{kl} $
}
}
}
\vspace{1mm}

\If{$p = P$}{ \hspace{95.3mm} \bl{\tcp{\tiny calculate beliefs for legacy and new POs}}
\vspace{-2mm}

\For{$k = 1 : \underline{K}$}{
Compute $\underline{w}^{\text{A}(j)}_{k} =\ist \underline{w}^{(1,j)}_{k} \hspace{.5mm} \prod^{M}_{\substack{l=1}} \rmv\rmv \Bigg(  \frac{\mu_{\mathrm{m}}\Big(\underline{\V{x}}^{(j)}_{k}\rmv\rmv,\underline{\V{e}}^{(j)}_{k}\Big) f\Big(\V{z}_{l}| \underline{\V{x}}^{(j)}_k\rmv\rmv, \underline{\V{e}}^{(j)}_{k}\Big)}{\mu_{\mathrm{fa}}  f_{\mathrm{fa}}\big(\V{z}_{l}\big)} \ist + \ist \tilde{\underline{\xi}}_{kl}^{(P)} \rmv \Bigg)$ and $\underline{w}^{\text{B}}_{k} =\ist \tilde{\underline{\alpha}}^{\mathrm{n}}_k \hspace{.8mm} \prod^{M}_{\substack{l=1}}  \tilde{\underline{\xi}}_{kl}^{(P)}$ \\[0mm]

Calculate $\underline{w}^{(j)}_{k} \ist=\ist \frac{\underline{w}^{\text{A}(j)}_{k}}{\underline{w}^{\text{B}}_{k}+\sum^{J}_{j'=1} \ist \underline{w}^{\text{A}(j')}_{k}}$
}

\For{$k = 1 : M$}{ \hspace{93mm} 
\vspace{-1mm}

Compute $\overline{w}^{\hspace{.1mm}\text{A}(j)}_{k} =\ist \overline{w}^{(1,j)}_{k} \ist \frac{\mu_{\mathrm{m}}\Big(\overline{\V{x}}^{(j)}_{k}\rmv\rmv,\overline{\V{e}}^{(j)}_{k}\Big) f\Big(\V{z}_{k}| \overline{\V{x}}^{(j)}_k\rmv\rmv, \overline{\V{e}}^{(j)}_{k}\Big)}{\mu_{\mathrm{fa}}  f_{\mathrm{fa}}\big(\V{z}_{k}\big)} \hspace{.7mm} \prod^{M}_{\substack{l=k+1}} \rmv\rmv \Bigg(  \frac{\mu_{\mathrm{m}}\Big(\overline{\V{x}}^{(j)}_{k}\rmv\rmv,\overline{\V{e}}^{(j)}_{k}\Big) f\Big(\V{z}_{l}| \overline{\V{x}}^{(j)}_k\rmv\rmv, \overline{\V{e}}^{(j)}_{k}\Big)}{\mu_{\mathrm{fa}}  f_{\mathrm{fa}}\big(\V{z}_{l}\big)} \ist + \ist \tilde{\overline{\xi}}_{kl}^{(P)} \rmv \Bigg)$ and $\overline{w}^{\text{B}}_{k} =\ist \tilde{\overline{\alpha}}^{\mathrm{n}}_k \hspace{.8mm} \prod^{M}_{\substack{l=k}}  \tilde{\overline{\xi}}_{kl}^{(P)}$ \\[0mm]

Calculate $\overline{w}^{(j)}_{k} \ist=\ist \frac{\overline{w}^{\hspace{.15mm}\text{A}(j)}_{k}}{\overline{w}^{\text{B}}_{k}+\sum^{J}_{j'=1} \ist \overline{w}^{\hspace{.15mm}\text{A}(j')}_{k}}$
}

}
\vspace{1mm}

}
\vspace{1mm}

Output: $\Big\{  \big\{ \big( \V{x}^{(j)}_{k}\rmv\rmv, \V{e}^{(j)}_{k}\rmv\rmv, w^{(j)}_{k} \big) \big\}_{j=1}^{J} \Big\}_{k=1}^{\underline{K} +M}$

\caption{Proposed Particle-Based EOT Method --- Single Time Step}
\label{al:totalEOT1}
\end{algorithm}

\pagebreak

\bibliographystyle{IEEEtran}
\bibliography{IEEEabrv,StringDefinitions,Books,Papers,Temp}

%% file: Bios/FlorianMeyerBio.tex
(S'12--M'15) received the Dipl.-Ing.\ (M.Sc.)\ and Ph.D.\ degrees (with highest honors) in electrical engineering from TU Wien, Vienna, Austria in 2011 and 2015, respectively. 
He is an Assistant Professor with the University of California San Diego, La Jolla, CA, jointly between the Scripps Institution of Oceanography and the Electrical and Computer Engineering Department. From 2017 to 2019 he was a Postdoctoral Fellow and Associate with Laboratory for Information \& Decision Systems at Massachusetts Institute of Technology, Cambridge, MA, and from 2016 to 2017 he was a Research Scientist with NATO Centre for Maritime Research and Experimentation, La Spezia, Italy. 

Dr. Meyer was a keynote speaker at the IEEE Aerospace Conference in 2020. He served on the technical program committees of several IEEE conferences and as a co-chair of the IEEE Workshop on Advances in Network Localization and Navigation at the IEEE International Conference on Communications in 2018, 2019, and 2020. He is an Associate Editor for the IEEE Transactions on Aerospace and Electronic Systems and the ISIF Journal of Advances in Information Fusion as well as an Erwin Schr{\"o}dinger Fellow.

%% file: Bios/JasonWilliams.tex
(S’01–M’07–SM’16) received degrees of BE(Electronics)/BInfTech from Queensland University of Technology, MSEE from the United States Air Force Institute of Technology, and PhD from Massachusetts Institute of Technology. He is currently a Senior Research Scientist in Robotic Perception at the Robotics and Autonomous Systems Group of Commonwealth Scientific and Industrial Research Organisation, Brisbane, Australia. His research interests include SLAM, computer vision, multiple object tracking and motion planning. He previously worked in sensor fusion and resource management as a Senior Research Scientist at the Defence Science and Technology Group, Australia, and in electronic warfare as an engineering officer in the Royal Australian Air Force.